\documentclass[12pt,a4paper]{article}

\usepackage[a4paper,margin=2.3cm]{geometry}

\usepackage[utf8]{inputenc}
\usepackage[T1]{fontenc}
\usepackage{setspace}
\usepackage{indentfirst}
\usepackage{placeins}
\usepackage{microtype}
\usepackage{enumitem}

\usepackage{amsmath,amssymb,amsfonts}
\usepackage{bm}

\usepackage{graphicx}
\usepackage{caption}
\usepackage{subcaption}
\usepackage{float}

\usepackage{booktabs}
\usepackage{array}
\usepackage{multirow}
\usepackage{threeparttable}
\usepackage{makecell}
\usepackage{longtable}

\usepackage{xcolor}

\usepackage[numbers,sort&compress]{natbib}

\usepackage{algorithm}
\usepackage{algpseudocode}

\usepackage[hidelinks]{hyperref}
\usepackage{url}

\newlength{\AlgIOIndent}
\newcommand{\AlgIO}[2]{%
	\Statex
	\hspace*{\dimexpr-\labelwidth-\labelsep\relax}%
	\begingroup
	\settowidth{\AlgIOIndent}{\textbf{#1:}\ }%
	\parbox[t]{\dimexpr\linewidth+\labelwidth+\labelsep\relax}{%
		\hangindent=\AlgIOIndent
		\hangafter=1
		\noindent\textbf{#1:} #2\par
	}%
	\endgroup
}

\newcommand{\keywordname}{\textbf{\textit{Keywords}}}
\newcommand{\keywords}[1]{%
    \par
    \addvspace{\medskipamount}%
    {%
        \rightskip=0pt plus 1cm
        \def\and{%
            \ifhmode\unskip\nobreak\fi
            \ $\cdot$ %
        }%
        \noindent
        \keywordname\enspace
        \ignorespaces#1\par
    }%
}

\makeatletter
\newcommand{\@toptitlebar}{%
    \hrule height 2pt
    \vskip 0.25in
    \vskip -\parskip
}
\newcommand{\@bottomtitlebar}{%
    \vskip 0.29in
    \vskip -\parskip
    \hrule height 2pt
    \vskip 0.09in
}
\renewcommand{\@maketitle}{%
    \vbox{%
        \hsize\textwidth
        \linewidth\hsize
        \vskip 0.1in
        \@toptitlebar
        \centering
        {\fontsize{18}{22}\selectfont
         \scshape
         \@title\par}
        \@bottomtitlebar
        \vskip 0.1in
        \begin{tabular}[t]{>{\centering\arraybackslash}p{0.94\textwidth}}
            \bfseries \@author
        \end{tabular}
        \vskip 0.4in
    }
}
\makeatother

\title{Variable-Projection Sparse Functional Principal Component Analysis: Interpretable Functional Dimensionality Reduction with Applications to Raman Spectral Data}

\author{
	Han Ying Lim\textsuperscript{1},
	Dharini Pathmanathan\textsuperscript{1,2,*},
	Philipp Otto\textsuperscript{1,3}
	and Sophie Dabo-Niang\textsuperscript{4}
	\\[0.5em]
	\textsuperscript{1}Institute of Mathematical Sciences, Faculty of Science, Universiti Malaya, 50603 Kuala Lumpur, Malaysia\\
	\textsuperscript{2}Centre of Research for Statistical Modelling and Methodology, Faculty of Science, Universiti Malaya, 50603 Kuala Lumpur, Malaysia\\
	\textsuperscript{3}School of Mathematics and Statistics, University of Glasgow, University Place, Glasgow, G12 8QQ, United Kingdom\\
	\textsuperscript{4}Univ. Lille, CNRS, UMR 8524 -- Laboratoire Paul Painlev\'e, Inria-Datavers, F-59000 Lille, France\\
	\textsuperscript{*}Corresponding author: \href{mailto:dharini@um.edu.my}{dharini@um.edu.my}
}

\begin{document}

\maketitle



\begin{abstract}
	Functional principal component analysis (FPCA) provides low-rank representations of functional data but generally produces dense components, making it difficult to identify the localised regions contributing to dominant modes of variation. This limitation is particularly relevant in Raman spectroscopy, where spectra are observed over an ordered domain and interpretation often focuses on chemically meaningful spectral regions. This study proposes variable-projection sparse FPCA (VP--SFPCA) for interpretable functional dimensionality reduction through sparse weight functions that promote localisation. The method formulates sparse FPCA as a regularised matrix-factorisation problem that incorporates the functional inner-product geometry and distinguishes sparse weight functions used to generate component scores from orthonormal loading functions used for reconstruction. Variable projection conditionally minimises over the loading functions, reducing the optimisation problem to the sparse weights. Performance was evaluated through simulation studies and empirical analyses of surface-enhanced Raman scattering (SERS) spectra, with conventional FPCA and SCAD--SFPCA serving as dense and sparse functional benchmarks, respectively. In the simulations, VP--SFPCA recovered localised functional structure while requiring substantially less computation than SCAD--SFPCA. In the empirical analysis, VP--SFPCA retained this computational advantage while yielding held-out reconstruction error close to that of conventional FPCA. Several prominent features of the estimated weight functions also coincided with established adenine SERS bands. Overall, VP--SFPCA provides a computationally practical approach to improving the interpretability of dominant functional modes through localisation.
\end{abstract}

\keywords{dimensionality reduction \and functional data analysis \and principal component analysis \and Raman spectroscopy \and variable projection}


\section{Introduction}
\label{sec:intro}

Functional observations arise in diverse scientific settings in which measurements are recorded over an ordered continuum, such as time, wavelength, or frequency. Although such observations may be treated as high-dimensional vectors, this representation does not explicitly account for smoothness, ordering, or local structure across the domain. Functional data analysis (FDA) preserves these features by representing each observational unit as a function, commonly through a finite system of smooth basis functions \citep{ramsay2005,shang2014}. Dimensionality reduction within this framework aims to summarise the dominant modes of variation while retaining as much relevant information as possible from the functional observations.

Functional principal component analysis (FPCA) extends multivariate principal component analysis (PCA) by identifying orthonormal functional principal components (FPCs) that successively maximise variation among the observed curves \citep{ramsay2005,shang2014}. However, conventional FPCs are generally dense, with nonzero values throughout the functional domain \citep{chen2015,lin2016,nie2020}. Although such components provide effective low-dimensional representations, they may combine contributions from across the entire domain, making the resulting components difficult to interpret \citep{zou2018}. Consequently, identifying the regions driving each mode of variation may require subjective judgement as to which component values are sufficiently large to be considered important \citep{lin2016}. A commonly used approach is simple thresholding, in which component values with absolute magnitudes below a prespecified threshold are set to zero \citep{zou2018}. This \textit{ad hoc} procedure, however, may result in misleading interpretations \citep{cadima1995}.

Sparse methods address this limitation by seeking a balance between dimensionality reduction and interpretability through sparsity-inducing penalties or constraints, whereby negligible coefficients are shrunk towards zero and some may be set exactly to zero \citep{filzmoser2012,zou2018}. In the univariate functional setting considered here, sparse FPCA seeks functional directions with localised support, thereby identifying subregions of the functional domain associated with the dominant modes of variation. Several sparse FPCA formulations have been proposed to improve the interpretability of FPCs. Localised functional principal component analysis (LFPCA) \citep{chen2015} promotes compactly supported FPCs through a penalised eigendecomposition and a sequential Deflated Fantope Localization procedure based on convex relaxation, with orthogonality enforced across the estimated components. Its sequential estimation strategy, however, becomes increasingly computationally demanding as the retained rank grows \citep{nie2020}. Interpretable functional principal component analysis (iFPCA) \citep{lin2016} targets localisation by penalising the length of the nonzero regions, but the resulting $\ell_0$-type optimisation problem is NP-hard and therefore requires a greedy approximation. 

To address these limitations, \citet{nie2020} formulated sparse FPCA as a penalised functional regression problem using the functional smoothly clipped absolute deviation (SCAD) penalty of \citet{lin2017}. This approach, referred to hereafter as SCAD--SFPCA, estimates multiple sparse FPCs simultaneously and avoids the NP-hard optimisation problem. \citet{nie2020} reported favourable recovery relative to LFPCA and iFPCA when the true FPCs were supported only on subregions of the domain. Nevertheless, estimation relies on alternating updates, and the resulting sparse FPCs are not guaranteed to be orthogonal \citep{nie2020}. Although the additional shape parameter of the functional SCAD penalty is commonly fixed at $3.7$ following \citet{fan2001}, thereby avoiding an additional tuning dimension, the nonconvexity of the penalty may still increase the computational burden of optimisation \citep{li2026}.

The sparse FPCA literature has also developed along several related directions. \citet{allen2019} combined sparsity and smoothness within a regularised singular value decomposition (SVD) framework, while \citet{li2016} incorporated auxiliary variables to guide the estimation of supervised sparse and smooth components. For high-dimensional multivariate functional processes, \citet{hu2022} proposed a thresholding-based estimator, whereas \citet{song2022} introduced group sparsity to select the functional variables represented in each component. \citet{zhang2023} further developed a multilevel multivariate formulation in which components can be both sparse across variates and localised within the domain of each functional variable. In contrast to these penalty- or thresholding-based approaches, \citet{battagliola2025} obtained localised orthogonal eigenfunctions by decomposing the underlying process into uncorrelated subprocesses with disjoint supports. Collectively, these developments show that sparsity in FPCA may refer to localisation within a functional domain, selection across multiple functional variables, or component estimation guided by auxiliary information. The present study focuses on unsupervised within-domain localisation for a single functional variable. The supervised and multivariate formulations are therefore complementary developments rather than direct alternatives to the proposed method.

Beyond the choice of formulation used to induce sparsity, an important modelling question is whether sparsity is imposed on the weights or the loadings \citep{guerra2021,park2024}. In PCA-type problems, weights define the linear combinations used to obtain component scores, whereas loadings determine how those scores represent or reconstruct the original variables \citep{park2024}. In conventional PCA, weights and loadings represent the same principal directions and coincide with the right singular vectors of the centred data matrix, up to sign and conventional scaling \citep{jolliffe2016,park2024}. This mathematical equivalence generally no longer holds once sparsity is imposed \citep{park2024}. Sparsity in the weights or loadings may therefore lead to different interpretations \citep{camacho2025,park2024}. Analogously, in the functional setting, sparse weight functions identify regions of the functional domain that contribute to component scores, whereas orthonormal loading functions define the component directions used to reconstruct the observed functions.

Variable projection \citep{golub2003} provides a framework for separating these roles. For an objective involving two blocks of variables, one block is eliminated through conditional minimisation, yielding a reduced value function that is optimised over the remaining block \citep{erichson2020,golub2003}. Building on the regression-based sparse PCA formulation of \citet{zou2006}, \citet{erichson2020} partially minimised over the orthonormal loading matrix and optimised the resulting value function over the sparse weight matrix. At each evaluation of the value function, the loading matrix is updated through a closed-form orthogonal Procrustes solution, whereas the reduced objective can be optimised using proximal methods, including proximal gradient algorithms \citep{parikh2014}. This formulation reduces the optimisation problem while retaining the joint estimation of multiple sparse components, providing a computationally efficient alternative to the original regression-based sparse PCA algorithm \citep{erichson2020}.

However, the variable-projection sparse PCA framework of \citet{erichson2020} is formulated for multivariate observations under standard Euclidean geometry. For functional observations expanded in a nonorthonormal B-spline basis, the $L^2$ inner product is represented in coefficient space by the Gram matrix $\mathbf{W}$ of the basis functions \citep{happ2018,ramsay2005}. Accordingly, $\mathbf{W}$ is incorporated into the component score calculation, reconstruction criterion, and functional orthonormality constraint. A Cholesky factorisation of $\mathbf{W}$ \citep{higham2009} converts the weighted orthonormality constraint on the loading functions into an equivalent Euclidean form, allowing the loading update to be obtained through a standard Procrustes solution. The sparse weight coefficients are not subject to this orthonormality constraint and are retained in the original B-spline basis, where sparsity is imposed directly.

The choice of representation therefore affects how sparsity is interpreted. B-spline basis functions have compact local support, so consecutive zero coefficients in the original basis can induce localised zero regions in the corresponding weight functions \citep{deboor2001,nie2020}. Although the same geometry induced by the Gram matrix can be expressed in Euclidean coordinates through a Cholesky-based transformation \citep{happ2018}, the transformed coordinates generally differ from the original representation \citep{kessy2018}. The $\ell_1$ penalty acts coordinatewise on the coefficients \citep{tibshirani1996} and is generally not preserved under a change of coordinates. Sparse PCA formulations based on basis rotation similarly demonstrate that the resulting sparsity can depend on the chosen representation \citep{chen2024}. Hence, an ordinary coordinatewise $\ell_1$ penalty applied after transformation would, in general, impose a different sparsity structure from that defined on the original B-spline coefficients. Zeros in the transformed coordinates would not necessarily correspond to zeros in the original locally supported B-spline coefficients. Accordingly, the Gram matrix and its Cholesky factorisation are used to accommodate the functional geometry and weighted orthonormality of the loading functions, while the $\ell_1$ penalty is applied directly to the original B-spline weight coefficients to preserve the intended interpretation of localised sparsity.

Spectroscopy provides a suitable empirical setting in which such localised functional representations are scientifically meaningful. Raman spectroscopy provides molecularly specific information through inelastic light scattering, whereas surface-enhanced Raman scattering (SERS) enhances weak Raman signals from molecules located on or near nanostructured surfaces \citep{han2021}. Each spectrum records intensity over a densely sampled and ordered Raman-shift domain, producing high-dimensional and strongly correlated measurements \citep{bin2013,houhou2021}. Interpretation may be further complicated by overlapping bands, baseline variation, measurement noise, and experimental heterogeneity \citep{han2021}. Dimensionality reduction can therefore help summarise the dominant spectral variation while retaining the spectral regions relevant to chemical interpretation.

Although Raman spectra are commonly analysed as high-dimensional vectors, their ordered and smooth structure also supports a functional representation \citep{houhou2021}. Individual spectral bands extend across neighbouring Raman-shift measurements, making the continuity of the spectral domain relevant to chemical interpretation \citep{han2021,houhou2021}. In a comparison of functional and discrete representations of Raman spectra, \citet{houhou2021} found that B-spline-based FPCA was particularly beneficial in the presence of measurement noise and small shifts in peak position. These findings support FPCA as a promising framework for reducing the dimensionality of Raman spectra while preserving their ordered spectral structure. 

In spectroscopy, variable-selection approaches have identified informative wavenumbers using principal variable selection based on modified Gram--Schmidt procedures \citep{skogholt2023} and peak extraction from PCA eigenvectors \citep{schmitt2025}. Although these methods demonstrate the importance of isolating informative wavenumbers or spectral peaks, they primarily represent spectra as collections of discrete variables.

The present study develops variable-projection sparse FPCA (VP--SFPCA) primarily to improve interpretability through localised sparse weight functions, rather than to improve reconstruction accuracy relative to conventional FPCA. The proposed framework extends variable-projection sparse PCA \citep{erichson2020} to functional observations, with sparsity imposed on the weight functions. The performance of VP--SFPCA is evaluated through simulation studies and applications to SERS spectra, with conventional FPCA \citep{ramsay2005} and SCAD--SFPCA \citep{nie2020} serving as the dense and sparse functional benchmarks, respectively. The spectroscopic relevance of the estimated weight functions is further studied by relating their dominant features to established SERS band assignments.

The remainder of this paper is organised as follows. Section~\ref{sec:method} presents conventional FPCA and the proposed VP--SFPCA framework. Section~\ref{sec:sim} describes the simulation design and presents the corresponding results. Section~\ref{sec:app} introduces the empirical application, followed by a discussion of the empirical findings in Section~\ref{sec:results}. Finally, Section~\ref{sec:conclusion} provides concluding remarks and outlines directions for future research.


\section{Methodology}
\label{sec:method}

\subsection{Functional Principal Component Analysis}
\label{subsec:method-fpca}

Let $X$ be a stochastic process on a compact domain $\mathcal{T} \subset \mathbb{R}$ satisfying $\mathrm{E} \lVert X \rVert^2 < \infty$, where $L^2(\mathcal{T})$ denotes the Hilbert space of square-integrable functions on $\mathcal{T}$. This space is equipped with the inner product $\langle f,g\rangle = \int_{\mathcal{T}}f(t)g(t)\,\mathrm{d}t$ and the corresponding norm $\lVert f\rVert = \sqrt{\langle f,f\rangle}$ for $f,g \in L^2(\mathcal{T})$. Let $\mu(t) = \mathrm{E}\{X(t)\}$ denote the mean function and $\Gamma(s,t) = \operatorname{Cov}\{X(s),X(t)\}$ the covariance function. The covariance operator $\mathcal{C}:L^2(\mathcal{T})\rightarrow L^2(\mathcal{T})$ is defined by
\begin{equation}
	(\mathcal{C}f)(t)
	=
	\int_{\mathcal{T}}\Gamma(s,t)f(s)\,\mathrm{d}s.
	\label{eq:fpca-covariance-operator}
\end{equation}

Under the finite-second-moment condition, the operator $\mathcal{C}$ is self-adjoint, positive semidefinite, and trace-class, and is therefore compact \citep{ramsay2005,shang2014}. By the spectral theorem, it admits a sequence of orthonormal eigenfunctions $\{\varphi_k\}_{k\geq1}$, referred to as FPCs, with corresponding nonnegative eigenvalues satisfying $\nu_1\geq\nu_2\geq\cdots\geq0$ and
\begin{equation}
	\mathcal{C}\varphi_k = \nu_k \varphi_k,
	\qquad
	\langle \varphi_j, \varphi_k \rangle = \delta_{jk},
	\label{eq:fpca-covariance-eigenproblem}
\end{equation}
where $\delta_{jk}$ is the Kronecker delta.

By the Karhunen--Lo\`eve theorem, the $i$th realisation admits the mean-square expansion
\begin{equation}
	X_i(t)
	=
	\mu(t) + \sum_{k=1}^{\infty} \zeta_{ik} \varphi_k(t),
	\label{eq:fpca-kl-expansion}
\end{equation}
where the corresponding $k$th FPC score $\zeta_{ik} = \langle X_i - \mu, \varphi_k \rangle$ has mean $\mathrm{E}(\zeta_{ik}) = 0$ and covariance $\operatorname{Cov}(\zeta_{ij},\zeta_{ik}) = \nu_k \delta_{jk}$. For dimensionality reduction, the process is approximated by retaining the leading $K$ components, with the adequacy of the truncation assessed by the proportion of total variation retained. Truncating the Karhunen--Lo\`eve expansion gives the population rank-$K$ approximation
\begin{equation}
	X_i^{(K)}(t)
	=
	\mu(t) + \sum_{k=1}^{K} \zeta_{ik} \varphi_k(t).
	\label{eq:fpca-population-truncated-kl}
\end{equation}
The retained rank may be selected as the smallest value of $K$ for which the first $K$ components account for a prespecified proportion of the total variation, such as $70\%$, $80\%$, or higher \citep{jolliffe2016,shang2014}.

In practice, the population quantities are replaced by their empirical counterparts. Let $\widehat{\varphi}_k$ denote the $k$th orthonormal eigenfunction of the empirical covariance operator $\widehat{\mathcal{C}}$, and define the corresponding score as $\widehat{\zeta}_{ik} = \left\langle X_i - \widehat{\mu}, \widehat{\varphi}_k \right\rangle$. The empirical rank-$K$ reconstruction is then
\begin{equation}
	\widehat{X}_i^{(K)}(t) = \widehat{\mu}(t) + \sum_{k=1}^{K} \widehat{\zeta}_{ik} \widehat{\varphi}_k(t).
	\label{eq:fpca-empirical-reconstruction}
\end{equation}
Thus, in conventional FPCA, the same empirical eigenfunction serves as both the projection direction used to calculate a component score and the functional direction used for reconstruction \citep{ramsay2005}.

When functions are observed with noise on a discrete grid, smoothness may be introduced by smoothing the individual curves, smoothing the estimated covariance function, or incorporating a roughness penalty directly into FPC estimation \citep{nie2020}. An alternative formulation was proposed by \citet{silverman1996}, in which a roughness penalty is incorporated into the inner product used to define the orthonormality constraints. Although conventional FPCA efficiently captures the dominant modes of functional variation, its estimated component functions are generally dense over $\mathcal{T}$ and may therefore be difficult to interpret \citep{chen2015,lin2016,nie2020}. This limitation motivates the sparse formulations considered subsequently.

\subsection{Sparse Functional Principal Component Analysis via Variable Projection}
\label{subsec:method-vp-sfpca}

For finite-dimensional estimation, the functions are represented using a set of $p$ cubic B-spline basis functions $\boldsymbol{\psi}(t) = (\psi_1(t), \ldots, \psi_p(t))^{\top}$, that is,
\begin{equation}
	X_i(t) \approx \mathbf{d}_i^{\top} \boldsymbol{\psi}(t),
	\qquad i = 1, \ldots, n,
	\label{eq:vp-basis-representation}
\end{equation}
where $\mathbf{d}_i \in \mathbb{R}^{p}$ is the basis-coefficient vector. The centred coefficient matrix is given by $\mathbf{C} = \bigl(\mathbf{c}_1, \ldots, \mathbf{c}_n\bigr)^{\top} \in \mathbb{R}^{n\times p}$, where $\mathbf{c}_i = \mathbf{d}_i - \overline{\mathbf{d}}$ and $\overline{\mathbf{d}} = n^{-1}\sum_{i=1}^{n}\mathbf{d}_i$.

Cubic B-splines are piecewise polynomials of degree three with local support over at most four consecutive knot intervals \citep{deboor2001}. This local-support property allows coefficient sparsity to induce localised zero regions, since a represented function is identically zero on a subinterval whenever the coefficients of all basis functions overlapping that subinterval are zero \citep{deboor2001,ramsay2005}. As the B-spline basis is generally nonorthogonal \citep{deboor2001}, the functional inner-product geometry is represented through the Gram matrix
\begin{equation}
	\mathbf{W}
	=
	\int_{\mathcal{T}} \boldsymbol{\psi}(t) \boldsymbol{\psi}(t)^{\top}\,\mathrm{d}t.
	\label{eq:vp-gram-matrix}
\end{equation}
If $f(t) = \mathbf{f}^{\top} \boldsymbol{\psi}(t)$ and $g(t) = \mathbf{g}^{\top} \boldsymbol{\psi}(t)$, then $\langle f,g \rangle = \mathbf{f}^{\top} \mathbf{W} \mathbf{g}$.

Within this basis-induced geometry, the component scores are obtained by projecting the centred functions onto the weight functions, whereas the loading functions define the directions used to reconstruct the centred functions. The two sets of functions are estimated separately, with sparsity imposed on the weights and orthonormality imposed on the loadings. Let $\mathbf{A} = (\mathbf{a}_1,\ldots,\mathbf{a}_K) \in \mathbb{R}^{p\times K}$ and $\mathbf{B} = (\mathbf{b}_1, \ldots, \mathbf{b}_K) \in \mathbb{R}^{p\times K}$ denote the coefficient matrices of the loading functions $u_k(t) = \mathbf{a}_k^{\top} \boldsymbol\psi(t)$ and the sparse weight functions $\beta_k(t) = \mathbf{b}_k^{\top}\boldsymbol\psi(t)$, respectively. The loading coefficients satisfy the functional orthogonality constraint $\mathbf{A}^{\top} \mathbf{W} \mathbf{A} = \mathbf{I}_K$, whereas no such constraint is imposed on $\mathbf{B}$. The score matrix $\mathbf{Z}$ and reconstructed centred coefficient matrix $\widehat{\mathbf{C}}$ are computed as
\begin{equation}
	\mathbf{Z} = \mathbf{C} \mathbf{W} \mathbf{B},
	\qquad
	\widehat{\mathbf{C}} = \mathbf{Z} \mathbf{A}^{\top} = \mathbf{C} \mathbf{W} \mathbf{B} \mathbf{A}^{\top}.
	\label{eq:vp-score-reconstruction}
\end{equation}

The proposed method extends the variable-projection sparse PCA framework of \citet{erichson2020} to functional observations represented in a nonorthonormal basis. For sparsity parameter $\lambda\geq0$ and ridge parameter $\tau\geq0$, VP--SFPCA solves
\begin{equation}
	\min_{\mathbf{A},\mathbf{B}}\,
	\frac{1}{2n}
	\left\|
	\left(\mathbf{C} - \mathbf{C} \mathbf{W} \mathbf{B} \mathbf{A}^{\top}\right) \mathbf{W}^{1/2}
	\right\|_F^2
	+
	\lambda\|\mathbf{B}\|_1
	+
	\frac{\tau}{2} \left\|\mathbf{W}^{1/2} \mathbf{B}\right\|_F^2,
	\label{eq:vp-objective}
\end{equation}
subject to $\mathbf{A}^{\top} \mathbf{W} \mathbf{A} = \mathbf{I}_K$, where 
\begin{equation}
	\left\| \mathbf{W}^{1/2} \mathbf{B} \right\|_F^2
	=
	\operatorname{tr}\!\left(\mathbf{B}^{\top} \mathbf{W} \mathbf{B}\right)
	=
	\sum_{k=1}^{K}\|\beta_k\|_{L^2(\mathcal{T})}^2.
	\label{eq:vp-ridge-functional-norm}
\end{equation}
The $\ell_1$ term promotes sparsity in the B-spline coefficients, whereas the ridge term controls the functional norms of the weight functions and improves numerical stability. Together, they constitute an elastic-net-type regularisation \citep{zou2005} adapted to the functional inner-product geometry. The additional ridge term in the elastic net is particularly beneficial when the basis coefficients are strongly correlated and may reduce the tendency of an $\ell_1$-only penalty to retain too few nonzero coefficients \citep{filzmoser2012,zou2005}.

Defining the empirical covariance matrix of the centred basis coefficients $\mathbf{M} = n^{-1}\mathbf{C}^{\top}\mathbf{C}$ and $\mathbf{Q} = \mathbf{W} \mathbf{M} \mathbf{W}$, the criterion in Equation~\eqref{eq:vp-objective} can be expressed equivalently in trace form as
\begin{equation}
	\begin{aligned}
		J(\mathbf{A},\mathbf{B})
		&=
		\frac{1}{2}\operatorname{tr}(\mathbf{M}\mathbf{W})
		-
		\operatorname{tr}\!\left(\mathbf{B}^{\top}\mathbf{Q}\mathbf{A}\right)
		+
		\frac{1}{2}\operatorname{tr}\!\left(\mathbf{B}^{\top}\mathbf{Q}\mathbf{B}\right)
        +
		\lambda\|\mathbf{B}\|_1
		+
		\frac{\tau}{2}\operatorname{tr}\!\left(\mathbf{B}^{\top}\mathbf{W}\mathbf{B}\right),
	\end{aligned}
	\label{eq:vp-trace-objective}
\end{equation}
subject to $\mathbf{A}^{\top} \mathbf{W} \mathbf{A} = \mathbf{I}_K$. When $\lambda = \tau = 0$, the formulation reduces to the conventional rank-$K$ FPCA reconstruction problem. Its minimisers span the same principal functional subspace as conventional FPCA, up to sign and rotational indeterminacies \citep{babamoradi2013,bro2008,jolliffe2016}.

\subsubsection{Variable-projection optimisation}
\label{subsubsec:method-vp-optimisation}

Variable projection \citep{golub2003} partially minimises the criterion over the orthogonally constrained loading matrix, conditional on the sparse weight matrix \citep{erichson2020}. The conditional minimiser of $\mathbf{A}$ is obtained through SVD, after which $\mathbf{B}$ is updated using a proximal-gradient step \citep{parikh2014}.

Consistent with the value-function formulation of \citet{erichson2020}, define the value function
\begin{equation}
	\nu(\mathbf{B}) 
	:=
	\min_{\mathbf{A}} \frac{1}{2n}
	\left\|
	\left(\mathbf{C} - \mathbf{C} \mathbf{W} \mathbf{B} \mathbf{A}^{\top}\right) \mathbf{W}^{1/2}
	\right\|_F^2,
	\label{eq:vp-value-function}
\end{equation}
subject to $\mathbf{A}^{\top} \mathbf{W} \mathbf{A} = \mathbf{I}_K$. The reduced problem of Equation~\eqref{eq:vp-trace-objective} is therefore
\begin{equation}
	\min_{\mathbf{B}}
	\left\{
	\nu(\mathbf{B})
	+
	\frac{\tau}{2}
	\operatorname{tr}(\mathbf B^{\top}\mathbf W\mathbf B)
	+
	\lambda\lVert\mathbf B\rVert_1
	\right\},
	\label{eq:vp-reduced-problem}
\end{equation}
where the first two terms constitute the differentiable part of the objective, and the final term is the nonsmooth penalty treated through the proximal step of the proximal-gradient algorithm.

Since $\mathbf{Q}$ is symmetric, $\operatorname{tr}(\mathbf{B}^{\top} \mathbf{Q} \mathbf{A}) = \operatorname{tr}(\mathbf{A}^{\top} \mathbf{Q} \mathbf{B})$. For fixed $\mathbf{B}$, minimisation with respect to $\mathbf{A}$ therefore reduces to the weighted orthogonal Procrustes problem, that is,
\begin{equation}
	\max_{\mathbf{A}}
	\operatorname{tr} \left(\mathbf{A}^{\top} \mathbf{Q} \mathbf{B}\right)
	\qquad
	\text{subject to}
	\qquad
	\mathbf{A}^{\top} \mathbf{W} \mathbf{A} = \mathbf{I}_K.
	\label{eq:vp-procrustes-problem}
\end{equation}
Since $\mathbf{W}$ is symmetric positive definite, let $\mathbf{W} = \mathbf{L} \mathbf{L}^{\top}$ denote its Cholesky factorisation, where $\mathbf{L}$ is lower triangular with positive diagonal entries \citep{higham2009}. Define the transformed loading matrix as $\widetilde{\mathbf{A}} = \mathbf{L}^{\top} \mathbf{A}$. The functional orthogonality constraint is then equivalent to $\widetilde{\mathbf{A}}^{\top} \widetilde{\mathbf{A}} = \mathbf{I}_K$, and
\begin{equation}
	\operatorname{tr}
	\left(\mathbf{A}^{\top} \mathbf{Q} \mathbf{B}\right)
	=
	\operatorname{tr}
	\left[
	\widetilde{\mathbf{A}}^{\top} \left(\mathbf{L}^{-1} \mathbf{Q} \mathbf{B}\right)
	\right].
	\label{eq:vp-transformed-trace}
\end{equation}

Thus, the transformed Procrustes matrix is defined as
\begin{equation}
	\mathbf{H}(\mathbf{B}) 
	=
	\mathbf{L}^{-1} \mathbf{Q} \mathbf{B}.
	\label{eq:vp-transformed-procrustes-matrix}
\end{equation}
Let $\mathbf{H}(\mathbf{B}) = \mathbf{U} \boldsymbol{\Sigma}_{H} \mathbf{V}^{\top}$ denote the SVD of $\mathbf{H}(\mathbf{B})$, where $\mathbf{U} \in \mathbb{R}^{p \times K}$ and $\mathbf{V} \in \mathbb{R}^{K \times K}$. The solution in the transformed coordinates is therefore $\widetilde{\mathbf{A}} = \mathbf{U}\mathbf{V}^{\top}$, which gives the loading update
\begin{equation}
	\mathbf{A}(\mathbf{B})
	=
	\left(\mathbf{L}^{\top}\right)^{-1} \mathbf{U} \mathbf{V}^{\top},
	\label{eq:vp-A-update}
\end{equation}
satisfying $\mathbf{A}(\mathbf{B})^{\top} \mathbf{W} \mathbf{A}(\mathbf{B}) = \mathbf{I}_K$.

At iteration $r$, let $\mathbf{A}^{(r)} = \mathbf{A}(\mathbf{B}^{(r)})$. By partial minimisation, the gradient of the smooth reduced criterion with respect to $\mathbf{B}$ at the current iterate can be obtained by differentiating the smooth $\mathbf{B}$-subproblem with $\mathbf{A}^{(r)}$ held fixed. Consider
\begin{equation}
	q_r(\mathbf{B})
	=
	\frac{1}{2}
	\operatorname{tr} \left(\mathbf{B}^{\top}\mathbf{Q}\mathbf{B}\right)
	-
	\operatorname{tr} \left(\mathbf{B}^{\top}\mathbf{Q}\mathbf{A}^{(r)}\right)
	+ 
	\frac{\tau}{2} \operatorname{tr}
	\left(\mathbf{B}^{\top}\mathbf{W}\mathbf{B}\right).
	\label{eq:vp-B-smooth}
\end{equation}
The first and third terms are quadratic in $\mathbf{B}$ and determine the curvature of the subproblem, whereas the second term is linear in $\mathbf{B}$ and affects the gradient but not its curvature. Since $\mathbf{Q}$ and $\mathbf{W}$ are symmetric, differentiation gives
\begin{equation}
	\nabla q_r(\mathbf{B})
	=
	(\mathbf{Q} + \tau \mathbf{W}) \mathbf{B} - \mathbf{Q} \mathbf{A}^{(r)}.
	\label{eq:vp-B-gradient}
\end{equation}
Evaluating this gradient at the current iterate $\mathbf{B}^{(r)}$ yields
\begin{equation}
	\mathbf{G}^{(r)}
	=
	\nabla q_r(\mathbf{B}^{(r)})
	=
	\mathbf{Q} \left(\mathbf{B}^{(r)} - \mathbf{A}^{(r)}\right) + \tau \mathbf{W} \mathbf{B}^{(r)}.
	\label{eq:vp-gradient}
\end{equation}
	
For any $\mathbf{B}_1$ and $\mathbf{B}_2$, the matrix-norm inequality yields
\begin{align}
	\left\| \nabla q_r(\mathbf{B}_1) - \nabla q_r(\mathbf{B}_2) \right\|_F
	&=
	\left\| (\mathbf{Q}+\tau\mathbf{W}) (\mathbf{B}_1-\mathbf{B}_2) \right\|_F \nonumber \\
	&\leq
	\left\| \mathbf{Q}+\tau\mathbf{W} \right\|_2 \left\| \mathbf{B}_1-\mathbf{B}_2 \right\|_F,
	\label{eq:vp-lipschitz-derivation}
\end{align}
where $\|\cdot\|_2$ denotes the spectral norm. It follows that $\nabla q_r$ is Lipschitz continuous with constant
\begin{equation}
	\mathcal{L}
	=
	\left\| \mathbf{Q} + \tau \mathbf{W} \right\|_2.
	\label{eq:vp-lipschitz}
\end{equation}

Furthermore, $\mathbf{M}$ is positive semidefinite and $\mathbf{W}$ is positive definite, implying that $\mathbf{Q} = \mathbf{W} \mathbf{M} \mathbf{W}$ is positive semidefinite. Consequently, $\mathbf{Q} + \tau \mathbf{W}$ is positive semidefinite for $\tau \geq 0$, and the smooth part of the fixed-$\mathbf{A}^{(r)}$ $\mathbf{B}$-subproblem is a convex quadratic with a Lipschitz-continuous gradient. Hence, a fixed step size satisfying $0 < h \leq \mathcal{L}^{-1}$ is valid for the conditional proximal-gradient update of $\mathbf{B}$ \citep{parikh2014}. The loading matrix is subsequently recomputed through the Procrustes update, following the variable-projection scheme of \citet{erichson2020}. This step-size argument applies to the conditional $\mathbf{B}$-update.

The sparse weight coefficients are updated according to
\begin{equation}
	\mathbf{B}^{(r+1)}
	=
	\mathcal{S}_{h\lambda} \left(\mathbf{B}^{(r)} - h\mathbf{G}^{(r)}\right),
	\label{eq:vp-B-update}
\end{equation}
where $\mathcal{S}_{c}$ is the elementwise soft-thresholding operator. For a scalar $x$,
\begin{equation}
	\mathcal{S}_{c}(x)
	=
	\begin{cases}
		x-c, & x>c,\\
		0,   & |x|\leq c,\\
		x+c, & x<-c,
	\end{cases}
	\qquad
	=
	\operatorname{sign}(x)(|x|-c)_+,
	\qquad c\geq0.
	\label{eq:vp-soft-thresholding}
\end{equation}
After updating $\mathbf{B}$, the loading matrix is recomputed as $\mathbf{A}^{(r+1)} = \mathbf{A}(\mathbf{B}^{(r+1)})$.

As described in Algorithm~\ref{alg:vp-sfpca}, the weight matrix is initialised as $\mathbf{B}^{(0)} = \left(\mathbf{L}^{\top}\right)^{-1} \mathbf{V}_{K}$, where $\mathbf{V}_{K}$ contains the first $K$ right singular vectors of the transformed centred coefficient matrix $\mathbf{C}\mathbf{L}$ \citep{erichson2020,park2024}. The corresponding loading matrix $\mathbf{A}^{(0)}$ is obtained from the Procrustes update conditional on $\mathbf{B}^{(0)}$. The proximal weight and Procrustes loading updates are then repeated until the relative improvement in the objective value satisfies $0 \leq \Delta^{(r+1)} \leq \epsilon$, where $\epsilon$ is the tolerance threshold. The algorithm is considered converged when the criterion is satisfied, with a maximum number of iterations imposed as an additional computational limit.

\begin{algorithm}[htbp]
	\caption{VP--SFPCA for given tuning parameters $\lambda$ and $\tau$.}
	\label{alg:vp-sfpca}
	\begin{algorithmic}[1]
		\AlgIO{Input}{Centred coefficient matrix $\mathbf{C}$, Gram matrix $\mathbf{W}$, rank $K$, sparsity parameter $\lambda$, ridge parameter $\tau$, tolerance $\epsilon$, iteration cap $R_{\max}$}
		
		\AlgIO{Output}{Estimated loading matrix $\widehat{\mathbf{A}}$ and sparse weight matrix $\widehat{\mathbf{B}}$}
		
		\State Compute $\mathbf{M} = n^{-1}\mathbf{C}^{\top}\mathbf{C}$ and $\mathbf{Q} = \mathbf{W}\mathbf{M}\mathbf{W}$.
		
		\State Compute the Cholesky factorisation $\mathbf{W} = \mathbf{L}\mathbf{L}^{\top}$.
		
		\State Initialise $\mathbf{B}^{(0)} = (\mathbf{L}^{\top})^{-1}\mathbf{V}_{K}$, where $\mathbf{V}_{K}$ contains the first $K$ right singular vectors of $\mathbf{C}\mathbf{L}$.
		
		\State Compute $\mathbf{A}^{(0)} = \mathbf{A}(\mathbf{B}^{(0)})$ using Equation~\eqref{eq:vp-A-update}.
		
		\State Set $\mathcal{L} = \lVert\mathbf{Q}+\tau\mathbf{W}\rVert_{2}$ and $h = \mathcal{L}^{-1}$ and evaluate $J^{(0)} = J(\mathbf{A}^{(0)},\mathbf{B}^{(0)})$.
		
		\For{$r = 0,\ldots,R_{\max}-1$}
		
		\State Compute $\mathbf{G}^{(r)} = \mathbf{Q}\bigl(\mathbf{B}^{(r)}-\mathbf{A}^{(r)}\bigr) + \tau\mathbf{W}\mathbf{B}^{(r)}$.
		
		\State Update $\mathbf{B}^{(r+1)} = \mathcal{S}_{h\lambda} \left( \mathbf{B}^{(r)}-h\mathbf{G}^{(r)} \right)$.
		
		\State Compute $\mathbf{A}^{(r+1)} = \mathbf{A}(\mathbf{B}^{(r+1)})$ using the SVD-based Procrustes update.
		
		\State Evaluate $J^{(r+1)} = J(\mathbf{A}^{(r+1)},\mathbf{B}^{(r+1)})$.
		
		\State Compute $\displaystyle \Delta^{(r+1)} = \frac{J^{(r)}-J^{(r+1)}}{J^{(r+1)}}$.
		
		\If{$0 \leq \Delta^{(r+1)} \leq \epsilon$}
		\State \textbf{break}
		\EndIf
		
		\EndFor
		
		\State \Return $\widehat{\mathbf{A}}=\mathbf{A}^{(r+1)}$ and $\widehat{\mathbf{B}}=\mathbf{B}^{(r+1)}$.
		
	\end{algorithmic}
\end{algorithm}

The proposed method imposes sparsity on the B-spline coefficients rather than directly estimating the support of each weight function. Nevertheless, the local support of cubic B-spline basis functions allows consecutive zero coefficients to induce localised zero regions in the estimated weight functions.

\subsection{Tuning Parameter Selection}
\label{subsec:method-tuning}

Following \citet{nie2020}, a two-stage tuning strategy was adopted to select the smoothing or roughness parameter $\gamma$, the sparsity parameter $\lambda$, and the ridge parameter $\tau$. In the first stage, $\gamma$ was selected by minimising the mean generalised cross-validation (GCV) criterion \citep{ramsay2005}.

In the second stage, $\lambda$ and $\tau$ were selected jointly for each sparse FPCA method, conditional on the selected value of $\gamma$. Selection was based on the reconstruction Akaike information criterion (AIC) as described in \citet{nie2020}:
\begin{equation}
	\mathrm{AIC}
	=
	n\log\left(\frac{\mathrm{RSS}}{n\lvert\mathcal T\rvert}\right) + 2\,\mathrm{df},
	\label{eq:method-aic}
\end{equation}
where
\begin{equation}
	\mathrm{RSS}
	=
	\sum_{i=1}^{n} \left\lVert X_i^c - \widehat X_i^c \right\rVert_{L^2(\mathcal{T})}^2
	\label{eq:method-rss}
\end{equation}
is the total integrated reconstruction error for the centred training curves, $X_i^c$ denotes the $i$th centred curve, $\widehat X_i^c$ is its rank-$K$ reconstruction, and $\lvert\mathcal T\rvert$ denotes the length of the functional domain. The degrees of freedom, $\mathrm{df}$, were defined as the number of nonzero B-spline coefficients. For VP--SFPCA, the count was computed from the sparse weight matrix $\widehat{\mathbf B}$, whereas for SCAD--SFPCA it was computed from the coefficient matrix of the estimated sparse FPCs.

Candidate values were considered over logarithmically spaced grids, denoted by
\begin{equation}
	\mathcal G_\Omega(a,b)
	=
	\left\{10^{\,a + \omega(b-a)/(\Omega-1)}: \omega = 0, \ldots, \Omega-1\right\},
	\label{eq:method-log-grid}
\end{equation}
where $a$ and $b$ denote the lower and upper exponents, respectively, $\Omega$ is the number of candidate values, and $\omega$ indexes the grid points. Logarithmically spaced grids were used because regularisation parameters are typically examined over several orders of magnitude in penalised estimation problems, and this scale tends to perform well in practice \citep{schipper2021,friedman2010}.

For VP--SFPCA, the sparsity and ridge parameters specified on the tuning grids were internally scaled by the spectral norm $\|\mathbf{Q}\|_2$ before entering the optimisation criterion. Specifically, if $\lambda_0$ and $\tau_0$ denote the candidate grid values, the effective penalties are $\lambda = \lambda_0 \|\mathbf{Q}\|_2$ and $\tau = \tau_0 \|\mathbf{Q}\|_2$, respectively.

Among candidates with finite AIC values, selection was restricted to fits satisfying the prescribed convergence criterion. The candidate with the smallest AIC within the admissible set was retained. For each final training--test evaluation, data-dependent tuning over the prespecified candidate grids was conducted using only the training partition. The basis dimensions, tuning grids, and iteration caps used in the simulation and empirical studies are reported in the corresponding sections.

\subsection{Adjusted Proportion of Variance Explained}
\label{subsec:method-adj-var}

In conventional FPCA, the component scores are mutually uncorrelated. However, the score vectors obtained from sparse FPCA need not be uncorrelated since the sparse functional directions are not constrained to be mutually orthogonal \citep{nie2020}. Consequently, the marginal variance of a later component may partly overlap with variation already represented by preceding components. To avoid double counting, the adjusted variance proposed by \citet{nie2020} was applied separately to VP--SFPCA and SCAD--SFPCA.

For VP--SFPCA, each estimated sparse weight function was first rescaled to unit $L^2$ norm for adjusted variance assessment. If $\mathbf b_k$ denotes the coefficient vector of the $k$th sparse weight function, the normalised coefficient vector is $\widetilde{\mathbf{b}}_k = \mathbf{b}_k / \sqrt{\mathbf{b}_k^\top \mathbf{W} \mathbf{b}_k}$, and the corresponding score vector used in this assessment is $\mathbf{z}_k = \mathbf{C} \mathbf{W} \widetilde{\mathbf{b}}_k$.

For $k \geq 2$, $\mathbf{z}_k$ is regressed on $\mathbf{z}_1, \ldots, \mathbf{z}_{k-1}$, and $\mathbf{r}_k$ denotes the resulting residual vector, with $\mathbf{r}_1 = \mathbf{z}_1$. Thus, $\mathbf{r}_k$ represents the variation in the $k$th score vector that is not linearly explained by the preceding score vectors. The adjusted proportion of variance explained (PVE) by the $k$th component is defined as
\begin{equation}
	\mathrm{AdjPVE}_k
	=
	\frac{\left\lVert\mathbf{r}_k\right\rVert_2^2}{
		\displaystyle
		\sum_{i=1}^{n} \left\lVert X_i^c\right\rVert_{L^2(\mathcal{T})}^2
	},
	\label{eq:method-adj-var}
\end{equation}
where $X_i^c$ is defined as above. The cumulative adjusted PVE represented by the first $k$ components is then calculated as
\begin{equation}
	\mathrm{CumAdjPVE}_k
	=
	\sum_{j=1}^{k} \mathrm{AdjPVE}_j.
	\label{eq:method-cum-adj-var}
\end{equation}

\subsection{Computational Environment}
\label{subsec:method-computation}

All analyses were conducted using R version 4.5.2 \citep{rcoreteam2025} on a 64-bit Windows 11 workstation (build 26200) equipped with a 12th-generation Intel Core i7-12700 processor, comprising 12 physical cores and 20 logical processors, and 7.7~GB of RAM. Computations were performed using a single thread without parallel processing. The main R packages used in the study were \texttt{fda} (version 6.3.0) \citep{ramsay2025} for functional estimation and \texttt{DiceKriging} (version 1.6.1) \citep{roustant2012} for the implementation of SCAD--SFPCA. Runtime comparisons therefore pertain to the present implementations and hardware.


\section{Simulation Study}
\label{sec:sim}

Two data-generating models were considered in the simulation study to compare the performance of VP--SFPCA relative to SCAD--SFPCA \citep{nie2020} and conventional FPCA \citep{ramsay2005}. Model~1 represents a direct sparse-component setting in which the sparse target functions are themselves the signal-generating directions. Model~2 represents a weight--loading factorisation setting in which prespecified sparse weight functions are retained as recovery targets distinct from the orthonormal loading functions used to generate and reconstruct the curves. This distinction is motivated by \citet{park2024}, who noted that weights, loadings, and right singular vectors cease to be mathematically equivalent under sparse regularisation.

\subsection{Functional Data Generation}
\label{subsec:sim-data}

The functional domain was set as $\mathcal{T} = [0,60]$. Each curve was observed at $20$ equally spaced points and represented using $p = 20$ cubic B-spline basis functions, $\boldsymbol{\psi}(t) = (\psi_1(t), \ldots, \psi_p(t))^{\top}$. The number of components was fixed at $K = 4$. For function-level evaluation, the true and estimated functions were evaluated on a finer equally spaced grid with increment $\Delta_t = 0.12$. 

Four sparse target functions were constructed as localised bump functions centred at $7.5$, $22.5$, $37.5$, and $52.5$, with supports $[0,15]$, $[15,30]$, $[30,45]$, and $[45,60]$, respectively. These non-overlapping regions allow direct evaluation of whether the methods recover localised functional variation rather than dense components over the entire domain.

Let $\beta_1(t), \ldots, \beta_K(t)$ denote the sparse target functions, with
\begin{equation}
	\beta_k(t)
	=
	\mathbf{b}_k^{\top} \boldsymbol{\psi}(t),
	\qquad k = 1, \ldots, K,
	\label{eq:sim-target-basis}
\end{equation}
and let $\mathbf{B}_0 = (\mathbf{b}_1, \ldots, \mathbf{b}_K)$ denote the resulting prespecified sparse-target coefficient matrix. Each target was normalised to unit functional norm, so that $\mathbf{b}_k^{\top} \mathbf{W} \mathbf{b}_k = 1$.

For each Monte Carlo replication, the score vectors were generated using Gaussian random variates with mean zero and covariance matrix $\boldsymbol{\Sigma} = \operatorname{diag}(30,20,10,3)$. The component variances therefore decreased from the first to the fourth signal direction. The noise function was generated in the same coefficient space as
\begin{equation}
	\varepsilon_i(t)
	=
	\sum_{m=1}^{p} e_{im} \psi_m(t),
	\qquad
	\mathbf{e}_i
	\stackrel{\mathrm{i.i.d.}}{\sim} N_p(\mathbf{0}, \sigma_{\varepsilon}^{2} \mathbf{I}_p),
	\qquad
	\sigma_{\varepsilon} = 1,
	\label{eq:sim-noise}
\end{equation}
where $\mathbf{e}_i=(e_{i1},\ldots,e_{ip})^{\top}$.

\subsection{Simulation Models}
\label{subsec:sim-models}

\subsubsection{Model 1: Direct sparse-component setting}
\label{subsubsec:sim-model1}

Under Model~1, the sparse target functions were used directly as the signal-generating directions,
\begin{equation}
	X_i(t)
	=
	\sum_{k=1}^{K}s_{ik}\beta_k(t) + \varepsilon_i(t),
	\label{eq:sim-model1}
\end{equation}
where $s_{ik}$ is the latent coefficient, $\beta_k(t)$ represents the $k$th sparse target function, and $\varepsilon_i(t)$ is the Gaussian noise function. This setting evaluates how well each method recovers sparse functional structure when the underlying signal itself is localised in distinct subregions of the domain.

\subsubsection{Model 2: Weight--loading factorisation setting}
\label{subsubsec:sim-model2}

Model~2 distinguishes the prespecified sparse weight functions $\beta_k(t)$ from the orthonormal loading functions $u_k(t)$. To construct the loadings, let $\mathbf{F}$ denote the B-spline coefficient matrix of the four dense functions
\begin{equation}
	\sin\left(\frac{\pi t}{60}\right),\quad
	\cos\left(\frac{\pi t}{60}\right),\quad
	\sin\left(\frac{2\pi t}{60}\right),\quad
	\cos\left(\frac{2\pi t}{60}\right).
	\label{eq:sim-dense-perturbations}
\end{equation}
Their component orthogonal to the span of $\mathbf{B}_0$ under the $\mathbf{W}$ inner product was obtained as
\begin{equation}
	\mathbf{F}_{\perp}
	=
	\mathbf{F} - \mathbf{B}_0 \left(\mathbf{B}_0^{\top} \mathbf{W} \mathbf{B}_0\right)^{-1} \mathbf{B}_0^{\top} \mathbf{W} \mathbf{F},
	\label{eq:sim-orthogonal-perturbation}
\end{equation}
after which its columns were normalised to unit $\mathbf{W}$-norm. A separation constant $\eta_A = 0.40$ was used to form
\begin{equation}
	\mathbf{A}_{*}
	=
	\mathbf{B}_0 + \eta_A \mathbf{F}_{\perp},
	\qquad
	\mathbf{A}_0
	=
	\mathbf{A}_{*} \left(\mathbf{A}_{*}^{\top} \mathbf{W} \mathbf{A}_{*}\right)^{-1/2}.
	\label{eq:sim-loading-construction}
\end{equation}
The columns of $\mathbf{A}_0 = (\mathbf{a}_{01}, \ldots, \mathbf{a}_{0K})$ therefore satisfy $\mathbf{A}_0^{\top}\mathbf{W}\mathbf{A}_0=\mathbf{I}_K$, so that $\mathbf{A}_0$ and $\mathbf{B}_0$ denote the true loading and weight coefficient matrices, respectively. The true loading functions are
\begin{equation}
	u_k(t)
	=
	\mathbf{a}_{0k}^{\top} \boldsymbol{\psi}(t),
	\qquad k = 1, \ldots, K.
	\label{eq:sim-true-loadings}
\end{equation}
The $i$th curve was generated as
\begin{equation}
	X_i(t)
	=
	\sum_{k=1}^{K} s_{ik} u_k(t) + \varepsilon_i(t).
	\label{eq:sim-model2}
\end{equation}
Equivalently, its basis-coefficient vector is $\mathbf{c}_i = \mathbf{A}_0 \mathbf{s}_i + \mathbf{e}_i$, where $\mathbf{s}_i = (s_{i1}, \ldots, s_{iK})^\top$ denotes the latent-score vector. The sparse functions $\beta_k(t)$ are retained as the prespecified recovery targets, whereas $u_k(t)$ are the orthonormal loading functions governing data generation and reconstruction. Model~2 therefore evaluates recovery of sparse weight functions that are distinct from the orthonormal loading functions used to generate and reconstruct the curves.

\subsection{Parameter Tuning}
\label{subsec:sim-tuning}

The two-stage tuning procedure in Section~\ref{subsec:method-tuning} was applied in each Monte Carlo replication. The smoothing parameter was selected by GCV, and the method-specific sparsity and ridge parameters were subsequently selected by AIC. The candidate grids are summarised in Table~\ref{tab:sim-tuning-grid}. The numerical ranges differ between the two sparse methods because their regularisation parameters enter the respective optimisation procedures on different scales. For VP--SFPCA, the values reported in Table~\ref{tab:sim-tuning-grid} correspond to the unscaled grid values described in Section~\ref{subsec:method-tuning}, whereas the SCAD--SFPCA sparsity and ridge parameters enter the regression updates directly \citep{erichson2020,nie2020}. Consequently, identical numerical grids would not represent equivalent regularisation across the two methods.

\begin{table}[htbp]
	\centering
	\caption{Tuning parameter grids used in the simulation study.}
	\label{tab:sim-tuning-grid}
	\begin{tabular*}{\textwidth}{@{\extracolsep\fill}lccc@{\extracolsep\fill}}%
		\toprule
		\textbf{Method}
		& $\boldsymbol{\gamma}$
		& $\boldsymbol{\lambda}$
		& $\boldsymbol{\tau}$ \\
		\midrule
		VP--SFPCA
		& $\mathcal{G}_{7}(-4,2)$
		& $\mathcal{G}_{10}(-4,-1)$
		& $\mathcal{G}_{7}(-4,-1)$ \\
		SCAD--SFPCA
		& $\mathcal{G}_{7}(-4,2)$
		& $\mathcal{G}_{10}(0,\log_{10}50)$
		& $\mathcal{G}_{7}(-1,2)$ \\
		\bottomrule
	\end{tabular*}
\end{table}

Pilot studies were conducted separately for each model at sample sizes $n \in \{20, 50, 100\}$, with $50$ Monte Carlo replications for each setting. Iteration caps of $30$, $40$, and $50$ were compared. The pilot results indicated no consistent improvement in the attained objective values when the iteration cap was increased beyond $30$. Accordingly, a maximum of $30$ iterations was retained to avoid unnecessary computational burden. 

Examination of the pilot tuning-parameter selections showed that the lower boundary was frequently selected for $\tau$ under the proposed VP--SFPCA. Nevertheless, sensitivity analyses using alternative grids extending to smaller $\tau$ values yielded essentially unchanged recovery performance, while lower-boundary selection remained frequent. The $\tau$ grid given in Table~\ref{tab:sim-tuning-grid} was therefore retained for the full simulation study. Further details are provided in the Supporting Information.

The full simulation study was then conducted separately for each model at $n \in \{50, 100\}$, with $100$ Monte Carlo replications for each setting and a maximum of $30$ iterations for each candidate fit.

\subsection{Performance Evaluation}
\label{subsec:sim-evaluation}

Recovery relative to the sparse target functions was evaluated using integrated error (IE). To account for possible changes in component order and sign ambiguity in PCA-type decompositions \citep{babamoradi2013,bro2008,jolliffe2016}, the estimated functions were first matched to the true targets by maximising the absolute functional inner product and were then sign-aligned. For each method, let $\widetilde{\beta}_{k}(t)$ denote the matched and sign-adjusted estimate corresponding to $\beta_k(t)$. The IE for the $k$th component was defined as
\begin{equation}
	\mathrm{IE}_{k}
	= 
	\int_{\mathcal{T}} \left\{\widetilde{\beta}_{k}(t) - \beta_k(t) \right\}^{2} 	\, \mathrm{d}t,
	\qquad k = 1, \ldots, K.
	\label{eq:sim-ie}
\end{equation}
Means and standard deviations were calculated across all $100$ Monte Carlo replications. Smaller values indicate more accurate recovery relative to the sparse target functions. In addition, the computational efficiency of the two sparse FPCA methods was evaluated. 

As a supplementary assessment of score recovery, Tucker congruence was computed between the true and estimated scores within each Monte Carlo replication. This statistic provides a cosine-type measure of similarity, with larger values indicating stronger agreement between the true and estimated component scores \citep{lorenzo2006,park2024}. Detailed results are provided in the Supporting Information.

\subsection{Simulation Results}
\label{subsec:sim-results}

\subsubsection{Model 1: Direct sparse-component setting}
\label{subsubsec:sim-model1-results}

\begin{figure*}[htbp]
	\centering
	\subfloat[]{%
		\includegraphics[width=0.80\textwidth]{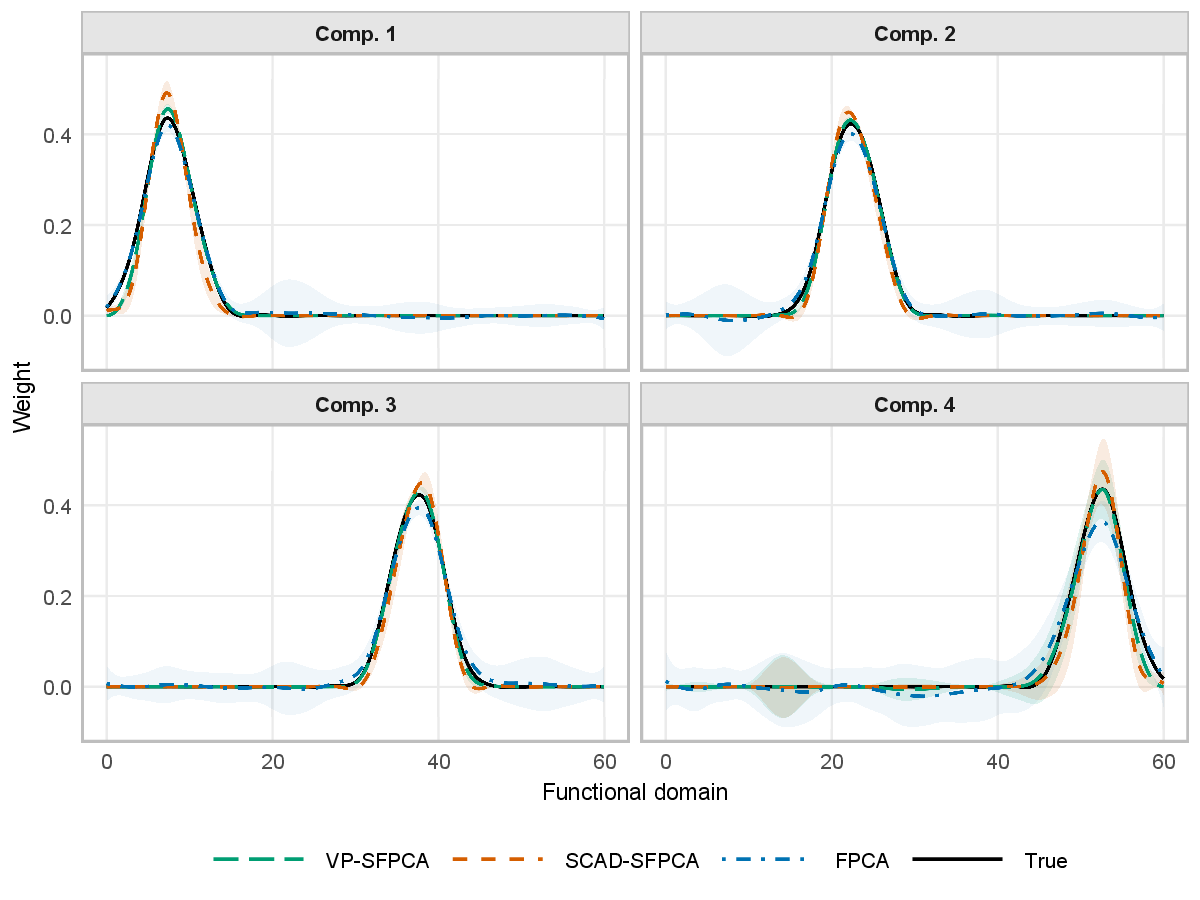}%
		\label{fig:sim-model1-weights-n50}%
	}\\[1em]
	\subfloat[]{%
		\includegraphics[width=0.80\textwidth]{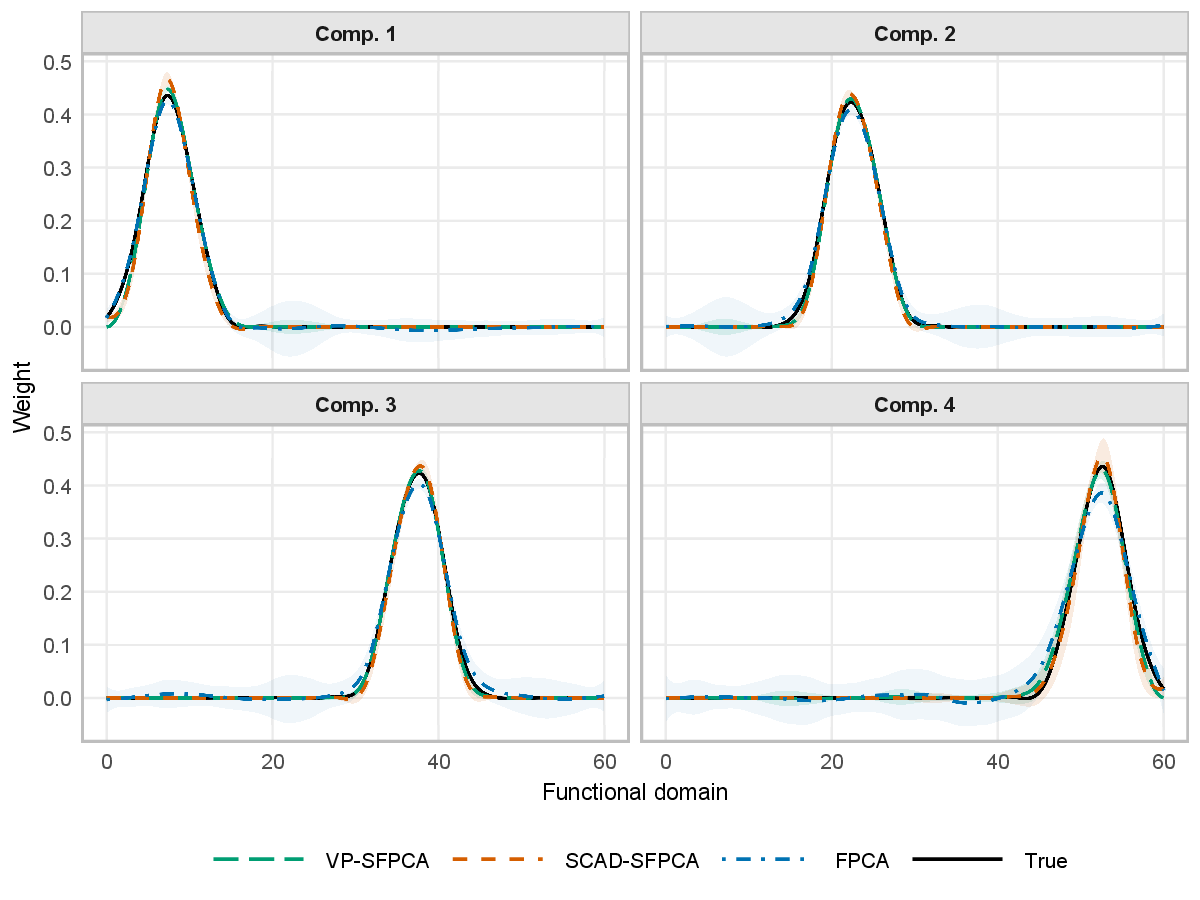}%
		\label{fig:sim-model1-weights-n100}%
	}
	\caption{Mean estimated functions under Model~1 for (a) $n = 50$ and (b) $n = 100$, obtained using the proposed VP--SFPCA (green long-dashed lines) and the benchmark methods, SCAD--SFPCA (orange dashed lines) and conventional FPCA (blue dot-dashed lines). The true sparse target functions are shown as black solid lines. Shaded regions represent variability in the corresponding estimated functions across $100$ Monte Carlo replications.}
	\label{fig:sim-model1-weights}
\end{figure*}

Figure~\ref{fig:sim-model1-weights} compares the true sparse target functions with the estimates obtained by VP--SFPCA, SCAD--SFPCA, and conventional FPCA. Both sparse methods recovered the localised target structure well, with reduced variability as the sample size increased. The fourth component remained the most variable, particularly at $n = 50$, consistent with its smaller latent variance. Conventional FPCA also recovered the general structure of the target functions but exhibited greater departures from the targets, especially for the later components.

The component-wise IE summaries are reported in Table~\ref{tab:sim-model-ie}. Recovery error was lower at $n = 100$ than at $n = 50$ for all three methods. Averaged across the four components, the mean IE decreased from approximately $0.027$ to $0.009$ for VP--SFPCA, from $0.046$ to $0.015$ for SCAD--SFPCA, and from $0.109$ to $0.053$ for conventional FPCA as the sample size increased from $n = 50$ to $n = 100$. Both sparse methods achieved lower recovery errors than conventional FPCA, with VP--SFPCA consistently yielding the lowest mean IE for each component at both sample sizes. The fourth component exhibited the largest mean IE and standard deviation for all three methods, consistent with its smaller latent variance and greater sensitivity to noise.

\begin{table*}[htbp]
	\centering
	\caption{Component-wise IEs under Models~1 and 2. Values are reported as means (standard deviations) across $100$ Monte Carlo replications.}
	\label{tab:sim-model-ie}
	\begin{tabular*}{\textwidth}{@{\extracolsep\fill}lllcccc@{\extracolsep\fill}}%
		\toprule
		\textbf{Setting}
		& $\boldsymbol{n}$
		& \textbf{Method}
		& $\boldsymbol{\mathrm{IE}_{1}}$
		& $\boldsymbol{\mathrm{IE}_{2}}$
		& $\boldsymbol{\mathrm{IE}_{3}}$
		& $\boldsymbol{\mathrm{IE}_{4}}$ \\
		\midrule
		\multirow{6}{*}{Model 1}
		& \multirow{3}{*}{50} & VP--SFPCA
		& 0.011 (0.006) & 0.004 (0.002) & 0.006 (0.008) & 0.087 (0.254) \\
		& & SCAD--SFPCA
		& 0.040 (0.033) & 0.016 (0.011) & 0.025 (0.030) & 0.103 (0.219) \\
		& & FPCA
		& 0.057 (0.060) & 0.074 (0.064) & 0.087 (0.045) & 0.219 (0.210) \\
		\cmidrule(lr){2-7}
		& \multirow{3}{*}{100} & VP--SFPCA
		& 0.007 (0.007) & 0.003 (0.007) & 0.004 (0.004) & 0.021 (0.024) \\
		& & SCAD--SFPCA
		& 0.012 (0.009) & 0.006 (0.004) & 0.009 (0.008) & 0.033 (0.030) \\
		& & FPCA
		& 0.028 (0.029) & 0.040 (0.030) & 0.045 (0.022) & 0.099 (0.053) \\
		\midrule
		\multirow{6}{*}{Model 2} 
		& \multirow{3}{*}{50} & VP--SFPCA
		& 0.023 (0.012) & 0.008 (0.009) & 0.008 (0.008) & 0.124 (0.308) \\
		& & SCAD--SFPCA
		& 0.059 (0.020) & 0.061 (0.047) & 0.041 (0.030) & 0.140 (0.277) \\
		& & FPCA
		& 0.200 (0.074) & 0.212 (0.074) & 0.191 (0.071) & 0.329 (0.197) \\
		\cmidrule(lr){2-7}
		& \multirow{3}{*}{100} & VP--SFPCA
		& 0.020 (0.012) & 0.008 (0.007) & 0.008 (0.006) & 0.027 (0.039) \\
		& & SCAD--SFPCA
		& 0.043 (0.015) & 0.033 (0.020) & 0.032 (0.020) & 0.070 (0.038) \\
		& & FPCA
		& 0.164 (0.035) & 0.168 (0.032) & 0.151 (0.041) & 0.192 (0.076) \\
		\bottomrule
	\end{tabular*}
\end{table*}

\begin{figure*}[htbp]
	\centerline{\includegraphics[width=0.80\textwidth]{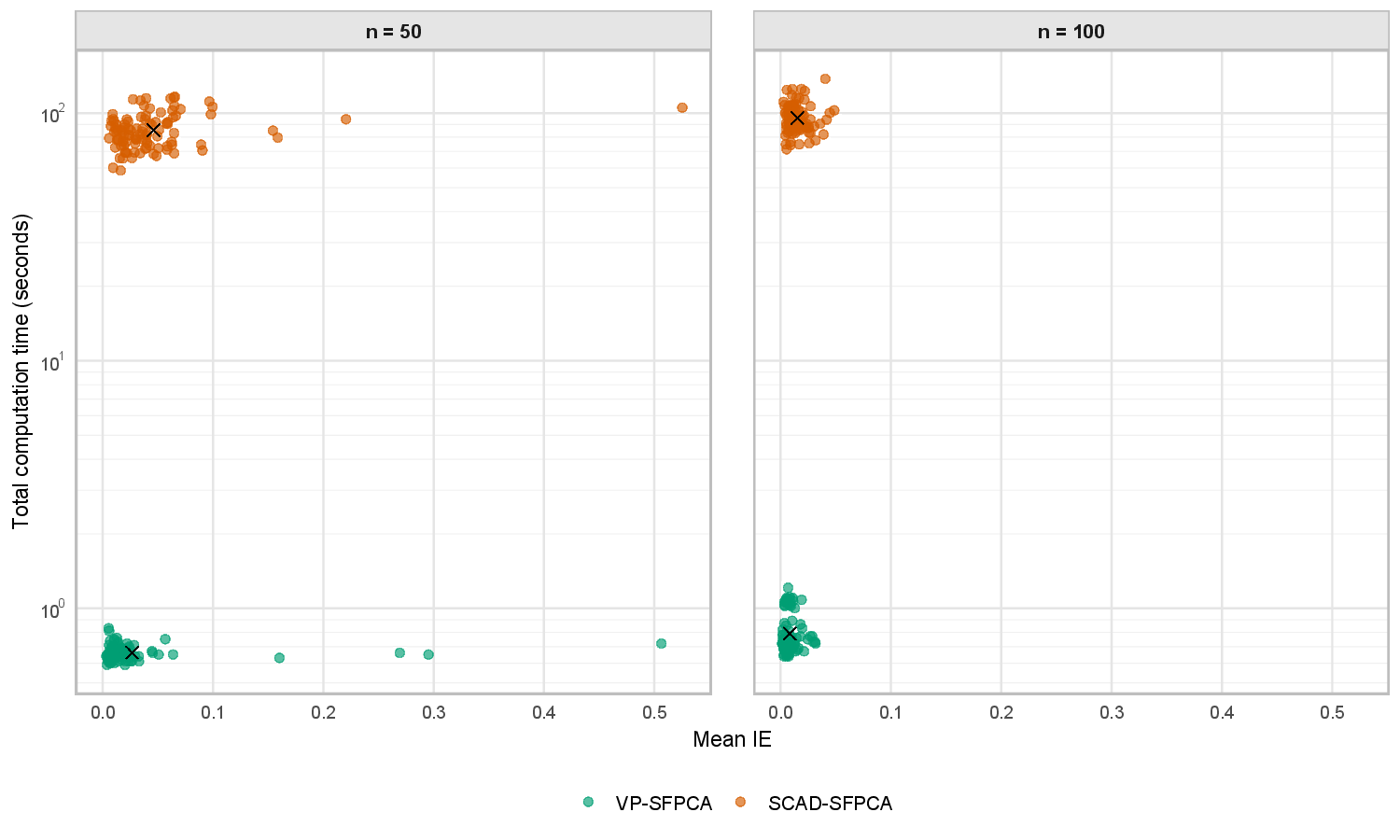}}
	\caption{Mean IE versus total computation time under Model~1. Each coloured point represents one Monte Carlo replication, with mean IE calculated across the $K = 4$ components. Black crosses denote method-specific averages across $100$ replications.}
	\label{fig:sim-model1-tradeoff}
\end{figure*}

Although both sparse methods achieved relatively low recovery errors in Model~1, they differed substantially in computation time. As shown in Figure~\ref{fig:sim-model1-tradeoff}, the mean total computation time for VP--SFPCA was approximately $0.66$ and $0.79$ seconds at $n = 50$ and $n = 100$, respectively, compared with $85.26$ and $95.50$ seconds for SCAD--SFPCA. This computational difference is consistent with the different optimisation structures of the two methods, whereby SCAD--SFPCA involves a nonconvex functional SCAD penalty \citep{fan2001,li2026,nie2020}, whereas VP--SFPCA exploits variable projection to partially minimise the objective with respect to the orthogonality-constrained loading functions \citep{erichson2020,golub2003}.

\subsubsection{Model 2: Weight--loading factorisation setting}
\label{subsubsec:sim-model2-results}

\begin{figure*}[htbp]
	\centering
	\subfloat[]{%
		\includegraphics[width=0.80\textwidth]{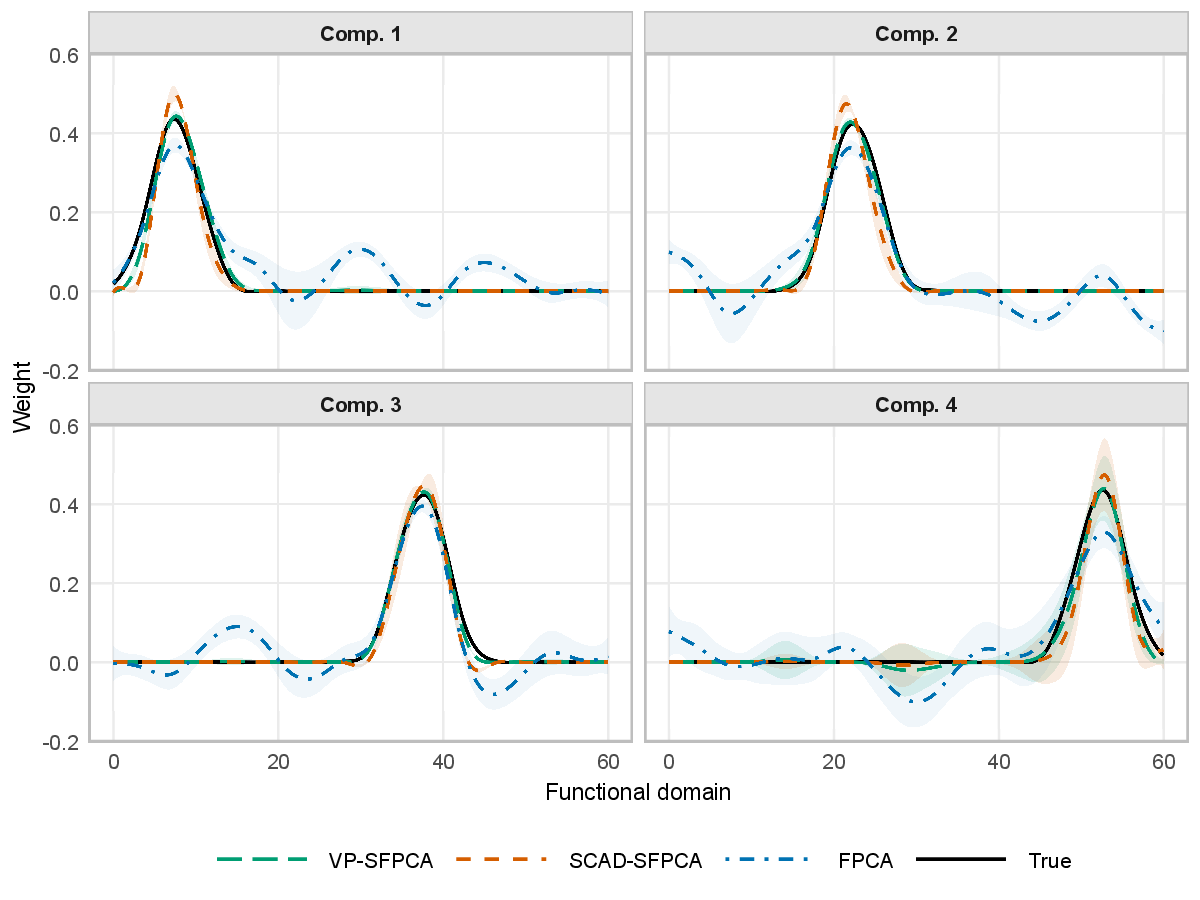}%
		\label{fig:sim-model2-weights-n50}%
	}\\[1em]
	\subfloat[]{%
		\includegraphics[width=0.80\textwidth]{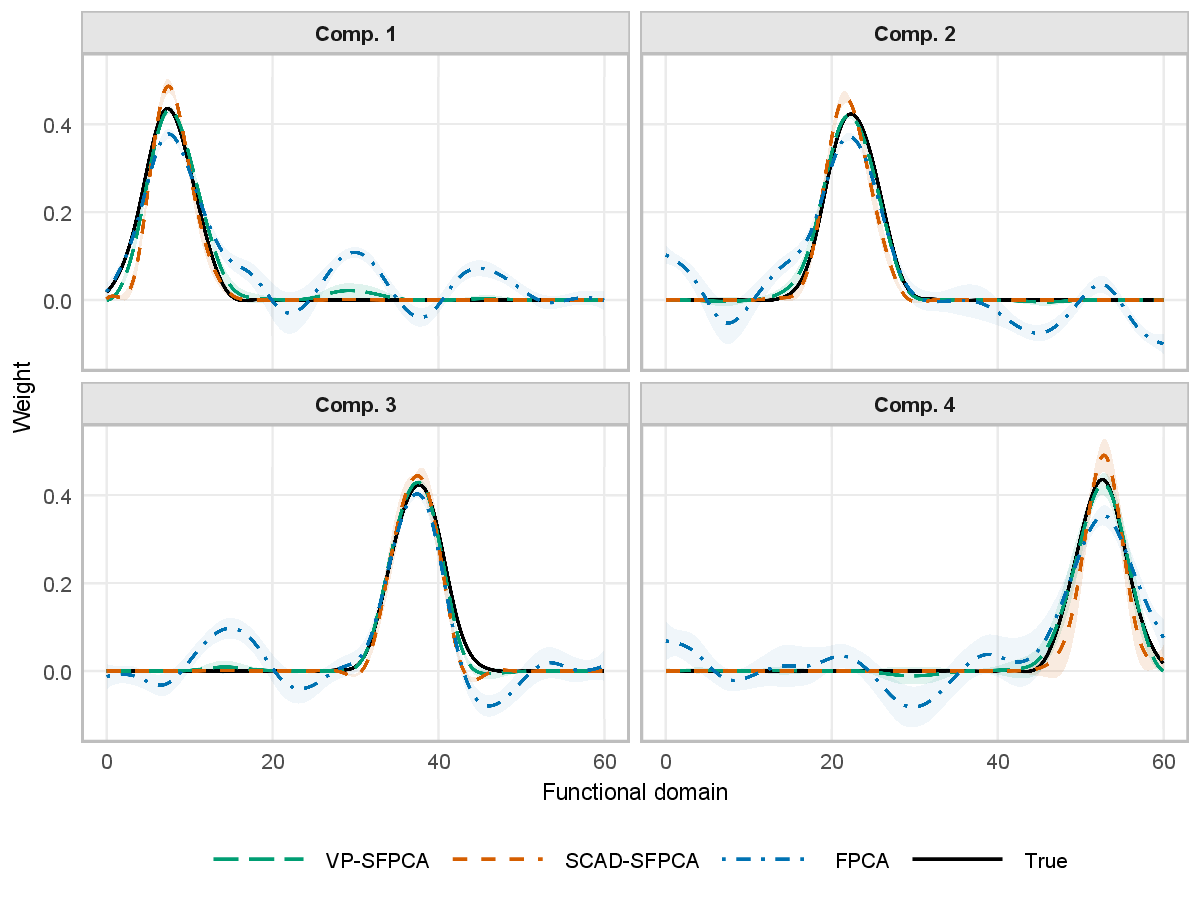}%
		\label{fig:sim-model2-weights-n100}%
	}
	\caption{Mean estimated functions under Model~2 for (a) $n = 50$ and (b) $n = 100$, obtained using the proposed VP--SFPCA (green long-dashed lines) and the benchmark methods, SCAD--SFPCA (orange dashed lines) and conventional FPCA (blue dot-dashed lines). The true sparse target functions are shown as black solid lines. Shaded regions represent variability in the corresponding estimated functions across $100$ Monte Carlo replications.}
	\label{fig:sim-model2-weights}
\end{figure*}

Figure~\ref{fig:sim-model2-weights} compares the true sparse weight functions with the estimates obtained by VP--SFPCA, SCAD--SFPCA, and conventional FPCA under Model~2. Recovery was more challenging than under Model~1 because the prespecified sparse weight functions were distinct from the orthonormal loading functions used to generate and reconstruct the curves. Nevertheless, both sparse methods recovered the principal localised structure of the target weight functions. The estimates were more variable at $n = 50$, especially for the fourth component, and became more stable as the sample size increased to $n = 100$. Conventional FPCA exhibited substantially greater departures from the sparse target weight functions. The larger departures are expected because, under Model~2, conventional FPCA targets the orthonormal signal-generating directions rather than the prespecified sparse weight targets used for recovery evaluation.

Table~\ref{tab:sim-model-ie} shows that the average recovery errors for the sparse methods were larger under Model~2 than under Model~1, reflecting the additional difficulty introduced by separating the prespecified sparse weight targets from the reconstruction loadings. The sparse methods nevertheless achieved more accurate recovery of the sparse target weight functions, with VP--SFPCA yielding the lowest recovery errors overall. The mean IE across the four components decreased from approximately $0.041$ at $n = 50$ to $0.016$ at $n = 100$ for VP--SFPCA, whereas the corresponding mean for SCAD--SFPCA decreased from approximately $0.075$ to $0.044$. For conventional FPCA, the mean IE decreased from approximately $0.233$ to $0.169$. At the component level, VP--SFPCA produced the lowest mean IE for each component at both sample sizes, followed by SCAD--SFPCA and conventional FPCA. The fourth component, which represented the weakest signal, again had the largest mean IE and standard deviation for both sparse methods.

\begin{figure*}[htbp]
	\centering
	\includegraphics[width=0.80\textwidth]{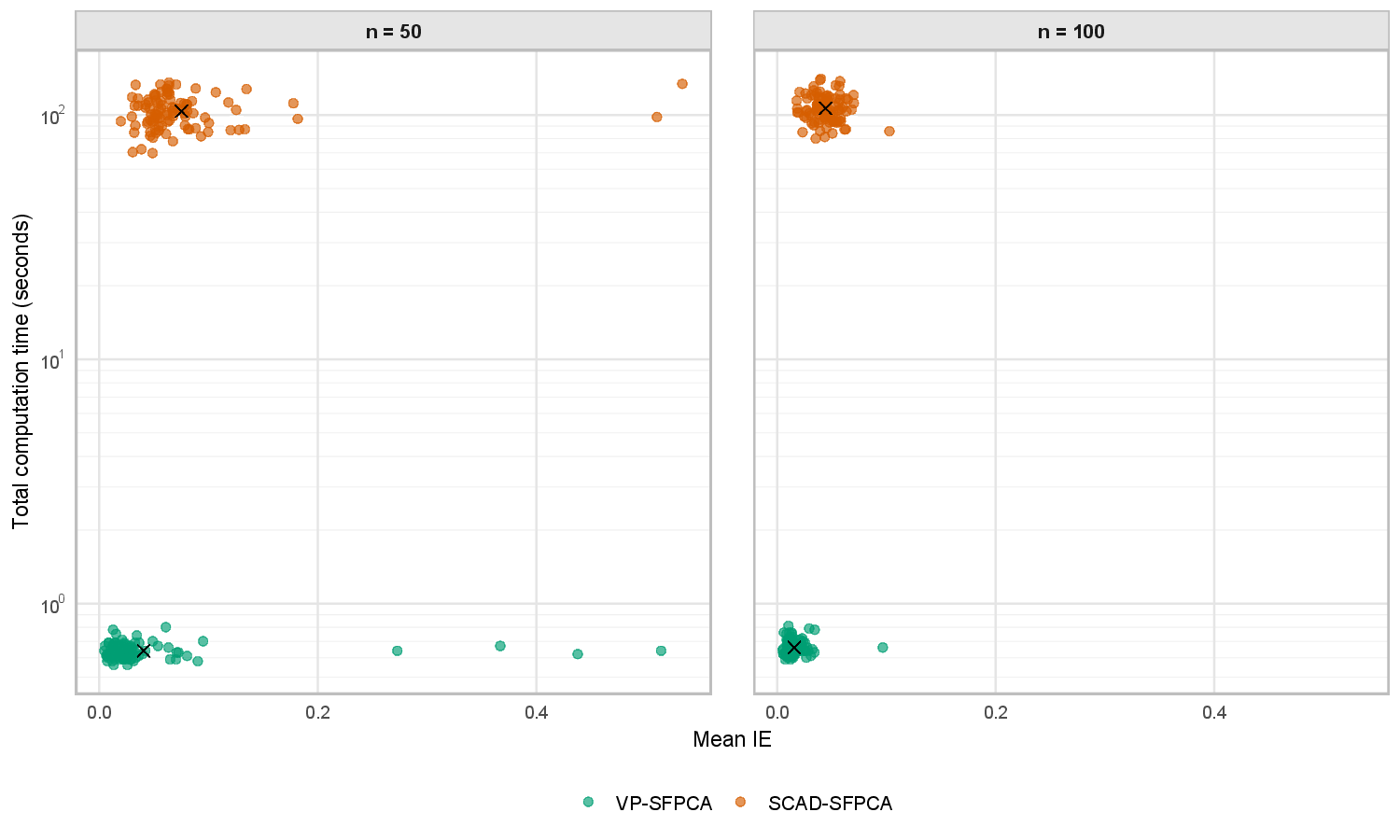}
	\caption{Mean IE versus total computation time under Model~2. Each coloured point represents one Monte Carlo replication, with mean IE calculated across the $K = 4$ components. Black crosses denote method-specific averages across $100$ replications.}
	\label{fig:sim-model2-tradeoff}
\end{figure*}

The computational advantage of VP--SFPCA was also evident under Model~2. As shown in Figure~\ref{fig:sim-model2-tradeoff}, the mean total computation time for VP--SFPCA was approximately $0.64$ and $0.66$ seconds at $n = 50$ and $n = 100$, respectively, compared with $103.59$ and $106.44$ seconds for SCAD--SFPCA. Taken together with the recovery results, these findings show that under Model~2, VP--SFPCA achieved lower sparse weight-function recovery errors and substantially lower computation time than SCAD--SFPCA.

Across both simulation models, the Tucker congruence analysis presented in the Supporting Information further showed that VP--SFPCA and SCAD--SFPCA achieved comparable recovery of the simulated score structure.


\section{Empirical Application}
\label{sec:app}

The empirical performance of the proposed VP--SFPCA method in identifying localised dominant modes of spectral variation was evaluated using two SERS datasets, with SCAD--SFPCA \citep{nie2020} and conventional FPCA \citep{ramsay2005} included as benchmark methods. The first dataset comprises adenine spectra \citep{fornasaro2020data,fornasaro2020} acquired under a common protocol across multiple laboratories, providing a controlled single-analyte setting with interlaboratory and technical-replicate variation. The second dataset, presented in the Supporting Information, contains serum spectra \citep{gurian2021data,gurian2021} from patients with hepatocellular carcinoma (HCC) and healthy blood donors, representing a biologically heterogeneous clinical setting.

\subsection{Dataset and Preprocessing}
\label{subsec:app-data}

The adenine dataset was derived from the European interlaboratory SERS study of \citet{fornasaro2020}. The original study evaluated six combinations of substrate type, metal, and excitation wavelength for the quantitative measurement of adenine in $0.01\,\mathrm{M}$ phosphate buffer at pH~$7.4$. Adenine was selected as a standard analyte because it is stable, non-toxic, commercially available, detectable on both silver and gold substrates, and exhibits intense, well-characterised SERS bands \citep{fornasaro2020}.

The present analysis was restricted to the sAg@785 protocol, in which solid silver substrates based on metal-coated silicon nanopillars were measured using $785\,\mathrm{nm}$ excitation. For each sample, measurements were collected at three random positions on each of three solid substrates, yielding nine spectra per laboratory--sample combination \citep{fornasaro2020}. After removal of the blank sample, the analysed data comprised $756$ spectra from six laboratories and $14$ nonblank sample codes, including nine calibration samples and five test samples. Since the present study focused on the unsupervised characterisation of spectral variation rather than on quantitative adenine calibration, the original calibration and test labels were not used to partition the data, and all nonblank spectra were included in the analysis.

The spectra were initially restricted to $400$--$1650\,\mathrm{cm}^{-1}$. Since the participating instruments did not all cover identical Raman-shift values, only Raman shifts with complete observations across all retained sAg@785 spectra were included. This resulted in a common grid of $355$ points from $517$ to $1579\,\mathrm{cm}^{-1}$ at $3\,\mathrm{cm}^{-1}$ intervals. An asymmetric least-squares baseline was estimated separately for each spectrum using a smoothing parameter of $10^{5}$, an asymmetry parameter of $0.01$, and $10$ iterations. Negative baseline-corrected intensities were set to zero, after which each spectrum was normalised by its trapezoidal integrated area. Any spectrum with a non-finite or near-zero integrated area was excluded before analysis.

\begin{figure*}[htbp]
	\centering
	\includegraphics[width=0.80\textwidth]{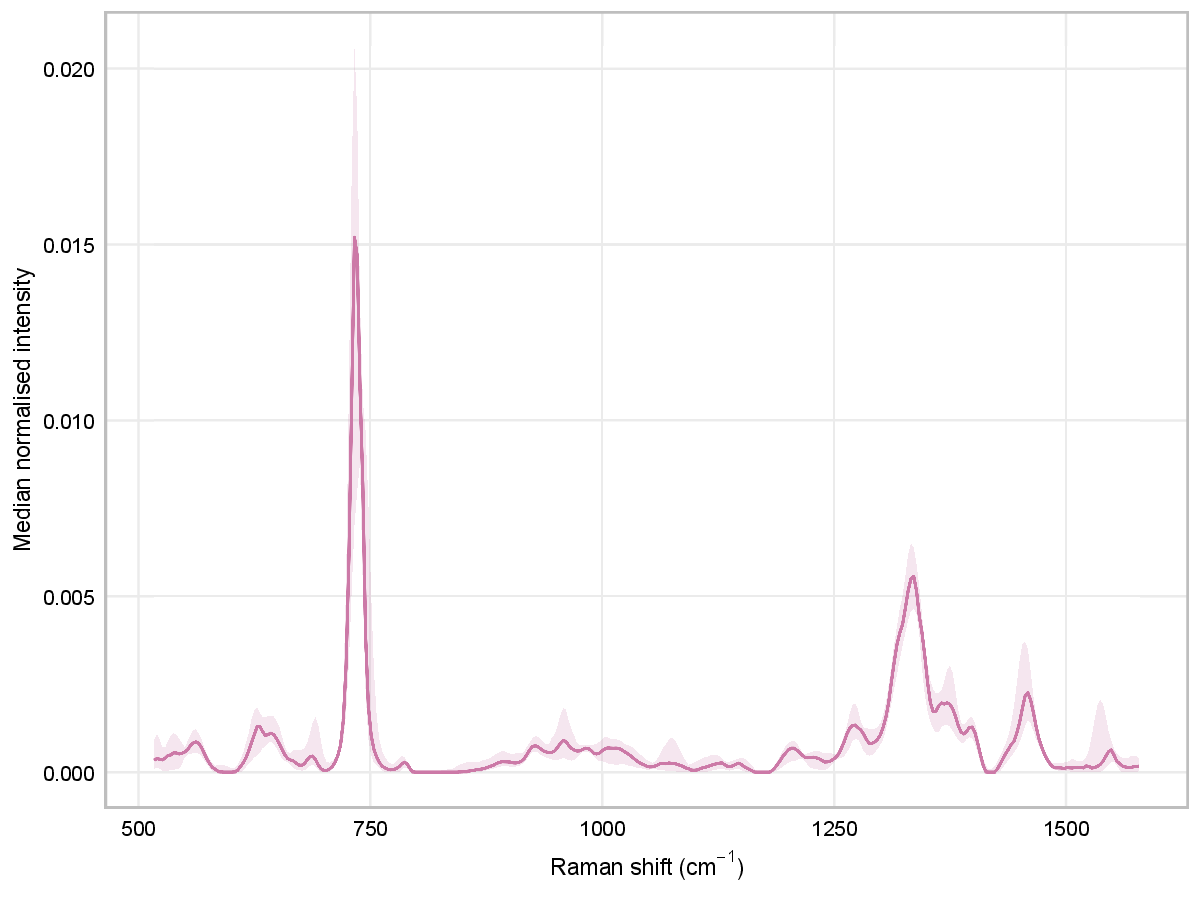}%
	\caption{Median preprocessed SERS spectrum for the adenine dataset, with the shaded region representing the interquartile range across spectra.}
	\label{fig:app-spectra}
\end{figure*}

All individual preprocessed spectra were retained for the subsequent analysis, without averaging or aggregating the technical replicates. Figure~\ref{fig:app-spectra} shows the median spectrum for the adenine dataset. The spectrum is dominated by a prominent band near $730\,\mathrm{cm}^{-1}$, together with several weaker features at higher Raman shifts, particularly between approximately $1250$ and $1550\,\mathrm{cm}^{-1}$. The shaded region represents the interquartile range across spectra, summarising between-spectrum variation in intensity at each Raman shift.

\subsection{Analysis Design and Parameter Tuning}
\label{subsec:app-tuning}

The adenine data were analysed using repeated grouped five-fold cross-validation, with the laboratory--sample combination as the resampling unit. This yielded $84$ groups, each containing nine technical-replicate spectra, with all spectra within a group assigned to the same fold. This grouping prevented technical replicates from being split between the training and held-out samples, thereby respecting the dependence structure of the replicated measurements and avoiding replicate leakage \citep{roberts2017}. The same grouped partitions were used for VP--SFPCA, SCAD--SFPCA, and conventional FPCA. Five folds were chosen as a practical balance between retaining sufficient data for model fitting and limiting the computational burden of repeated validation. This choice is consistent with the common use of five- to ten-fold cross-validation for estimation problems, although the appropriate number of folds remains problem dependent \citep{arlot2010}. With $50$ repetitions and five folds, the final design comprised $250$ training--test evaluations per method. The resulting held-out reconstruction errors therefore provide an internal assessment of reconstruction performance for laboratory--sample combinations not included in the corresponding model fit.

A rank of $K = 3$ was retained for all three methods to ensure matched dimensionality. This was the smallest number of components accounting for at least $80\%$ of the total variation under conventional FPCA fitted to the complete data. As a diagnostic, the same criterion was applied separately to conventional FPCA fitted to each training sample, with $K = 3$ selected in $95.6\%$ of the evaluations and $K = 4$ in the remaining $4.4\%$. The predominance of $K = 3$ across the training samples indicates that the fixed-rank choice was representative of the rank selected from the training data in the large majority of evaluations.

The number of basis functions and the iteration cap were examined through a preliminary sensitivity analysis using five cross-validation repetitions, with the corresponding results provided in the Supporting Information. Basis dimensions of $p \in \{30,40,50\}$ were compared using an iteration cap of $40$. Across these settings, the relative performance of the methods in terms of mean held-out integrated squared error (ISE), as defined in Equation~\eqref{eq:app-ise}, and computation time remained broadly similar, with VP--SFPCA retaining a substantial computational advantage over SCAD--SFPCA. Among the basis dimensions considered, $p = 50$ yielded the lowest mean held-out reconstruction ISE for both sparse methods, although this was accompanied by increased computation time. The full analysis therefore used $p = 50$. Iteration caps of $30$, $40$, and $50$ were also compared with the basis dimension fixed at $p = 40$. Since mean held-out reconstruction ISE varied little across the three caps, a cap of $30$ was selected to avoid potentially greater computational cost with little corresponding improvement in reconstruction accuracy.

For each cross-validation split, data-dependent smoothing and centring were based on the corresponding training sample. The training spectra were smoothed using the selected cubic B-spline basis, with the smoothing parameter chosen from $\mathcal{G}_{9}(-4,4)$ by minimising the mean GCV criterion across the training spectra. The resulting functions were centred using the mean function estimated from the training sample. The held-out spectra were then represented using the same basis with the training-selected smoothing parameter, followed by centring with the corresponding training mean function.

Conditional on the selected smoothing parameter, the sparsity and ridge parameters were jointly selected within each training sample by minimising AIC, following the tuning procedure described in Section~\ref{subsec:method-tuning}. The empirical tuning grids are summarised in Table~\ref{tab:app-tuning-grid}, where the reported VP--SFPCA grid values are unscaled. Given the different parameter scaling of the two sparse methods, their tuning grids were specified separately. For SCAD--SFPCA, the orders of magnitude and grid ranges were guided by the smallest and largest conventional FPCA eigenvalues among the retained components, following \citet{nie2020}.

The final tuning grids were selected following a preliminary comparison of several candidate ranges based on five cross-validation repetitions. Broader grids reduced boundary selection, but further expansion produced no consistent improvement. Additional details are provided in the Supporting Information.

\begin{table}[htbp]
	\centering
	\caption{Tuning parameter grids used in the adenine application.}
	\label{tab:app-tuning-grid}
	\begin{tabular*}{\textwidth}{@{\extracolsep\fill}lccc@{\extracolsep\fill}}%
		\toprule
		\textbf{Method}
		& $\boldsymbol{\gamma}$
		& $\boldsymbol{\lambda}$
		& $\boldsymbol{\tau}$ \\
		\midrule
		VP--SFPCA
		& $\mathcal{G}_{9}(-4,4)$
		& $\mathcal{G}_{15}(-5,-1)$
		& $\mathcal{G}_{15}(-7,-1)$ \\
		SCAD--SFPCA
		& $\mathcal{G}_{9}(-4,4)$
		& $\mathcal{G}_{15}(-6,-3)$
		& $\mathcal{G}_{15}(-6,-3)$ \\
		\bottomrule
	\end{tabular*}
\end{table}

\subsection{Evaluation Criteria}
\label{subsec:app-evaluation}

The methods were compared with respect to the localisation and spectroscopic interpretability of Raman-shift regions, held-out reconstruction accuracy, coefficient-level sparsity, adjusted PVE, and computation time.

For each held-out spectrum, reconstruction accuracy was assessed using ISE,
\begin{equation}
	\mathrm{ISE}_i
	=
	\int_{\mathcal{T}} \left\{X_i(t) - \widehat{X}_i(t)\right\}^{2}\, \mathrm{d}t,
	\label{eq:app-ise}
\end{equation}
where $X_i(t)$ denotes the held-out spectrum and $\widehat{X}_i(t)$ denotes its reconstruction.

For the sparse methods, coefficient-level sparsity was defined from the B-spline coefficients of the estimated functions. Let $\widehat{b}_{mk}$ denote the $m$th coefficient of the $k$th sparse function. Sparsity was calculated as
\begin{equation}
	\mathrm{Sparsity}
	=
	1 - \frac{\#\left\{(m,k):|\widehat{b}_{mk}|>\epsilon_0\right\}}{pK},
	\qquad
	\epsilon_0=10^{-3}.
	\label{eq:app-sparsity}
\end{equation}
Here $p = 50$ is the number of cubic B-spline basis functions and $K$ is the number of retained components. Each nonzero coefficient was mapped to the support of its corresponding B-spline basis function to identify the Raman-shift regions represented by the estimated sparse functions.

For component-specific summaries across repeated fits, the estimated components were first permutation-matched within each method to a common reference by maximising the total absolute functional similarity across component permutations. The matched components were subsequently sign-aligned to the corresponding reference functions. This alignment was used for cross-fit aggregation of component-specific functions and localisation summaries.

Since the sparse score vectors are not mutually orthogonal, the sequentially adjusted PVE was computed using Equation~\eqref{eq:method-adj-var} for both VP--SFPCA and SCAD--SFPCA to account for variation shared among successive score vectors. Conventional FPCA was summarised using the conventional PVE.

For each method, computation time was measured as the total runtime required for parameter tuning and fitting within each training--test evaluation. Runtime comparisons therefore pertain to the present implementation and computational settings.


\section{Results and Discussion}
\label{sec:results}

The primary objective of the empirical application was to assess whether sparse functional representations could enhance the interpretability of the dominant modes of between-spectrum variation. This was evaluated by examining the localisation of Raman-shift regions contributing to each component and relating prominent extrema of the estimated functions to established adenine SERS band assignments.

\begin{figure*}[htbp]
	\centering
	\includegraphics[width=0.80\textwidth]{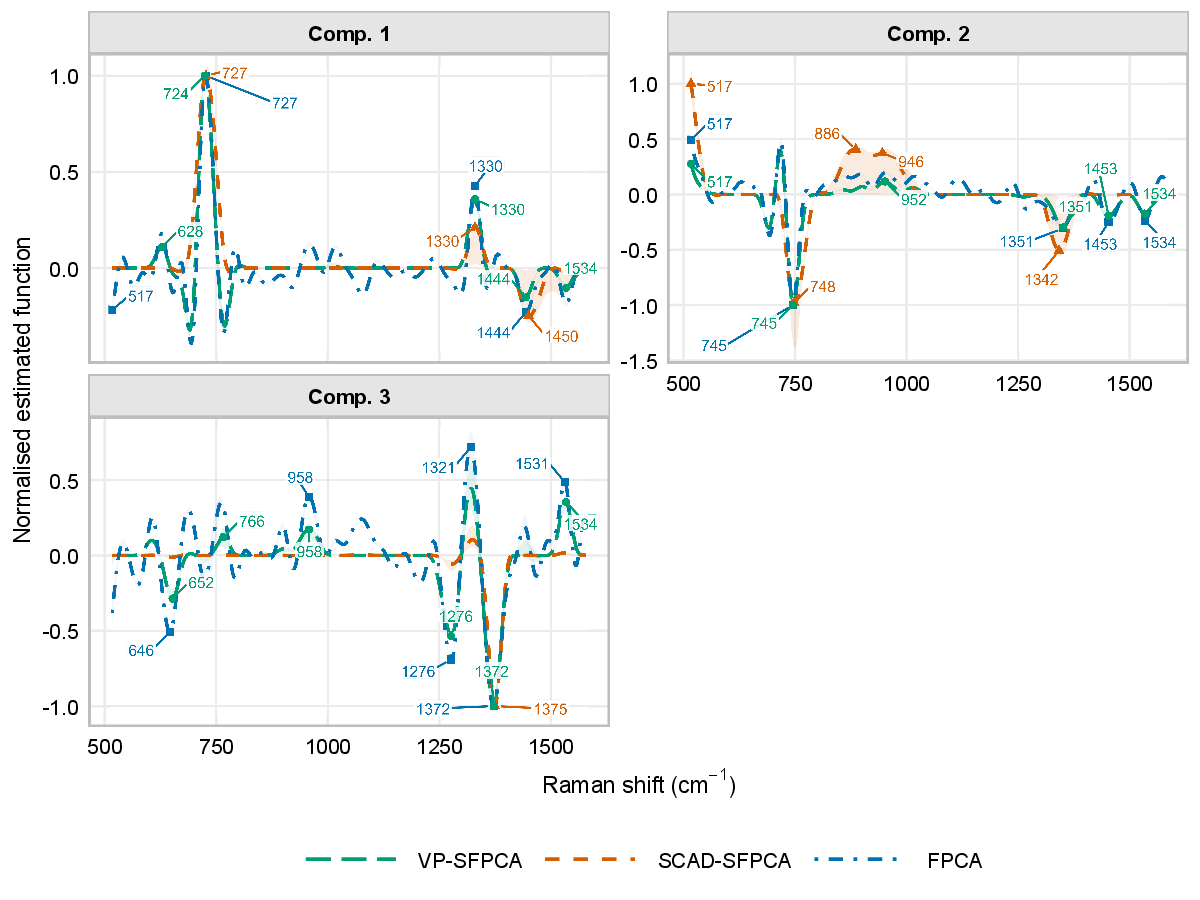}
	\caption{Median estimated functions across $250$ training--test fits from $50$ repetitions of grouped five-fold cross-validation for the adenine dataset, with a common rank of $K = 3$ selected as the smallest number of components accounting for at least $80\%$ of the total variation under conventional FPCA. The proposed VP--SFPCA (green long-dashed lines) and the benchmark methods, SCAD--SFPCA (orange dashed lines) and conventional FPCA (blue dot-dashed lines), are shown for comparison. Labels indicate local extrema of the absolute median estimated functions. Shaded regions indicate variability across fits.}
	\label{fig:app-adenine-weights}
\end{figure*}

Figure~\ref{fig:app-adenine-weights} compares the median estimated functions obtained across $250$ training--test fits. For VP--SFPCA, these correspond to the sparse score-generating weight functions, whereas for SCAD--SFPCA and conventional FPCA, they correspond to the sparse and dense FPCs, respectively. Hereafter, the term \emph{estimated functions} refers collectively to these quantities. Owing to possible changes in component order and sign across repeated
fits \citep{babamoradi2013}, the estimated functions were
permutation-matched to a common within-method reference and subsequently
sign-aligned before the medians were calculated. The labelled Raman shifts denote local extrema identified from the absolute median functions using the \texttt{detectPeaks} function in the \texttt{MALDIquant} package (version 1.22.3) \citep{gibb2012}.

The first component exhibited a dominant absolute extremum near $724$--$727\,\mathrm{cm}^{-1}$ across all three methods. Further agreement was observed around $1330\,\mathrm{cm}^{-1}$ and $1444$--$1450\,\mathrm{cm}^{-1}$. The VP--SFPCA weight function also showed a smaller extremum near $628\,\mathrm{cm}^{-1}$, together with a feature at $1534\,\mathrm{cm}^{-1}$. Compared with the dense conventional FPC, the estimated functions from both sparse methods were zero or close to zero over substantial portions of the Raman-shift domain, resulting in a more localised representation of the first component.

For the second component, the dominant absolute extremum occurred around $745\,\mathrm{cm}^{-1}$ for VP--SFPCA and conventional FPCA, and near $748\,\mathrm{cm}^{-1}$ for SCAD--SFPCA. All three methods also showed extrema at approximately $517$ and $1342$--$1351\,\mathrm{cm}^{-1}$. Greater differences were observed among the secondary features. SCAD--SFPCA showed positive extrema near $886$ and $946\,\mathrm{cm}^{-1}$, whereas several extrema identified by VP--SFPCA corresponded closely to those of conventional FPCA, particularly in the higher-wavenumber region.

Differences in localisation were more evident in the third component. The VP--SFPCA weight function retained several localised features, with extrema corresponding closely to those of conventional FPCA around $958$, $1276$, $1372$, and $1531$--$1534\,\mathrm{cm}^{-1}$. SCAD--SFPCA produced a more strongly localised third sparse FPC. Its dominant feature occurred near $1375\,\mathrm{cm}^{-1}$, with values close to zero over most of the remaining domain. Although SCAD--SFPCA yielded stronger localisation for this component, VP--SFPCA retained a broader range of features also identified by conventional FPCA while reducing the extent of the nonzero domain.

The prominent extrema around $724$--$748\,\mathrm{cm}^{-1}$ are consistent with the well-established adenine ring-breathing band, which has been reported at approximately $730$--$760\,\mathrm{cm}^{-1}$ under different experimental conditions \citep{ervin1980,kim1986,koglin1980,muniz2010,tzeng2020}. This region also overlaps the interval from $715$ to $750\,\mathrm{cm}^{-1}$ used by \citet{fornasaro2020} to obtain integrated SERS intensity for quantitative calibration. In addition, bands around $1250$--$1280\,\mathrm{cm}^{-1}$ have been assigned to mixed ring- and C--N-stretching motions with hydrogen-bending contributions \citep{lang2011,muniz2010}. The extrema near $1330$--$1351\,\mathrm{cm}^{-1}$ coincide with the prominent adenine ring-skeletal band reported around $1330$--$1340\,\mathrm{cm}^{-1}$ \citep{ervin1980,kim1986,koglin1980}. Bands in the higher-wavenumber region around $1400$--$1460\,\mathrm{cm}^{-1}$ have also been associated with in-plane motions involving ring stretching and C--H bending \citep{lang2011,muniz2010,safar2023}.

Based on the band assignments, the first two VP--SFPCA weight functions showed their most prominent features in the ring-breathing region, with additional features in higher-wavenumber adenine bands. The third weight function placed greater emphasis on the higher-wavenumber region, including features associated with ring-skeletal and ring-stretching motions. These correspondences provide spectroscopic context for interpreting the estimated sparse weight functions.

Table~\ref{tab:app-adenine-performance} summarises the overall empirical performance of the three methods across $250$ training--test fits. Both sparse methods yielded substantial coefficient sparsity under the examined settings. For VP--SFPCA, the mean proportions of nonzero B-spline coefficients were $25.4\%$, $40.1\%$, and $36.5\%$ for the three components, respectively. The corresponding proportions for SCAD--SFPCA were $31.4\%$, $41.7\%$, and $36.1\%$. Across the three components, the mean coefficient sparsity was $66.0\%$ for VP--SFPCA and $63.6\%$ for SCAD--SFPCA.

\begin{table*}[htbp]
	\centering
	\caption{Performance comparison of VP--SFPCA, SCAD--SFPCA, and conventional FPCA for the adenine dataset across $250$ training--test fits.}
	\label{tab:app-adenine-performance}
	\begin{tabular*}{\textwidth}{@{\extracolsep\fill}lcccc@{\extracolsep\fill}}
		\toprule
		\textbf{Method}
		& \textbf{Sparsity (\%)}
		& \textbf{Cumulative PVE (\%)}
		& \textbf{Held-out ISE $\boldsymbol{(\times 10^{-3})}$}
		& \textbf{Runtime (s)} \\
		\midrule
		VP--SFPCA   & $66.0$ & $69.8$ & $1.093$ & $9.01$ \\
		SCAD--SFPCA & $63.6$ & $42.5$ & $1.372$ & $1536.37$ \\
		FPCA        & $0.0$  & $81.7$ & $1.084$ & $0.38$ \\
		\bottomrule
	\end{tabular*}
	\begin{tablenotes}[flushleft]
		\item[] \textit{Note:} A common rank of $K = 3$, selected as the smallest number of components accounting for at least $80\%$ of the total variation under conventional FPCA, was retained for all methods. All reported values are means across $250$ training--test fits. For VP--SFPCA and SCAD--SFPCA, cumulative adjusted PVE is reported to account for correlation among the sparse component scores, whereas the conventional cumulative PVE is reported for FPCA.
	\end{tablenotes}
\end{table*}

Despite this level of sparsity, VP--SFPCA retained a considerable proportion of the variation represented by conventional FPCA. The first three conventional FPCs explained mean proportions of $48.2\%$, $26.5\%$, and $7.1\%$, respectively, giving a cumulative mean of $81.7\%$. The corresponding adjusted proportions for VP--SFPCA were $41.8\%$, $22.1\%$, and $5.9\%$, yielding a cumulative mean of $69.8\%$. The first two VP--SFPCA components therefore had a mean cumulative adjusted proportion of approximately $63.8\%$, with their individual contributions remaining reasonably close to those of the first two conventional FPCs.

By comparison, the three SCAD--SFPCA components explained adjusted proportions of $28.3\%$, $10.5\%$, and $3.7\%$, respectively, yielding a cumulative mean of $42.5\%$. Under the adjusted PVE defined in Equation~\eqref{eq:method-adj-var}, variation in each component score that is linearly explained by preceding scores is excluded from its contribution \citep{nie2020}. Taken together, these results indicate that, under the tuning and selection framework used in this study, VP--SFPCA accounted for a larger proportion of nonredundant variation than SCAD--SFPCA, despite the two methods exhibiting similar overall levels of coefficient sparsity.

Held-out reconstruction performance is also summarised in Table~\ref{tab:app-adenine-performance}. VP--SFPCA achieved a mean held-out reconstruction ISE of $1.093 \times 10^{-3}$, close to the $1.084 \times 10^{-3}$ obtained by conventional FPCA, corresponding to an increase of approximately $0.8\%$. SCAD--SFPCA, however, produced a higher mean held-out ISE of $1.372 \times 10^{-3}$. VP--SFPCA therefore achieved a lower held-out reconstruction error than SCAD--SFPCA under the common grouped partitions, with its overall mean ISE approximately $20.3\%$ lower. 

Adjusted PVE and reconstruction ISE nevertheless quantify different aspects of performance, with the former measuring the nonredundant variation represented by the component scores and the latter evaluating reconstruction of held-out spectra using the estimated low-rank representation. The higher held-out ISE of the sparse methods relative to conventional FPCA is consistent with the trade-off introduced by sparsity regularisation, whereby greater localisation may be accompanied by some reduction in reconstruction accuracy. Overall, this trade-off was less pronounced for VP--SFPCA than for SCAD--SFPCA.

The examined methods also differed substantially in computational efficiency. Although sparse regularisation can improve the interpretability of principal component representations, it introduces additional tuning parameters and iterative optimisation, thereby increasing the computational burden relative to conventional FPCA \citep{erichson2020,filzmoser2012}. As reported in Table~\ref{tab:app-adenine-performance}, the mean runtime was $9.01$ seconds for VP--SFPCA and $1536.37$ seconds for SCAD--SFPCA. Thus, VP--SFPCA was approximately $171$ times faster than SCAD--SFPCA under the present implementation and computational settings. Conventional FPCA remained the least computationally demanding method, with a mean runtime of $0.38$ seconds, consistent with its unpenalised eigendecomposition.

The substantial runtime difference between the two sparse methods is consistent with their distinct optimisation strategies. VP--SFPCA exploits variable projection to separate the orthogonality-constrained loading update from the proximal update of the sparse weights. The computational efficiency and scalability of this approach have been demonstrated in the multivariate setting \citep{erichson2020}. In contrast, SCAD--SFPCA relies on an alternating procedure involving local quadratic approximations to the functional SCAD penalty and repeated penalised functional regressions \citep{nie2020}. Although both sparse methods require tuning over multiple sparsity and ridge parameter combinations, the more computationally intensive iterative updates of SCAD--SFPCA correspond to the much longer runtimes observed in the present analysis.

\begin{figure*}[htbp]
	\centering
	\includegraphics[width=0.80\textwidth]{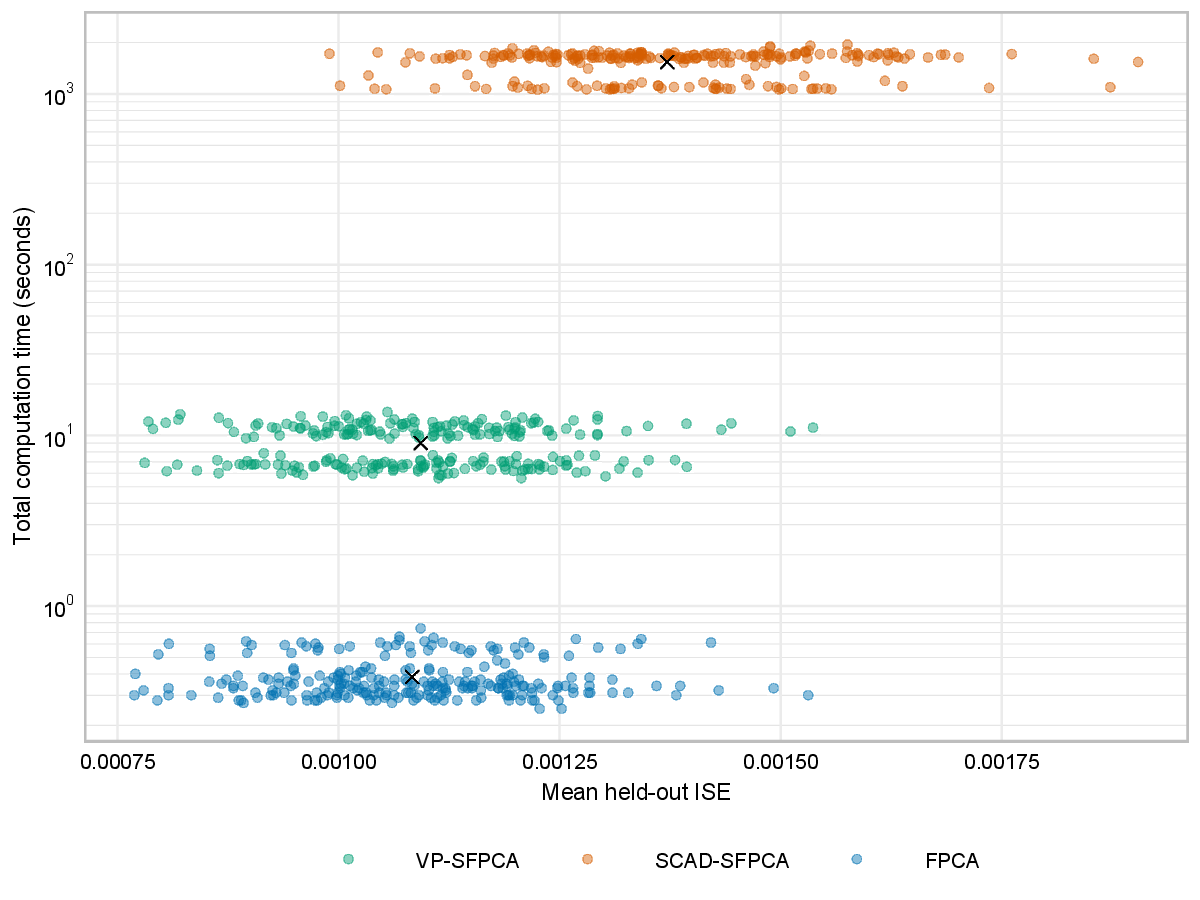}
	\caption{Mean held-out ISE versus computation time for the adenine dataset. Each coloured point represents one training--test fit, and black crosses denote method-specific means across $250$ fits.}
	\label{fig:app-adenine-tradeoff}
\end{figure*}

Figure~\ref{fig:app-adenine-tradeoff} illustrates the separation among the three methods in total computation time. VP--SFPCA remained close to conventional FPCA in held-out reconstruction accuracy while requiring considerably less computation than SCAD--SFPCA. Meanwhile, SCAD--SFPCA had the highest mean held-out ISE and runtime among the three methods, with its mean ISE approximately $25.5\%$ higher than that of VP--SFPCA. Conventional FPCA remained the fastest method and achieved slightly lower reconstruction error, but its dense components represented global dominant modes rather than localised spectral regions.

\section{Conclusion}
\label{sec:conclusion}

This study proposed variable-projection sparse functional principal component analysis for interpretable dimensionality reduction of functional data. The method formulates sparse FPCA as a regularised matrix-factorisation problem that incorporates the functional inner-product geometry induced by the basis representation. It separates sparse score-generating weight functions from orthonormal reconstruction loadings. Through variable projection, the orthogonality-constrained loading block is minimised conditionally on the sparse weights, leaving a reduced optimisation problem over the sparse weights.

The simulation studies showed that both sparse methods recovered the localised functional structure under the settings considered, with recovery generally improving as the sample size increased. Across both models and sample sizes, VP--SFPCA yielded lower component-wise mean IE than SCAD--SFPCA, whereas conventional FPCA showed larger errors relative to the sparse recovery targets. The two sparse methods nevertheless achieved closely comparable recovery of the simulated scores. VP--SFPCA also required substantially less computation than SCAD--SFPCA while achieving lower sparse-function recovery errors.

In the empirical SERS analyses, the localised component weights identified Raman-shift regions that contribute strongly to the dominant modes of spectral variation. Several prominent features of the VP--SFPCA weight functions coincided with established SERS bands for the molecule under study. These identified regions should nevertheless be interpreted as statistical sources of spectral variation rather than as definitive evidence of molecular or chemical mechanisms.

In sum, VP--SFPCA achieved a favourable balance among localisation, reconstruction accuracy, and computational efficiency. Conventional FPCA remained the fastest method and achieved the lowest held-out reconstruction error in the empirical analysis, but its dense components provided limited localisation. SCAD--SFPCA offered an established sparse functional alternative with a similar overall level of coefficient sparsity, although it was substantially more computationally demanding and produced higher held-out reconstruction error. The lower computational burden of VP--SFPCA may also make repeated tuning more practical than for SCAD--SFPCA. More broadly, the primary role of sparsity in the proposed framework is to enhance the interpretability of dominant functional modes through localisation while preserving reconstruction accuracy close to conventional FPCA.

Despite these advantages, several limitations remain. The degree of estimated sparsity depends on the basis dimension, tuning grids, and model-selection criterion. Moreover, sparsity is imposed on the B-spline coefficients rather than directly on intervals of the functional domain. Since neighbouring B-spline basis functions have overlapping support, individual zero coefficients do not necessarily produce exactly zero-valued regions of the estimated function. Future research could investigate alternative penalties that promote sparsity more directly over contiguous regions of the functional domain. Robust extensions could reduce sensitivity to outliers and atypical curves, while randomised linear-algebra techniques may improve computational scalability to larger functional datasets.


\subsection*{Funding}
\noindent This work was supported by Universiti Malaya through the Universiti Malaya Research Excellence Grant (Grant Number: UMREG044-2024). 


\bibliographystyle{plainnat}
\bibliography{referencesMain}

\end{document}


\maketitle
\vspace{-1.0em}

\tableofcontents

\clearpage


\section{Additional Results for the Simulation Study}
\label{sec:supp-sim}

This section provides additional results for the simulation study presented in Section~3 of the article. For each of the two simulation models, pilot studies were carried out at sample sizes $n \in \{20, 50, 100\}$, with $50$ Monte Carlo replications for each setting. The pilot analyses were used to select the iteration cap and to assess the sensitivity of the VP--SFPCA ridge-parameter grid. The full simulation study was then carried out at $n \in \{50, 100\}$, with $100$ Monte Carlo replications for each setting.

\FloatBarrier

\subsection{Pilot Study: Iteration-Cap Selection}
\label{subsec:supp-sim-pilot-iter}

Figure~\ref{fig:supp-sim-pilot-iter} compares the best attained objective values under iteration caps of $30$, $40$, and $50$. Since the objective functions are method-specific, comparisons are made across iteration caps within each method rather than between VP--SFPCA and SCAD--SFPCA. Increasing the iteration cap beyond $30$ produced no consistent improvement. The VP--SFPCA results were nearly unchanged across the three caps, while the SCAD--SFPCA distributions showed substantial overlap. These results support the use of a $30$-iteration cap in the full simulation study.

\begin{figure*}[htbp]
	\centering
	\subfloat[]{%
		\includegraphics[width=0.80\textwidth]{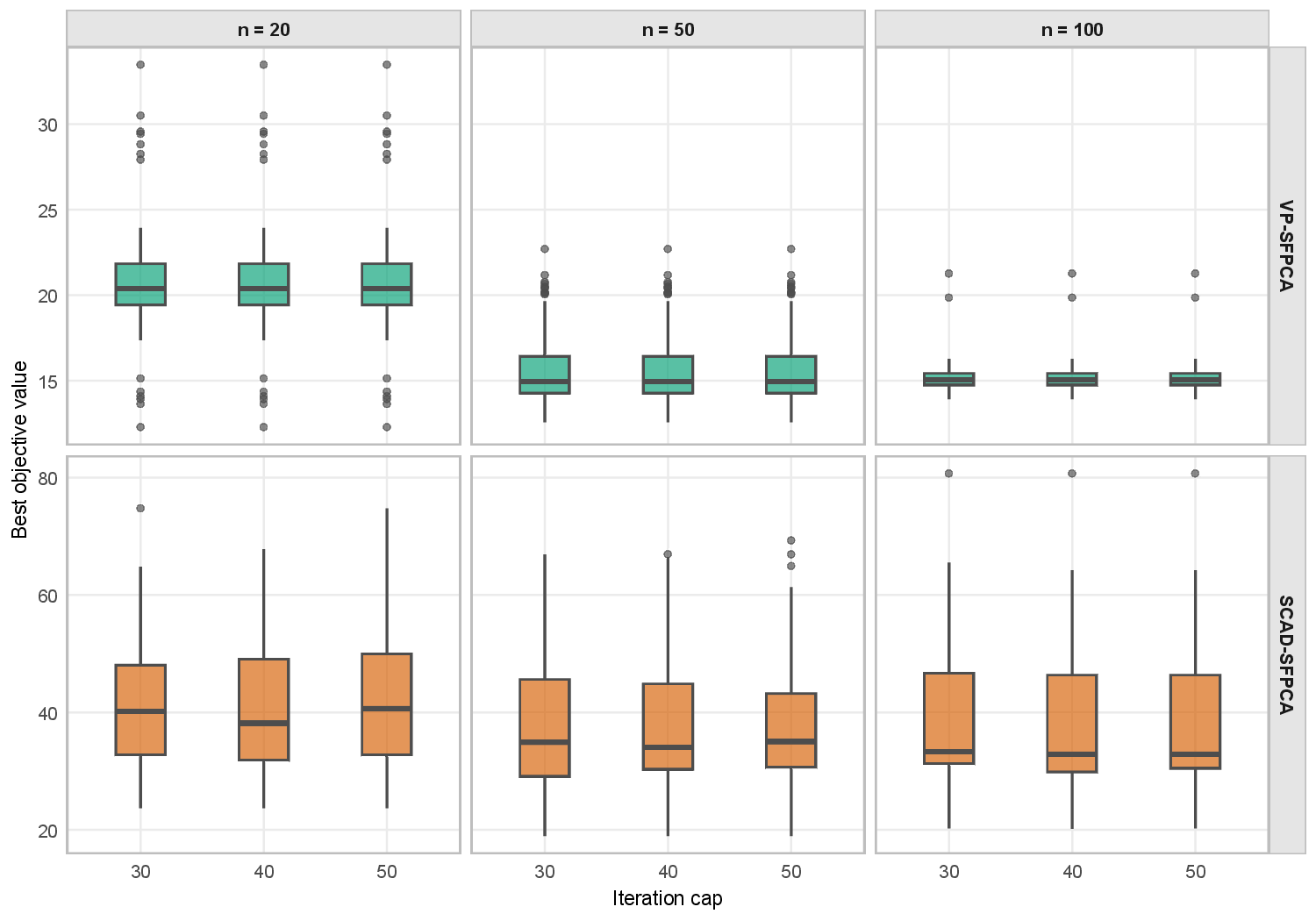}%
		\label{fig:supp-sim-pilot-iter-model1}%
	}
	\caption{Sensitivity of the best attained objective values to the iteration cap under (a) Model~1 and (b) Model~2, based on $50$ Monte Carlo replications.}
	\label{fig:supp-sim-pilot-iter}
\end{figure*}

\begin{figure*}[htbp]
	\ContinuedFloat
	\centering
	\subfloat[]{%
		\includegraphics[width=0.80\textwidth]{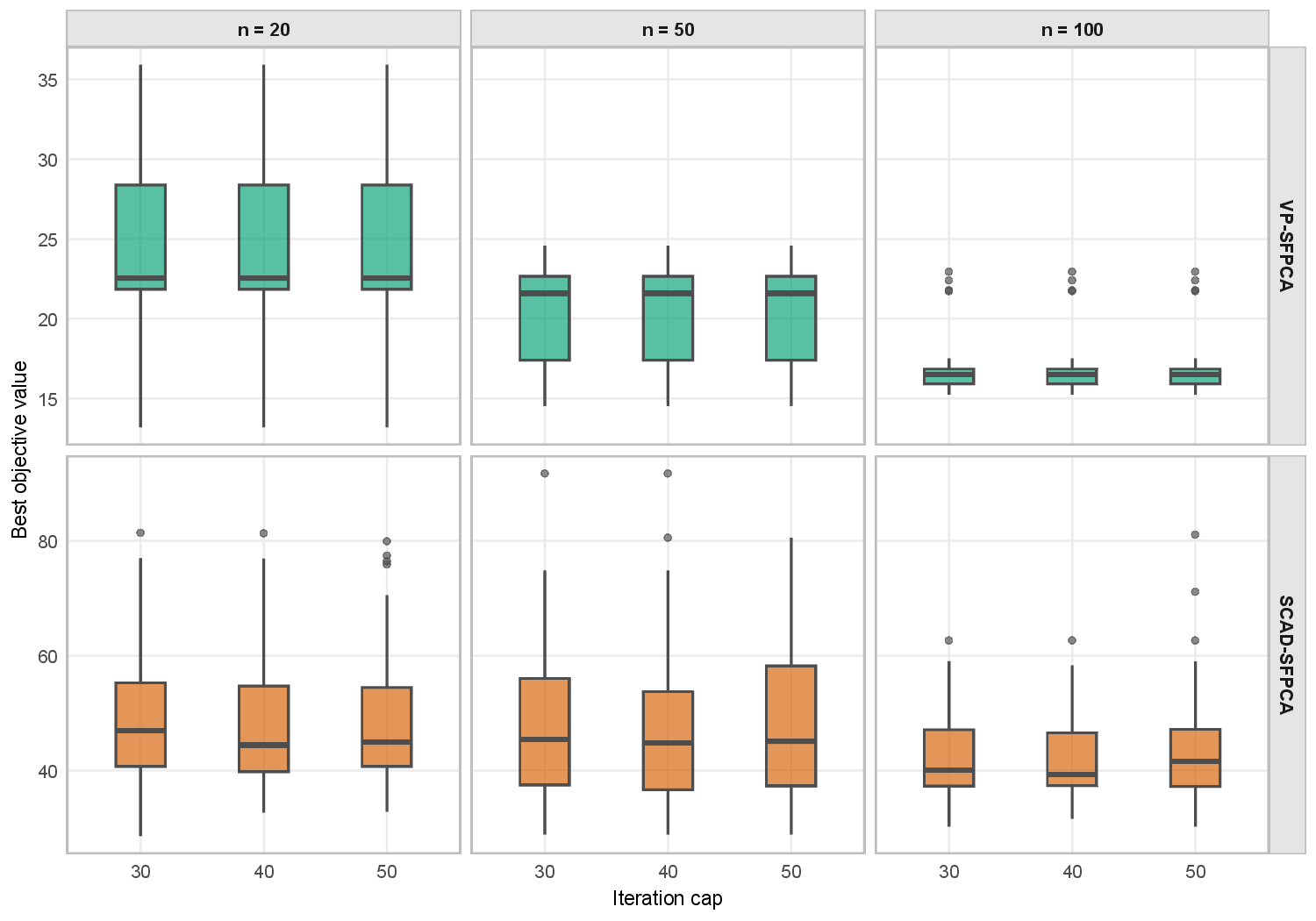}%
		\label{fig:supp-sim-pilot-iter-model2}%
	}
	\caption[]{(Continued)}
\end{figure*}

\FloatBarrier

\subsection{Pilot Study: Tuning-Grid Assessment}
\label{subsec:supp-sim-pilot-grid}

The tuning-parameter selections were also examined in the pilot study to assess whether the candidate grids constrained the selected values at their boundaries. Attention was given to the VP--SFPCA ridge parameter $\tau$, for which frequent lower-boundary selection was observed. The smoothing and sparsity grids were held fixed at $\gamma \in \mathcal{G}_{7}(-4,2)$ and $\lambda \in \mathcal{G}_{10}(-4,-1)$, respectively, while four candidate grids covering different ranges of $\tau$ were examined. The pilot analyses were conducted separately for each model at $n \in \{20,50,100\}$, using $50$ Monte Carlo replications and an iteration cap of $30$. Table~\ref{tab:supp-sim-pilot-ridge} summarises the candidate $\tau$ tuning grids considered in the sensitivity analysis.

\begin{table}[H]
	\centering
	\caption{Candidate VP--SFPCA $\tau$ tuning grids examined in the simulation pilot analysis.}
	\label{tab:supp-sim-pilot-ridge}
	\begin{tabular*}{\textwidth}{@{\extracolsep\fill}lc@{\extracolsep\fill}}%
		\toprule
		\textbf{Set} & $\boldsymbol{\tau}$ \\
		\midrule
		1 & $\mathcal{G}_{7}(-4,-1)$ \\
		2 & $\mathcal{G}_{7}(-6,-2)$ \\
		3 & $\mathcal{G}_{7}(-7,-1)$ \\
		4 & $\mathcal{G}_{7}(-10,-1)$ \\
		\bottomrule
	\end{tabular*}
\end{table}

Despite considering grids that included substantially smaller $\tau$ values, frequent selection of the lower boundary persisted across all four sets. Table~\ref{tab:supp-sim-pilot-boundary} reports the corresponding lower-boundary selection frequencies. Under Model~1, the frequencies ranged from $88\%$ to $94\%$, while under Model~2 they ranged from $96\%$ to $100\%$. Thus, allowing substantially smaller $\tau$ values did not systematically shift the selected values into the interior.

\begin{table}[H]
	\centering
	\caption{Lower-boundary selection frequencies (\%) for VP--SFPCA $\tau$ across tuning grids.}
	\label{tab:supp-sim-pilot-boundary}
	\begin{tabular*}{\textwidth}{@{\extracolsep\fill}lccccc@{\extracolsep\fill}}%
		\toprule
		\textbf{Setting} & $\boldsymbol{n}$ & \textbf{Set~1} & \textbf{Set~2} & \textbf{Set~3} & \textbf{Set~4} \\
		\midrule
		\multirow{3}{*}{Model 1}
		& 20  & 90 & 88 & 90 & 92 \\
		& 50  & 90 & 92 & 94 & 92 \\
		& 100 & 90 & 90 & 94 & 90 \\
		\midrule
		\multirow{3}{*}{Model 2}
		& 20  & 96  & 96  & 96  & 100 \\
		& 50  & 98  & 98  & 100 & 98  \\
		& 100 & 100 & 100 & 100 & 100 \\
		\bottomrule
	\end{tabular*}
\end{table}

Recovery performance was also essentially unchanged across the four $\tau$ grids. Table~\ref{tab:supp-sim-pilot-ie} reports the mean integrated error (IE) averaged across the four components. The differences were negligible even when the lower endpoint of the $\tau$ grid was reduced from $10^{-4}$ to $10^{-10}$.

\begin{table}[H]
	\centering
	\caption{Mean IE across the four components under the candidate VP--SFPCA $\tau$ grids.}
	\label{tab:supp-sim-pilot-ie}
	\begin{tabular*}{\textwidth}{@{\extracolsep\fill}lccccc@{\extracolsep\fill}}%
		\toprule
		\textbf{Setting} & $\boldsymbol{n}$ & \textbf{Set~1} & \textbf{Set~2} & \textbf{Set~3} & \textbf{Set~4} \\
		\midrule
		\multirow{3}{*}{Model 1}
		& 20  & 0.1035 & 0.1035 & 0.1038 & 0.1035 \\
		& 50  & 0.0135 & 0.0135 & 0.0135 & 0.0138 \\
		& 100 & 0.0078 & 0.0078 & 0.0078 & 0.0078 \\
		\midrule
		\multirow{3}{*}{Model 2}
		& 20  & 0.1470 & 0.1470 & 0.1470 & 0.1470 \\
		& 50  & 0.0233 & 0.0233 & 0.0233 & 0.0233 \\
		& 100 & 0.0155 & 0.0155 & 0.0155 & 0.0155 \\
		\bottomrule
	\end{tabular*}
\end{table}

The persistent lower-boundary selection, together with the nearly unchanged recovery errors, indicates that further reductions of the lower $\tau$ bound had little influence on recovery under the examined simulation settings. Set~1 was therefore retained for the full simulation study.

\FloatBarrier

\subsection{Score Recovery Assessed by Tucker Congruence}
\label{subsec:supp-sim-tucker}

Recovery of the low-dimensional score structure was additionally assessed using Tucker congruence between the true and estimated scores for $K = 4$ components within each Monte Carlo replication. Tucker congruence provides a cosine-type measure of similarity, with larger values indicating stronger agreement between the true and estimated component scores \citep{lorenzo2006,park2024}. Congruence values between $0.85$ and $0.94$ are commonly interpreted as indicating fair similarity \citep{lorenzo2006}.

Table~\ref{tab:supp-sim-tucker} reports the mean and standard deviation of Tucker congruence across $100$ Monte Carlo replications for each setting. Across both models and sample sizes, the three methods produced broadly similar mean Tucker congruence values. VP--SFPCA and SCAD--SFPCA yielded mean congruence values within the range commonly interpreted as indicating fair similarity in all settings. FPCA showed slightly lower mean congruence at $n=50$, particularly under Model~1, while its values increased at $n=100$. The standard deviations generally decreased with increasing sample size, indicating more stable score recovery. Overall, VP--SFPCA and SCAD--SFPCA showed closely comparable recovery of the underlying score structure, with FPCA exhibiting broadly similar agreement.

\begin{table}[H]
	\centering
	\caption{Tucker congruence coefficients for score recovery under Models~1 and~2. Values are reported as means (standard deviations) across $100$ Monte Carlo replications.}
	\label{tab:supp-sim-tucker}
	\begin{tabular*}{\textwidth}{@{\extracolsep\fill}lcccc@{\extracolsep\fill}}%
		\toprule
		\textbf{Setting} & $\boldsymbol{n}$ & \textbf{VP--SFPCA} & \textbf{SCAD--SFPCA} & \textbf{FPCA} \\
		\midrule
		\multirow{2}{*}{Model 1}
		& 50  & 0.868 (0.028) & 0.871 (0.027) & 0.845 (0.031) \\
		& 100 & 0.870 (0.014) & 0.872 (0.014) & 0.857 (0.018) \\
		\midrule
		\multirow{2}{*}{Model 2}
		& 50  & 0.853 (0.028) & 0.856 (0.025) & 0.848 (0.029) \\
		& 100 & 0.856 (0.016) & 0.858 (0.015) & 0.860 (0.018) \\
		\bottomrule
	\end{tabular*}
\end{table}

\FloatBarrier

\subsection{Computation Time by Tuning Stage}
\label{subsec:supp-sim-time-stage}

Figure~\ref{fig:supp-sim-time-stage} shows the distributions of computation time by tuning stage across $100$ Monte Carlo replications for each model and sample size. Tuning of $\gamma$ required little computation for either sparse method, with mean times of approximately $0.02$--$0.04$ seconds across the examined settings. A more pronounced computational difference occurred during the joint tuning of $\lambda$ and $\tau$. Under Model~1, the mean joint-tuning times were $0.62$ and $0.74$ seconds for VP--SFPCA at $n=50$ and $n=100$, respectively, compared with $85.17$ and $95.44$ seconds for SCAD--SFPCA. Under Model~2, the corresponding mean times were $0.59$ and $0.62$ seconds for VP--SFPCA, compared with $103.50$ and $106.37$ seconds for SCAD--SFPCA. The stage-specific results indicate that the overall runtime difference between the two sparse methods arose predominantly during the joint tuning of $\lambda$ and $\tau$. These timings pertain to the present implementations and computational settings.

\begin{figure*}[htbp]
	\centering
	\subfloat[]{%
		\includegraphics[width=0.80\textwidth]{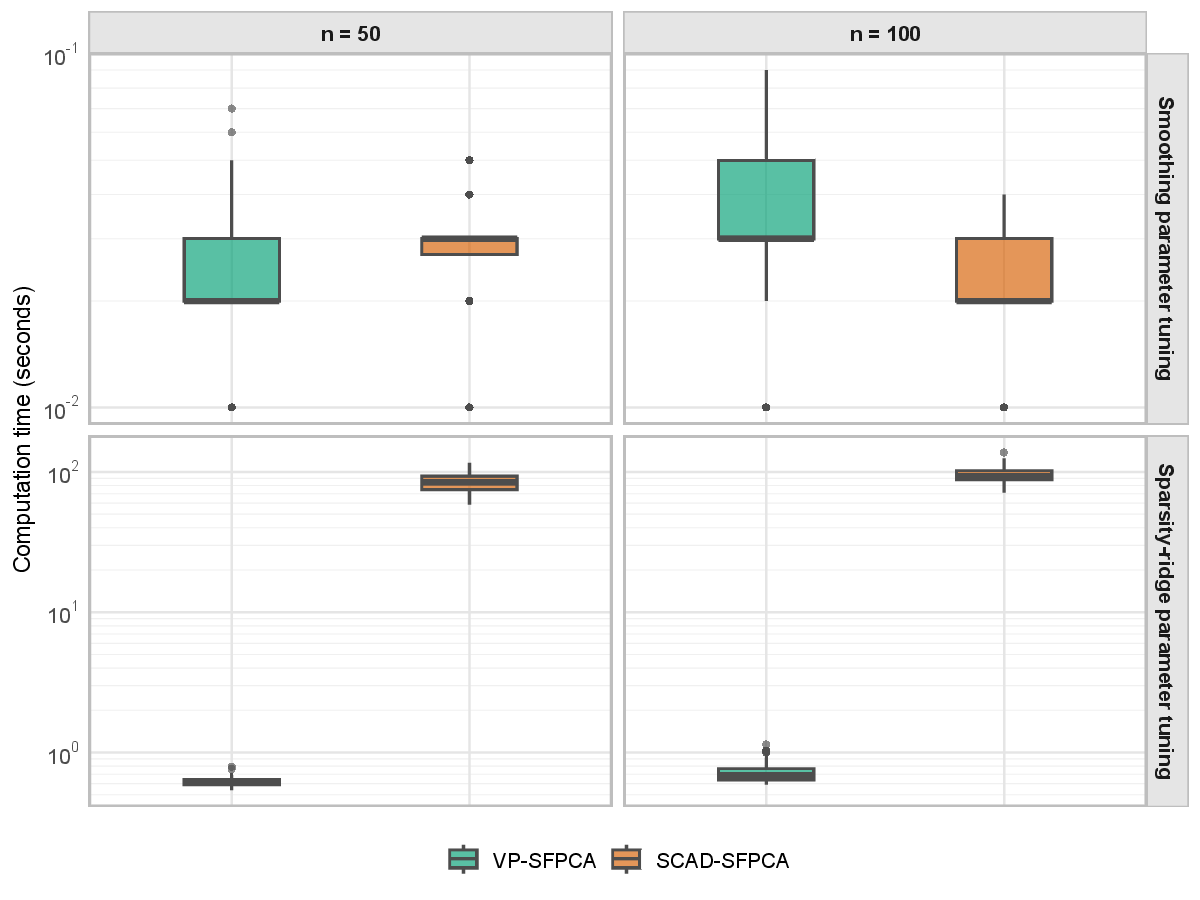}%
		\label{fig:supp-sim-time-stage-model1}%
	}
	\caption{Distributions of computation time by tuning stage under (a) Model~1 and (b) Model~2 for $n = 50$ and $n = 100$, based on $100$ Monte Carlo replications.}
	\label{fig:supp-sim-time-stage}
\end{figure*}

\begin{figure*}[htbp]
	\ContinuedFloat
	\centering
	\subfloat[]{%
		\includegraphics[width=0.80\textwidth]{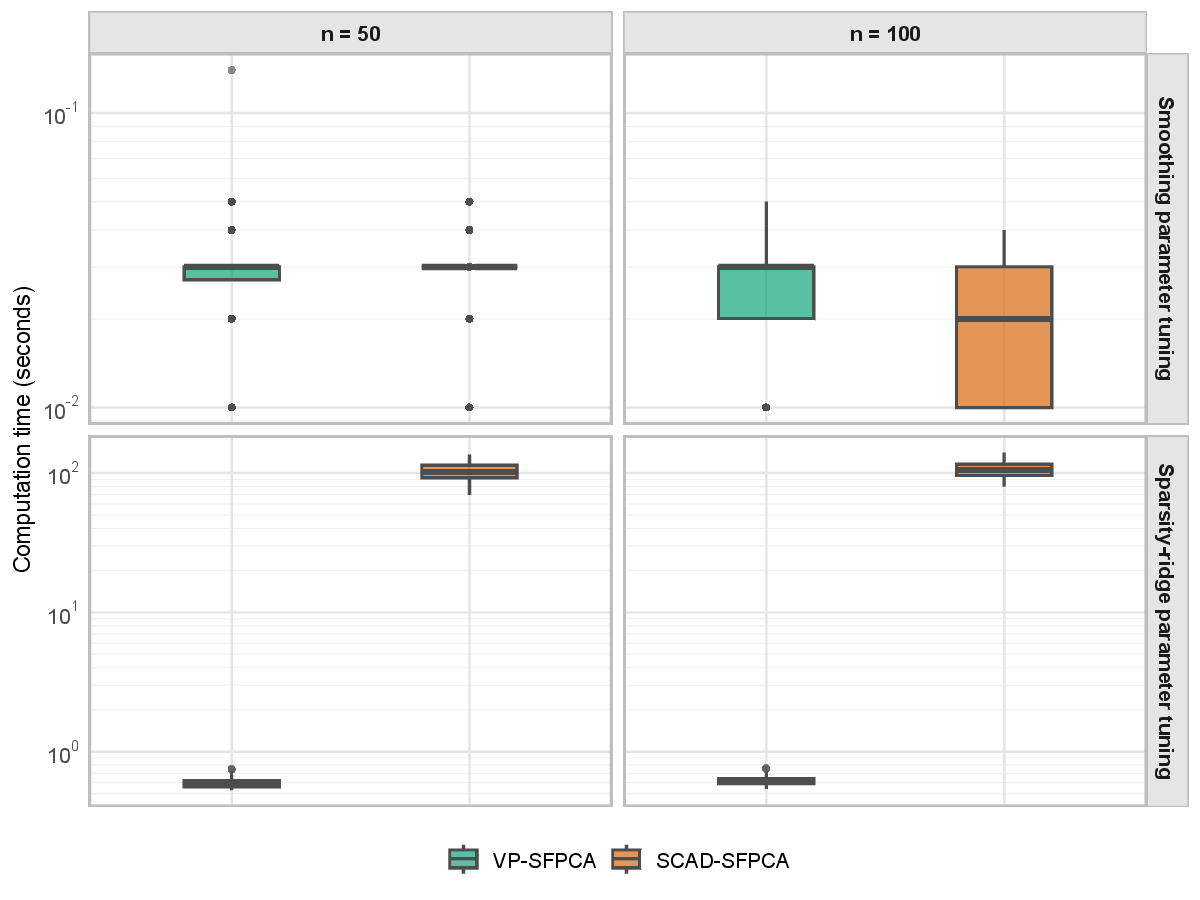}%
		\label{fig:supp-sim-time-stage-model2}%
	}
	\caption[]{(Continued)}
\end{figure*}

\clearpage


\section{Additional Results for the Adenine Application}
\label{sec:supp-adenine}

This section provides additional results for the adenine application presented in Section~4 of the article. Preliminary sensitivity analyses were conducted using five repetitions of grouped five-fold cross-validation. The basis dimension and iteration cap were assessed first, followed by a comparison of three candidate tuning-grid sets. The selected settings were then fixed before the full analysis with $50$ cross-validation repetitions. Additional results are provided for localisation stability and computation time by tuning stage.

\FloatBarrier

\subsection{Pilot Study: Basis-Dimension and Iteration-Cap Assessment}
\label{subsec:supp-adenine-pilot-iter}

A preliminary sensitivity analysis was first conducted to assess the basis dimension and iteration cap across the three candidate tuning-grid sets given in Table~\ref{tab:supp-adenine-pilot-grids}. 

\begin{table}[H]
	\centering
	\caption{Candidate tuning grids examined in the adenine pilot analysis.}
	\label{tab:supp-adenine-pilot-grids}
	\begin{tabular*}{\textwidth}{@{\extracolsep\fill}llccc@{\extracolsep\fill}}%
		\toprule
		\textbf{Set} & \textbf{Method} & $\boldsymbol{\gamma}$ & $\boldsymbol{\lambda}$ & $\boldsymbol{\tau}$ \\
		\midrule
		\multirow{2}{*}{1}
		& VP--SFPCA
		& $\mathcal{G}_{7}(-4,2)$
		& $\mathcal{G}_{15}(-5,-1)$
		& $\mathcal{G}_{15}(-5,-1)$ \\
		& SCAD--SFPCA
		& $\mathcal{G}_{7}(-4,2)$
		& $\mathcal{G}_{15}(-6,-4)$
		& $\mathcal{G}_{15}(-6,-4)$ \\
		\midrule
		\multirow{2}{*}{2}
		& VP--SFPCA
		& $\mathcal{G}_{9}(-4,4)$
		& $\mathcal{G}_{15}(-5,-1)$
		& $\mathcal{G}_{15}(-7,-1)$ \\
		& SCAD--SFPCA
		& $\mathcal{G}_{9}(-4,4)$
		& $\mathcal{G}_{15}(-6,-3)$
		& $\mathcal{G}_{15}(-6,-3)$ \\
		\midrule
		\multirow{2}{*}{3}
		& VP--SFPCA
		& $\mathcal{G}_{9}(-4,4)$
		& $\mathcal{G}_{15}(-4,-1)$
		& $\mathcal{G}_{15}(-4,-1)$ \\
		& SCAD--SFPCA
		& $\mathcal{G}_{9}(-4,4)$
		& $\mathcal{G}_{15}(-5,-2)$
		& $\mathcal{G}_{15}(-5,-2)$ \\
		\bottomrule
	\end{tabular*}
\end{table}

Figure~\ref{fig:supp-adenine-pilot-iter} presents the resulting accuracy--computation trade-offs for VP--SFPCA and SCAD--SFPCA in terms of mean held-out integrated squared error (ISE) and mean computation time. The overall sensitivity patterns were similar across the three grid specifications.

For the basis-dimension assessment, $p \in \{30,40,50\}$ was examined with the iteration cap fixed at $40$. Under each candidate grid set, the mean held-out ISE decreased as the basis dimension increased for both sparse methods, with $p = 50$ yielding the lowest value. This setting required greater computation time, particularly for SCAD--SFPCA, but yielded the lowest mean held-out ISE among the candidate basis dimensions. The basis dimension was therefore fixed at $p = 50$ for the subsequent tuning-grid assessment and full analysis.

For the iteration-cap assessment, caps of $30$, $40$, and $50$ were examined with the basis dimension fixed at $p = 40$. Differences in mean held-out ISE across the three caps were relatively small. Increasing the cap did not provide a consistent improvement across the sparse methods and candidate grid sets. An iteration cap of $30$ was therefore selected for the subsequent tuning-grid assessment and full analysis.

\begin{figure*}[htbp]
	\centering
	\subfloat[]{%
		\includegraphics[width=0.80\textwidth]{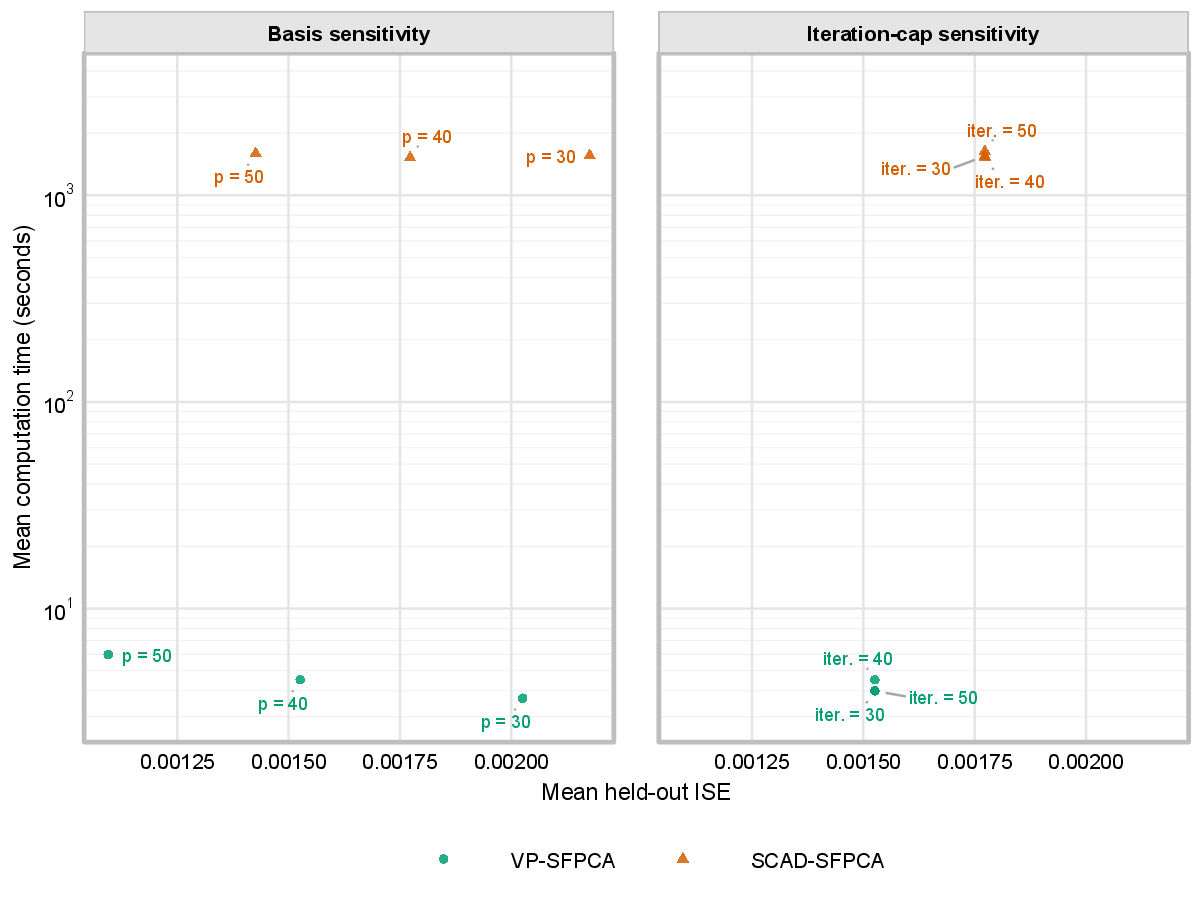}%
		\label{fig:supp-adenine-pilot-iter-set1}%
	}\\[1em]
	\subfloat[]{%
		\includegraphics[width=0.80\textwidth]{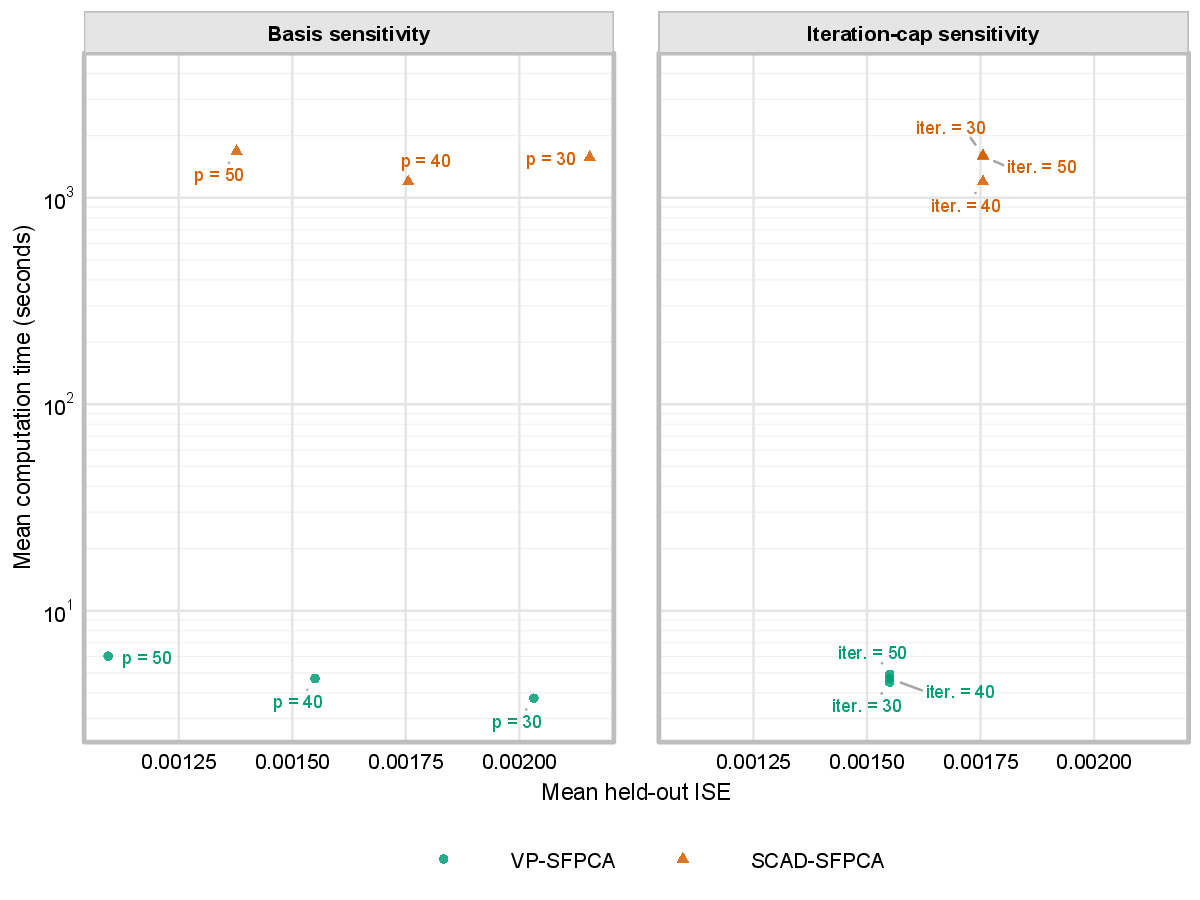}%
		\label{fig:supp-adenine-pilot-iter-set2}%
	}	
	\caption{Accuracy--computation trade-offs for the basis-dimension and iteration-cap sensitivity analyses for the adenine dataset under the candidate tuning-grid sets: (a) Set~1, (b) Set~2, and (c) Set~3, based on five repetitions of grouped five-fold cross-validation.}
	\label{fig:supp-adenine-pilot-iter}
\end{figure*}

\begin{figure*}[htbp]
	\ContinuedFloat
	\centering
	\subfloat[]{%
		\includegraphics[width=0.80\textwidth]{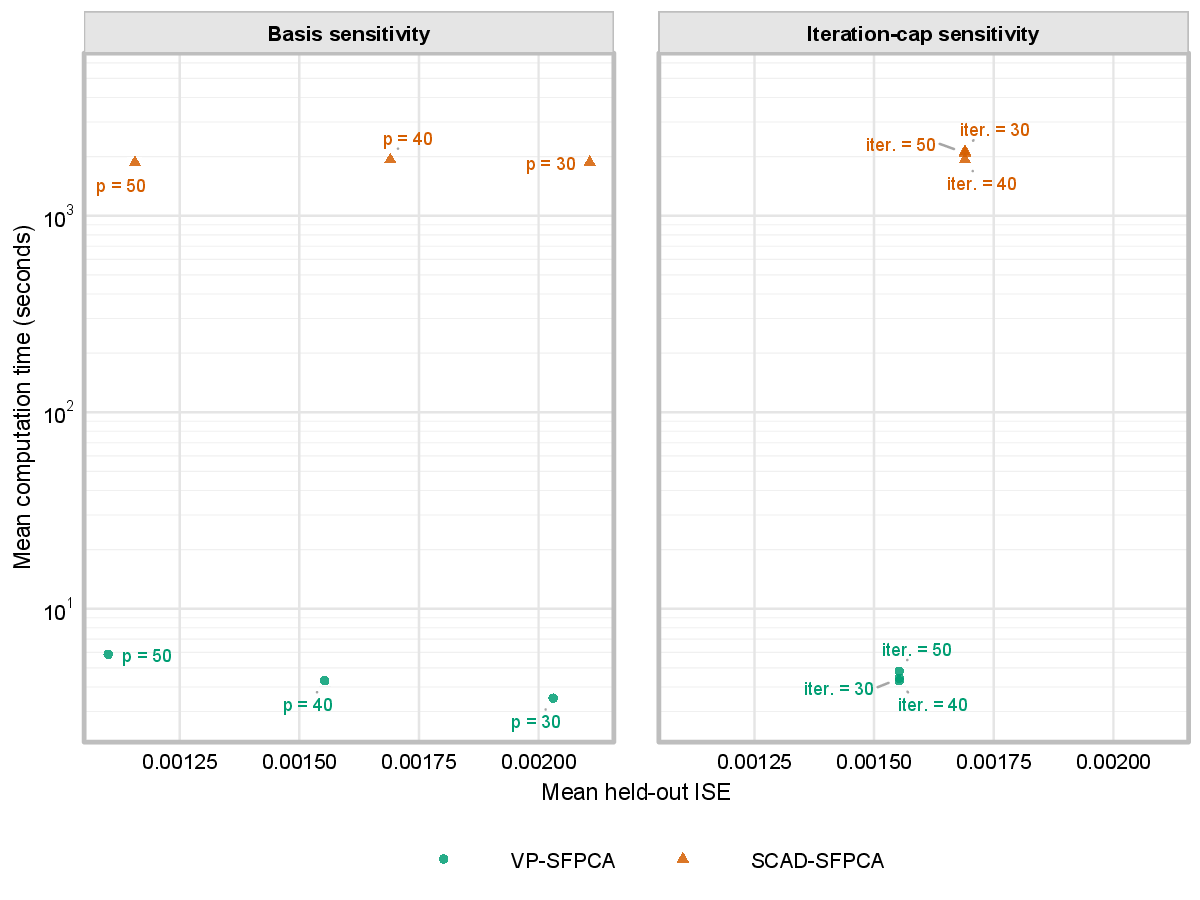}%
		\label{fig:supp-adenine-pilot-iter-set3}%
	}
	\caption[]{(Continued)}
\end{figure*}

\FloatBarrier

\subsection{Pilot Study: Tuning-Grid Assessment}
\label{subsec:supp-adenine-pilot-grid}

With the basis dimension fixed at $p = 50$ and the iteration cap at $30$, the three candidate tuning-grid sets given in Table~\ref{tab:supp-adenine-pilot-grids} were assessed in terms of boundary-constrained tuning-parameter selection and held-out reconstruction performance. The locations of the selected tuning parameters within each candidate grid are summarised in Table~\ref{tab:supp-adenine-pilot-boundary}.

Under Set~1, $\gamma$ was selected at its upper boundary in all pilot fits for both sparse methods. For SCAD--SFPCA, $\tau$ was also selected at the upper boundary in $96\%$ of fits, while $\lambda$ was selected at one of its boundaries in $88\%$ of fits. For Set~2, all $\gamma$ selections moved to the interior, and the upper-boundary selection frequency for SCAD--SFPCA $\tau$ decreased to $84\%$. The proportion of interior selections for SCAD--SFPCA $\lambda$ also increased from $12\%$ to $52\%$. For VP--SFPCA, $\lambda$ remained in the interior throughout, whereas $\tau$ continued to be selected at its lower boundary.

Further modification of the tuning ranges in Set~3 did not provide a consistent improvement in the selection behaviour. Although SCAD--SFPCA $\lambda$ was selected in the interior in $80\%$ of the pilot fits, $\tau$ was selected at its upper boundary in every fit. For VP--SFPCA, $\tau$ likewise remained at its lower boundary throughout all three grid sets.

\begin{table}[H]
	\centering
	\caption{Boundary and interior selection frequencies (\%) across tuning grids for the adenine dataset.}
	\label{tab:supp-adenine-pilot-boundary}
	\begin{tabular*}{\textwidth}{@{\extracolsep\fill}lllccc@{\extracolsep\fill}}%
		\toprule
		\textbf{Set} & \textbf{Method} & \textbf{Parameter} & \textbf{Lower boundary} & \textbf{Interior} & \textbf{Upper boundary} \\
		\midrule
		\multirow{6}{*}{1}
		& \multirow{3}{*}{VP--SFPCA}
		& $\gamma$ & 0   & 0   & 100 \\
		& & $\lambda$ & 0   & 100 & 0 \\
		& & $\tau$ & 100 & 0   & 0 \\
		\cmidrule(lr){2-6}
		& \multirow{3}{*}{SCAD--SFPCA}
		& $\gamma$ & 0   & 0   & 100 \\
		& & $\lambda$ & 36  & 12  & 52 \\
		& & $\tau$ & 0   & 4   & 96 \\
		\midrule
		\multirow{6}{*}{2}
		& \multirow{3}{*}{VP--SFPCA}
		& $\gamma$ & 0   & 100 & 0 \\
		& & $\lambda$ & 0   & 100 & 0 \\
		& & $\tau$ & 100 & 0   & 0 \\
		\cmidrule(lr){2-6}
		& \multirow{3}{*}{SCAD--SFPCA}
		& $\gamma$ & 0   & 100 & 0 \\
		& & $\lambda$ & 0   & 52  & 48 \\
		& & $\tau$ & 0   & 16  & 84 \\
		\midrule
		\multirow{6}{*}{3}
		& \multirow{3}{*}{VP--SFPCA}
		& $\gamma$ & 0   & 100 & 0 \\
		& & $\lambda$ & 0   & 100 & 0 \\
		& & $\tau$ & 100 & 0   & 0 \\
		\cmidrule(lr){2-6}
		& \multirow{3}{*}{SCAD--SFPCA}
		& $\gamma$ & 0   & 100 & 0 \\
		& & $\lambda$ & 20  & 80  & 0 \\
		& & $\tau$ & 0   & 0   & 100 \\
		\bottomrule
	\end{tabular*}
\end{table}

Held-out reconstruction performance is summarised in Table~\ref{tab:supp-adenine-pilot-grid-ise}. For VP--SFPCA, the mean held-out ISE was unchanged between Sets~1 and~2, and increased slightly under Set~3. For SCAD--SFPCA, the mean held-out ISE decreased across the three sets, although the lower value under Set~3 coincided with upper-boundary selection of $\tau$ in all pilot fits. The boundary diagnostics and held-out reconstruction results therefore indicated that Set~2 provided a more balanced tuning specification, whereas the further modification in Set~3 did not provide a consistent improvement across the two methods. Set~2 was therefore retained for the full adenine analysis.

\begin{table}[H]
	\centering
	\caption{Mean held-out ISE ($\times 10^{-3}$) under the candidate tuning grids for the adenine dataset.}
	\label{tab:supp-adenine-pilot-grid-ise}
	\begin{tabular*}{\textwidth}{@{\extracolsep\fill}lcc@{\extracolsep\fill}}%
		\toprule
		\textbf{Set} & \textbf{VP--SFPCA} & \textbf{SCAD--SFPCA} \\
		\midrule
		1 & 1.095 & 1.426 \\
		2 & 1.095 & 1.378 \\
		3 & 1.101 & 1.157 \\
		\bottomrule
	\end{tabular*}
\end{table}

\FloatBarrier

\subsection{Stability of Localisation}
\label{subsec:supp-adenine-localisation}

The stability of localisation was assessed using component-specific selection frequencies across $250$ training--test fits. Before aggregation, the same within-method component mapping used for the estimated-function summaries was applied to the component labels. For the adenine data, the identity permutation was obtained in all $250$ fits for both VP--SFPCA and SCAD--SFPCA, so no changes to the component ordering were required. 

\begin{figure*}[htbp]
	\centering
	\includegraphics[width=0.80\textwidth]{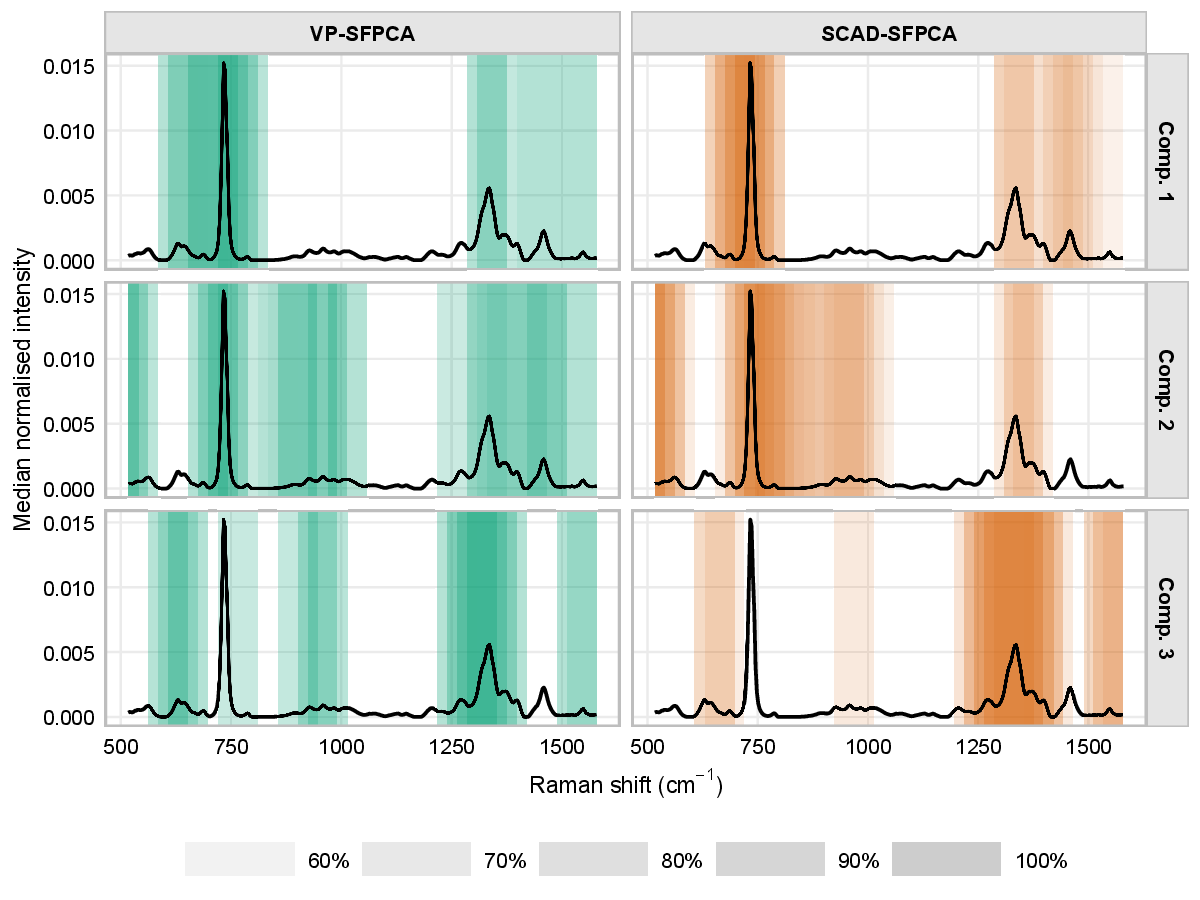}
    \caption{Selection frequencies of localised regions identified by VP--SFPCA and SCAD--SFPCA across $250$ training--test fits for the adenine dataset. Shaded regions denote Raman-shift intervals selected in at least $60\%$ of the fits, with darker shading representing higher selection frequencies. Black curves show the median preprocessed adenine spectrum.}
	\label{fig:supp-adenine-localisation}
\end{figure*}

Figure~\ref{fig:supp-adenine-localisation} shows the frequency with which Raman-shift regions were selected, together with the median preprocessed adenine spectrum for each component. Darker shading indicates more frequent selection. Both sparse methods identified regions that were repeatedly selected across the cross-validation fits, although the extent and location of these regions differed across methods and components.

\FloatBarrier

\subsection{Computation Time by Tuning Stage}
\label{subsec:supp-adenine-time-stage}

\begin{figure*}[htbp]
	\centering
	\includegraphics[width=0.80\textwidth]{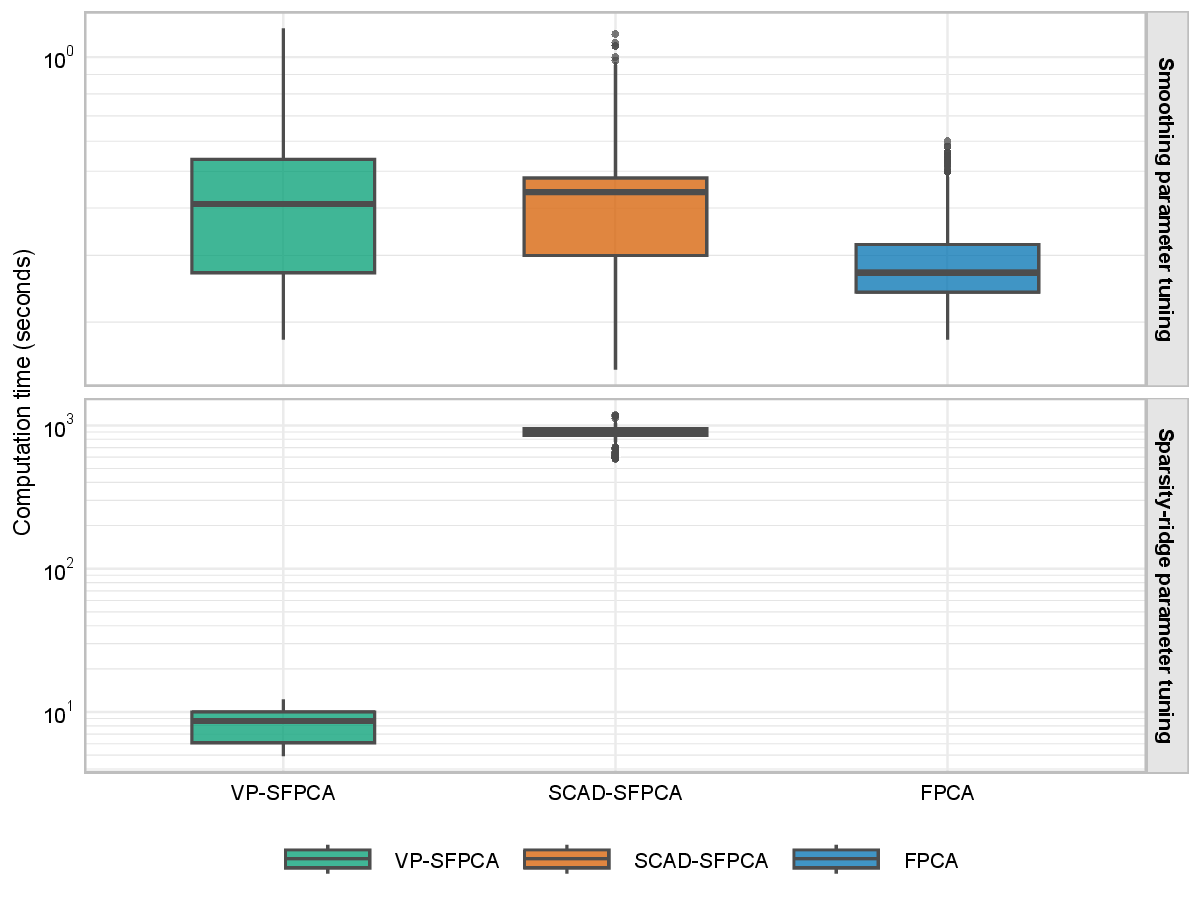}
	\caption{Distributions of computation time by tuning stage across $250$ fits of the repeated grouped five-fold cross-validation analysis for the adenine dataset.}
	\label{fig:supp-adenine-time-stage}
\end{figure*}

Figure~\ref{fig:supp-adenine-time-stage} summarises computation time by tuning stage across $250$ fits of the repeated grouped five-fold cross-validation analysis. Smoothing-parameter selection required relatively little computation for all three methods. Conventional FPCA does not involve sparsity--ridge tuning and is therefore shown only for smoothing-parameter selection. The main computational burden for the sparse methods arose during the joint tuning of $\lambda$ and $\tau$. VP--SFPCA required a mean of $8.11$ seconds for this stage, compared with $860.23$ seconds for SCAD--SFPCA. The stage-specific timings indicate that the large runtime difference between the two sparse methods arose predominantly during the joint tuning of $\lambda$ and $\tau$. These timings pertain to the present implementations and computational settings.

\clearpage


\section{Serum Application}
\label{sec:supp-hcc}

This section presents an additional empirical application of VP--SFPCA to serum SERS spectra from patients with hepatocellular carcinoma (HCC, labelled H0T) and healthy blood donors (labelled CTR). The analysis follows the same unsupervised comparison framework used for the adenine application, with conventional FPCA and SCAD--SFPCA included as benchmark methods. Repeated stratified five-fold cross-validation was used to preserve the balance between the H0T and CTR groups. Preliminary sensitivity analyses based on five repetitions were used to assess the basis dimension, iteration cap, and tuning grids. The selected settings were fixed before the full $50$-repetition analysis. The results below examine estimated component functions, held-out reconstruction, coefficient sparsity, variance representation, localisation stability, and computation time.

\FloatBarrier

\subsection{Dataset and Preprocessing}
\label{subsec:supp-hcc-data}

The serum dataset was obtained from the publicly available data \citep{gurian2021data} associated with the study of \citet{gurian2021}. It comprises label-free SERS spectra from $144$ male participants, including $72$ patients with HCC and $72$ healthy blood donors. Serum samples from the HCC group were collected at diagnosis before treatment. Fasting blood samples were obtained from all participants, and the separated serum was stored at $-80\,^{\circ}\mathrm{C}$ until SERS analysis.

For spectral acquisition, $5\,\mu\mathrm{L}$ of serum was deposited directly onto silver nanoparticle-coated plasmonic paper and allowed to dry before measurement. Spectra were acquired using $785\,\mathrm{nm}$ excitation, with three technical replicate spectra collected for each participant and averaged before preprocessing and analysis. Measurements were carried out over five acquisition days using three substrate batches. The acquisition design was stratified so that both H0T and CTR samples, as well as samples from each substrate batch, were represented on each measurement day, thereby reducing potential confounding between disease group and the acquisition conditions \citep{gurian2021}.

Preprocessing followed the procedure described by \citet{gurian2021}. The spectra were first restricted to $400$--$1800\,\mathrm{cm}^{-1}$ and interpolated by local polynomial regression onto a common Raman-shift grid with $2\,\mathrm{cm}^{-1}$ spacing. Baseline correction was then performed using modified polynomial fitting with polynomial degree $4$. The spectral domain was subsequently restricted to $430$--$1730\,\mathrm{cm}^{-1}$ to remove possible edge artefacts arising from baseline correction, after which the spectra were vector-normalised. Following averaging of the technical replicates and preprocessing, the dataset comprised $144$ participant-level spectra evaluated at $651$ Raman-shift values.

\begin{figure*}[htbp]
	\centering
	\includegraphics[width=0.80\textwidth]{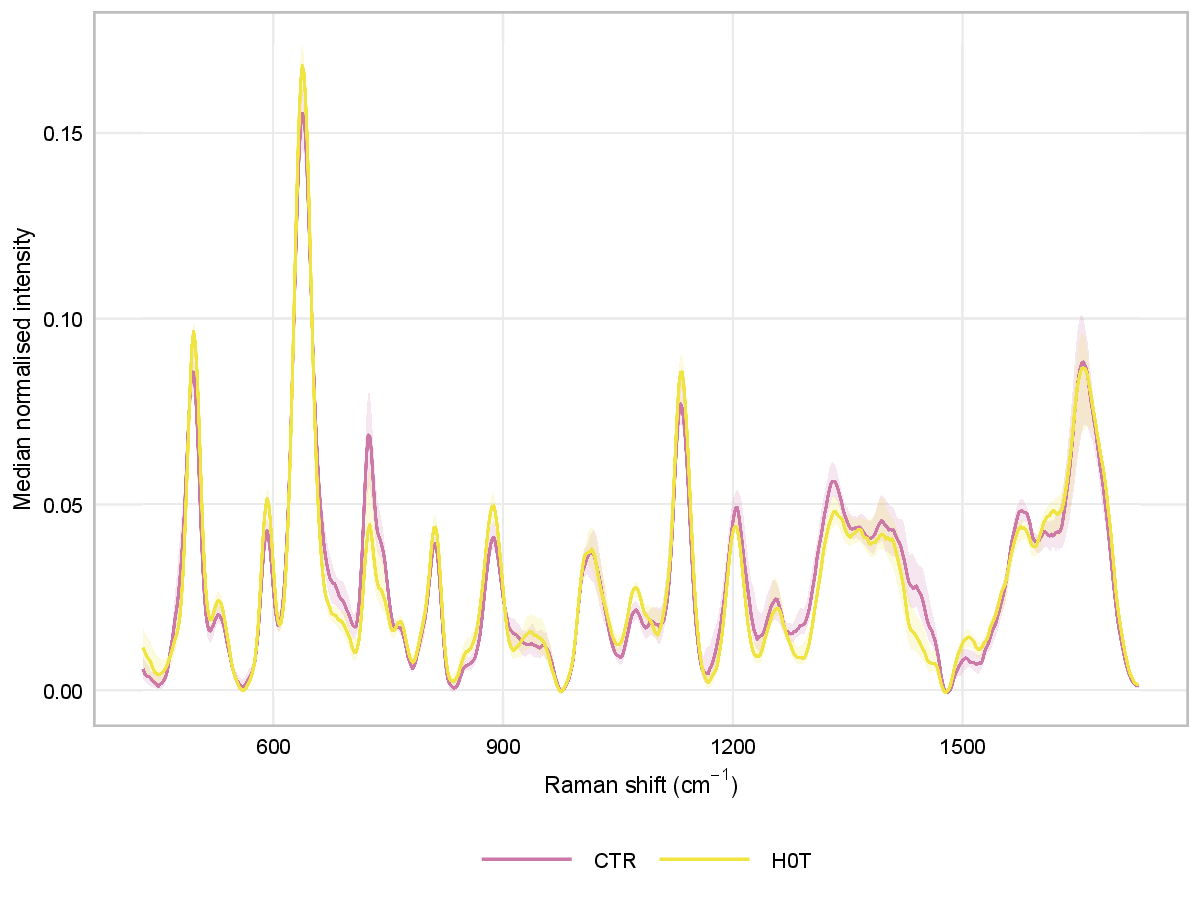}
	\caption{Median preprocessed SERS spectra for the serum dataset. Solid lines represent the group-specific medians, and shaded regions represent the corresponding interquartile ranges.}
	\label{fig:supp-hcc-spectra}
\end{figure*}

Figure~\ref{fig:supp-hcc-spectra} presents the preprocessed spectra for the two study groups. The group-wise median spectra show broadly similar overall spectral profiles, with differences in relative intensity at several Raman shifts. The corresponding interquartile ranges illustrate the between-participant variation within each group.

\FloatBarrier

\subsection{Analysis Design and Rank Selection}
\label{subsec:supp-hcc-design}

The spectra were analysed using repeated stratified five-fold cross-validation. Stratification preserved the proportions of the H0T and CTR groups across folds. The same partitions were applied to FPCA, VP--SFPCA, and SCAD--SFPCA. Within each training--test evaluation, data-dependent smoothing and method-specific tuning were based on the training set, while the corresponding held-out spectra were used for evaluation.

A common rank was used for all three methods to ensure a rank-matched comparison. Under conventional FPCA fitted to the complete dataset, $K = 4$ was the smallest number of components accounting for at least $80\%$ of the total variation, with the first four components accounting for $80.06\%$. The same criterion was applied separately to conventional FPCA fitted to each of the $250$ training samples as a diagnostic. It selected $K = 4$ in $72.0\%$ of the training samples and $K = 5$ in the remaining $28.0\%$. Accordingly, a common rank of $K = 4$ was retained for all methods and cross-validation fits in the serum analysis.

\FloatBarrier

\subsection{Pilot Study: Basis-Dimension and Iteration-Cap Assessment}
\label{subsec:supp-hcc-pilot-iter}

A preliminary sensitivity analysis based on five repetitions of stratified five-fold cross-validation was conducted to assess the basis dimension and iteration cap across the three candidate tuning-grid sets given in Table~\ref{tab:supp-hcc-pilot-grids}. 

\begin{table}[H]
	\centering
	\caption{Candidate tuning grids examined in the serum pilot analysis.}
	\label{tab:supp-hcc-pilot-grids}
	\begin{tabular*}{\textwidth}{@{\extracolsep\fill}llccc@{\extracolsep\fill}}%
		\toprule
		\textbf{Set} & \textbf{Method} & $\boldsymbol{\gamma}$ & $\boldsymbol{\lambda}$ & $\boldsymbol{\tau}$ \\
		\midrule
		\multirow{2}{*}{1}
		& VP--SFPCA
		& $\mathcal{G}_{7}(-4,2)$
		& $\mathcal{G}_{15}(-5,-1)$
		& $\mathcal{G}_{15}(-5,-1)$ \\
		& SCAD--SFPCA
		& $\mathcal{G}_{7}(-4,2)$
		& $\mathcal{G}_{15}(-3,1)$
		& $\mathcal{G}_{15}(-3,1)$ \\
		\midrule
		\multirow{2}{*}{2}
		& VP--SFPCA
		& $\mathcal{G}_{9}(-4,4)$
		& $\mathcal{G}_{15}(-5,-1)$
		& $\mathcal{G}_{15}(-7,-1)$ \\
		& SCAD--SFPCA
		& $\mathcal{G}_{9}(-4,4)$
		& $\mathcal{G}_{15}(-5,-1)$
		& $\mathcal{G}_{15}(-5,-1)$ \\
		\midrule
		\multirow{2}{*}{3}
		& VP--SFPCA
		& $\mathcal{G}_{9}(-4,4)$
		& $\mathcal{G}_{15}(-4,-1)$
		& $\mathcal{G}_{15}(-4,-1)$ \\
		& SCAD--SFPCA
		& $\mathcal{G}_{9}(-4,4)$
		& $\mathcal{G}_{15}(-5,-1)$
		& $\mathcal{G}_{15}(-5,-1)$ \\
		\bottomrule
	\end{tabular*}
\end{table}

Figure~\ref{fig:supp-hcc-pilot-iter} presents the resulting accuracy--computation trade-offs for VP--SFPCA and SCAD--SFPCA in terms of mean held-out ISE and mean computation time. The sensitivity patterns were broadly consistent across the three grid specifications. For the basis-dimension assessment, $p \in \{30,40,50\}$ was examined with the iteration cap fixed at $40$. Across the three candidate grid sets, $p = 30$ produced substantially higher mean held-out ISE for both sparse methods. For VP--SFPCA, the mean held-out ISE was similar for $p = 40$ and $p = 50$, whereas for SCAD--SFPCA, $p = 50$ yielded the lowest mean held-out ISE. The larger basis dimension required greater computation time, particularly for SCAD--SFPCA.

For the iteration-cap assessment, caps of $30$, $40$, and $50$ were examined with the basis dimension fixed at $p = 40$. Differences in mean held-out ISE across the three caps were relatively small. Increasing the iteration cap beyond $30$ did not produce a consistent reduction in held-out ISE across the methods and grid sets. Based on these results, $p = 50$ and an iteration cap of $30$ were retained for the subsequent tuning-grid assessment and full analysis.

\begin{figure*}[htbp]
	\centering
	\subfloat[]{%
		\includegraphics[width=0.80\textwidth]{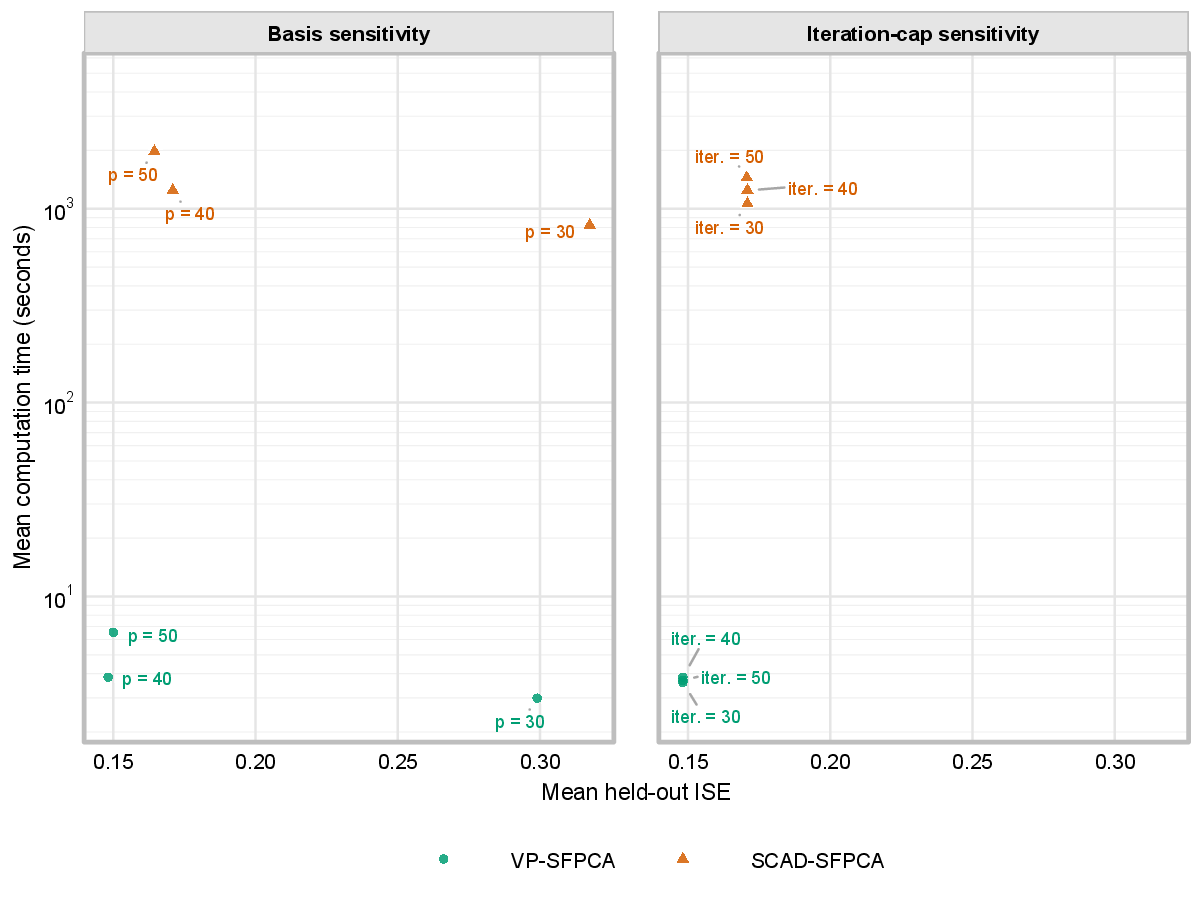}%
		\label{fig:supp-hcc-pilot-iter-set1}%
	}\\[1em]
    \subfloat[]{%
		\includegraphics[width=0.80\textwidth]{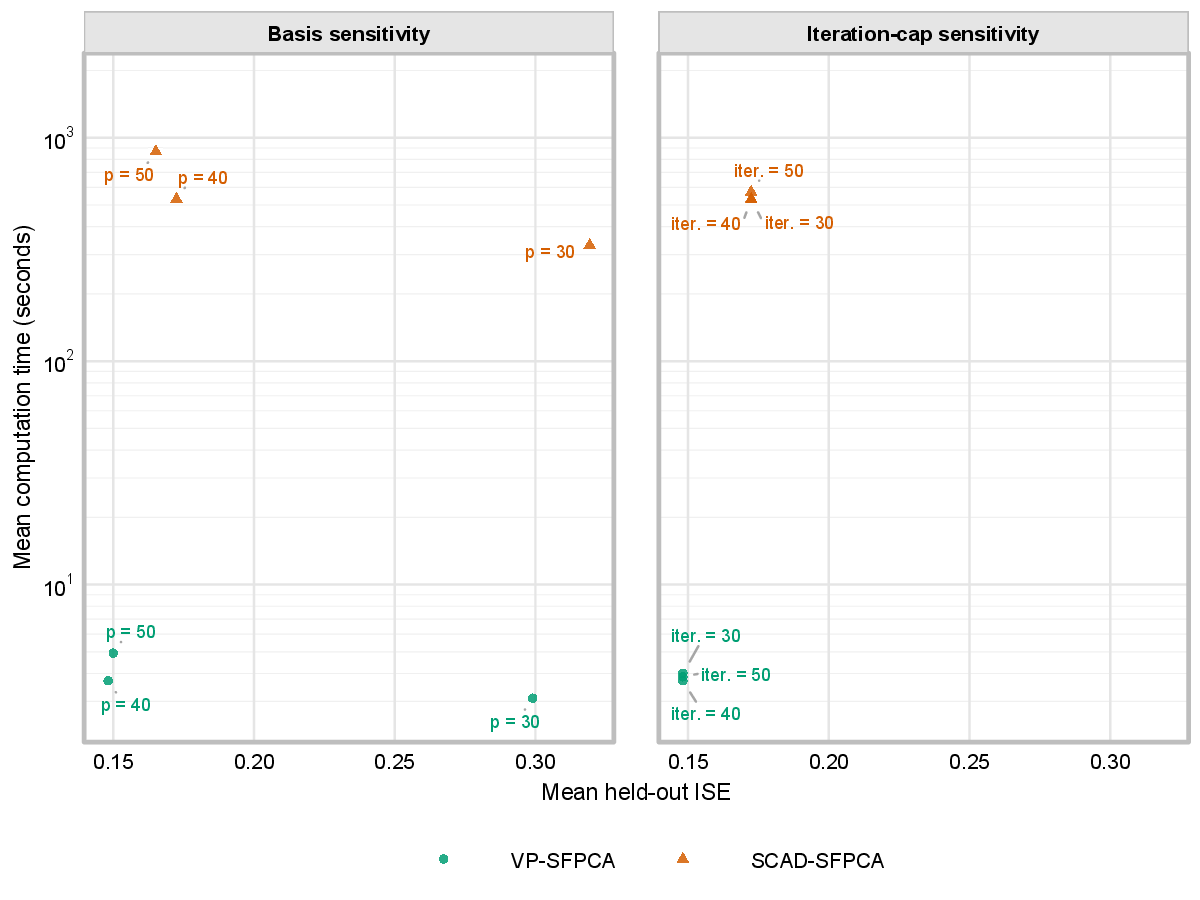}%
		\label{fig:supp-hcc-pilot-iter-set2}%
	}
	\caption{Accuracy--computation trade-offs for the basis-dimension and iteration-cap sensitivity analyses for the serum dataset under the candidate tuning-grid sets: (a) Set~1, (b) Set~2, and (c) Set~3, based on five repetitions of stratified five-fold cross-validation.}
	\label{fig:supp-hcc-pilot-iter}
\end{figure*}

\begin{figure*}[htbp]
	\ContinuedFloat
	\centering
	\subfloat[]{%
		\includegraphics[width=0.80\textwidth]{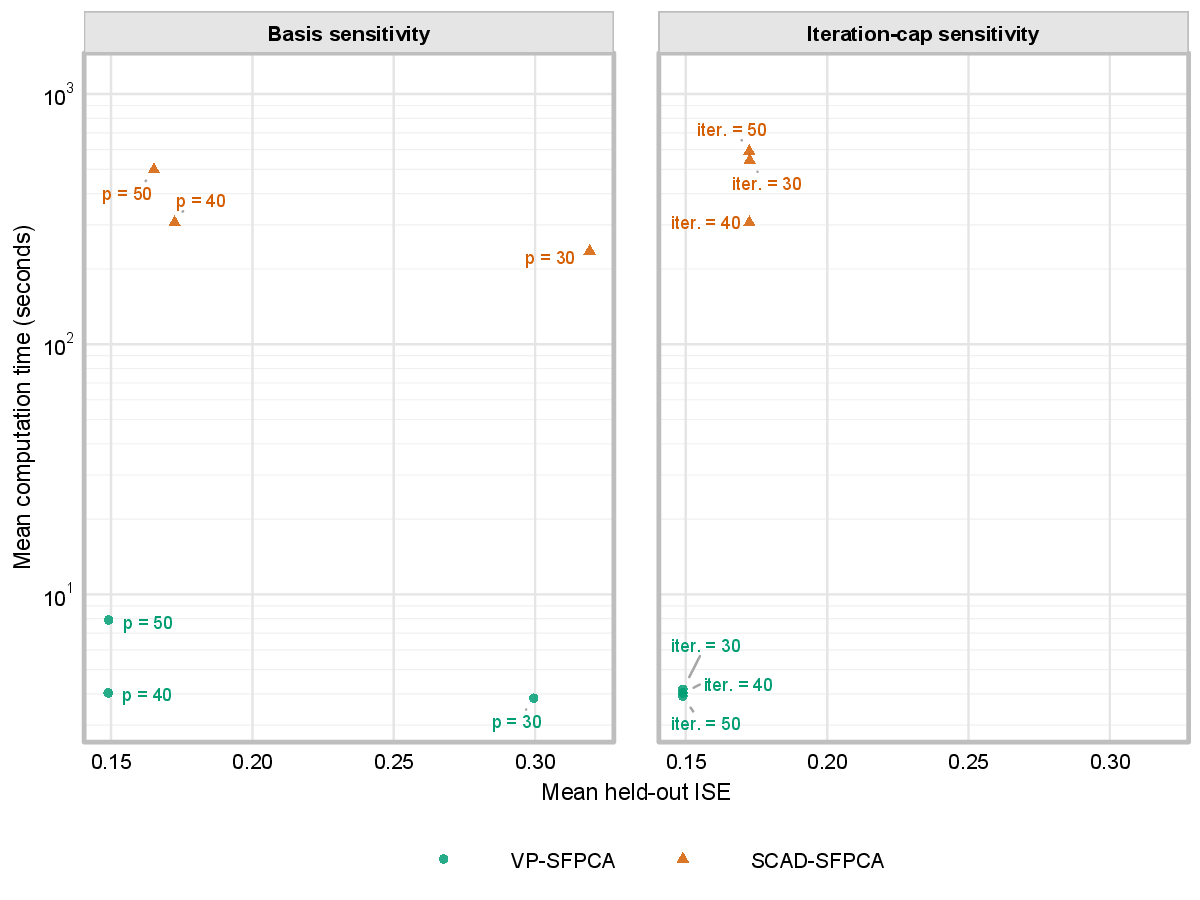}%
		\label{fig:supp-hcc-pilot-iter-set3}%
	}
	\caption[]{(Continued)}
\end{figure*}

\FloatBarrier

\subsection{Pilot Study: Tuning-Grid Assessment}
\label{subsec:supp-hcc-pilot-grid}

With the basis dimension fixed at $p = 50$ and the iteration cap at $30$, the three candidate tuning-grid sets given in Table~\ref{tab:supp-hcc-pilot-grids} were further assessed using five repetitions of stratified five-fold cross-validation, giving $25$ fits per method for each set. The assessment focused on boundary-constrained tuning-parameter selection and held-out reconstruction performance. 

The locations of the selected tuning parameters within each candidate grid are summarised in Table~\ref{tab:supp-hcc-pilot-boundary}. Under Set~1, $\gamma$ was selected at its upper boundary in all pilot fits for both sparse methods. Although $\lambda$ was selected in the interior throughout, $\tau$ was selected at its lower boundary in $92\%$ of the VP--SFPCA fits and $72\%$ of the SCAD--SFPCA fits.

Under Set~2, all $\gamma$ selections moved to the interior. For SCAD--SFPCA, both $\lambda$ and $\tau$ were selected in the interior in all pilot fits. VP--SFPCA $\lambda$ likewise remained in the interior throughout, whereas $\tau$ was selected at its lower boundary in every fit despite extending the grid towards substantially smaller values. For Set~3, $\gamma$ and $\lambda$ remained in the interior for both sparse methods, and SCAD--SFPCA $\tau$ was also selected in the interior throughout. VP--SFPCA $\tau$ continued to be selected predominantly at its lower boundary, with $96\%$ of the selections occurring there.

\begin{table}[H]
	\centering
	\caption{Boundary and interior selection frequencies (\%) across tuning grids for the serum dataset.}
	\label{tab:supp-hcc-pilot-boundary}
	\begin{tabular*}{\textwidth}{@{\extracolsep\fill}lllccc@{\extracolsep\fill}}%
		\toprule
		\textbf{Set} & \textbf{Method} & \textbf{Parameter} &
		\textbf{Lower boundary} & \textbf{Interior} & \textbf{Upper boundary} \\
		\midrule
		\multirow{6}{*}{1}
		& \multirow{3}{*}{VP--SFPCA}
		& $\gamma$  & 0  & 0   & 100 \\
		& & $\lambda$ & 0  & 100 & 0 \\
		& & $\tau$    & 92 & 8   & 0 \\
		\cmidrule(lr){2-6}
		& \multirow{3}{*}{SCAD--SFPCA}
		& $\gamma$  & 0  & 0   & 100 \\
		& & $\lambda$ & 0  & 100 & 0 \\
		& & $\tau$    & 72 & 28  & 0 \\
		\midrule
		\multirow{6}{*}{2}
		& \multirow{3}{*}{VP--SFPCA}
		& $\gamma$  & 0   & 100 & 0 \\
		& & $\lambda$ & 0   & 100 & 0 \\
		& & $\tau$    & 100 & 0   & 0 \\
		\cmidrule(lr){2-6}
		& \multirow{3}{*}{SCAD--SFPCA}
		& $\gamma$  & 0 & 100 & 0 \\
		& & $\lambda$ & 0 & 100 & 0 \\
		& & $\tau$    & 0 & 100 & 0 \\
		\midrule
		\multirow{6}{*}{3}
		& \multirow{3}{*}{VP--SFPCA}
		& $\gamma$  & 0  & 100 & 0 \\
		& & $\lambda$ & 0  & 100 & 0 \\
		& & $\tau$    & 96 & 4   & 0 \\
		\cmidrule(lr){2-6}
		& \multirow{3}{*}{SCAD--SFPCA}
		& $\gamma$  & 0 & 100 & 0 \\
		& & $\lambda$ & 0 & 100 & 0 \\
		& & $\tau$    & 0 & 100 & 0 \\
		\bottomrule
	\end{tabular*}
\end{table}

Held-out reconstruction performance is summarised in Table~\ref{tab:supp-hcc-pilot-grid-ise}. The mean held-out ISE was broadly similar across the three candidate grid sets for both sparse methods, indicating that the modifications to the tuning ranges had little influence on reconstruction performance in the pilot analysis.

\begin{table}[H]
	\centering
	\caption{Mean held-out ISE under the candidate tuning grids for the serum dataset.}
	\label{tab:supp-hcc-pilot-grid-ise}
	\begin{tabular*}{\textwidth}{@{\extracolsep\fill}lcc@{\extracolsep\fill}}%
		\toprule
		\textbf{Set} & \textbf{VP--SFPCA} & \textbf{SCAD--SFPCA} \\
		\midrule
		1 & 0.1500 & 0.1645 \\
		2 & 0.1493 & 0.1664 \\
		3 & 0.1491 & 0.1651 \\
		\bottomrule
	\end{tabular*}
\end{table}

Taken together, Set~1 showed clear boundary constraints. Set~2 removed the boundary selection for $\gamma$ and the SCAD--SFPCA tuning parameters, although extending the VP--SFPCA $\tau$ grid towards smaller values did not resolve its lower-boundary selection. Set~3 retained the improved selection behaviour for $\gamma$ and SCAD--SFPCA while using a more compact VP--SFPCA tuning range and yielding similar held-out reconstruction performance. As a result, the tuning grids in Set~3 were used for the full serum analysis.

\FloatBarrier

\subsection{Estimated Component Functions}
\label{subsec:supp-hcc-estimation}

\begin{figure*}[htbp]
	\centering
	\includegraphics[width=0.80\textwidth]{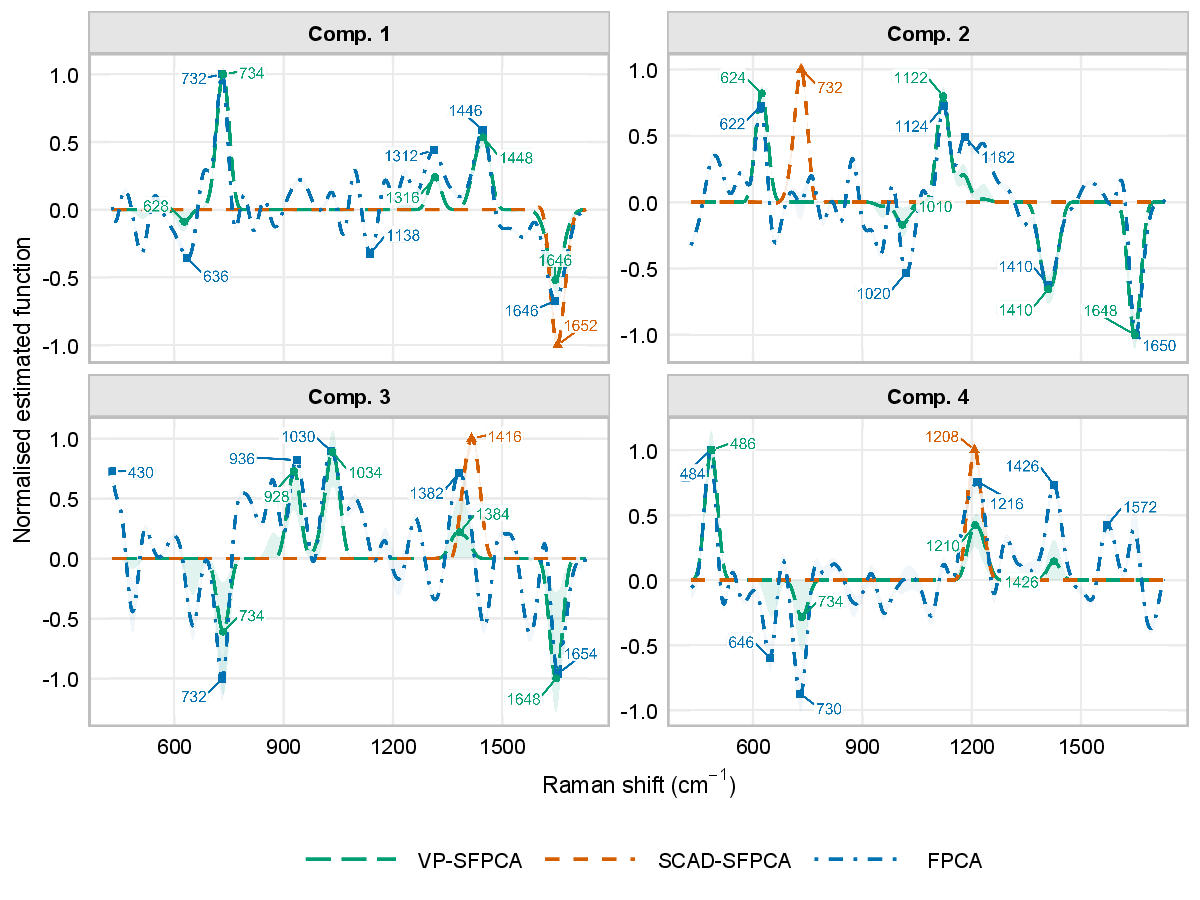}
	\caption{Median estimated functions across $250$ training--test fits from $50$ repetitions of stratified five-fold cross-validation for the serum dataset, with a common rank of $K = 4$ selected as the smallest number of components accounting for at least $80\%$ of the total variation under conventional FPCA. The proposed VP--SFPCA (green long-dashed lines) and the benchmark methods, SCAD--SFPCA (orange dashed lines) and conventional FPCA (blue dot-dashed lines), are shown for comparison. Labels indicate local extrema of the absolute median estimated functions. Shaded regions indicate variability across fits.}
	\label{fig:supp-hcc-estimated-functions}
\end{figure*}

Figure~\ref{fig:supp-hcc-estimated-functions} compares the median estimated functions obtained across $250$ training--test fits. For VP--SFPCA, these correspond to the sparse score-generating weight functions, whereas for SCAD--SFPCA and conventional FPCA they correspond to the sparse and dense FPCs, respectively. Before aggregation, the estimated components were permutation-matched within each method to a common reference by maximising the total absolute functional similarity across component permutations and were subsequently sign-aligned. Non-identity permutations occurred in $14.4\%$ of FPCA fits, $21.2\%$ of VP--SFPCA fits, and $30.8\%$ of SCAD--SFPCA fits. The labelled Raman shifts denote local extrema identified from the absolute median functions.

Across the four components, VP--SFPCA generally retained localised features that corresponded to extrema in the conventional FPCs, including recurring features around $628$--$636$, $730$--$734$, $1122$--$1124$, $1410$--$1448$, and approximately $1650\,\mathrm{cm}^{-1}$. SCAD--SFPCA produced more concentrated component functions, with dominant features near $1652$, $732$, $1416$, and $1208\,\mathrm{cm}^{-1}$ for components~1--4, respectively.

Several of the localised regions lie close to serum SERS bands highlighted by \citet{gurian2021} in their comparison of HCC and control spectra. Their difference-spectrum analysis associated positive bands near $594$, $638$, $812$, $888$, and $1132\,\mathrm{cm}^{-1}$ with uric acid and a negative band near $724\,\mathrm{cm}^{-1}$ with hypoxanthine. In the present analysis, estimated features around $628$--$636$, $1122$--$1138$, and $730$--$734\,\mathrm{cm}^{-1}$ lie close to these reported regions. The recurrent feature near $730\,\mathrm{cm}^{-1}$ is particularly notable, since \citet{gurian2021} also identified hypoxanthine- and uric-acid-associated bands in the principal-component loadings used to distinguish between the HCC and control groups. Recent experimental work has provided further evidence that uric acid and hypoxanthine are the primary contributors to the characteristic SERS spectrum of serum \citep{gobbato2025}.

Other estimated regions correspond closely to bands discussed by \citet{gurian2021} in relation to ergothioneine and glutathione. Their difference-spectrum analysis associated negative bands near $480$, $1220$, $1442$, and $1582\,\mathrm{cm}^{-1}$ with ergothioneine and bands near $664$ and $912\,\mathrm{cm}^{-1}$ tentatively with glutathione. In the present estimated functions, features were observed near $484$--$486$, $1210$--$1216$, and $1446$--$1448\,\mathrm{cm}^{-1}$, with an additional conventional FPCA feature around $1572\,\mathrm{cm}^{-1}$. \citet{gurian2021} further reported that ergothioneine- and tentatively glutathione-associated bands were less intense in HCC sera and discussed these differences in relation to possible changes in oxidative status. They also noted the proposed antioxidant role of ergothioneine, while acknowledging that its biological role remains uncertain.

These biochemical associations provide spectroscopic context for the localised regions identified by the sparse methods. However, the FPCA and sparse FPCA models were fitted without using the class labels. Accordingly, the resulting components characterise dominant between-spectrum variation rather than direct HCC--control differences or group-specific changes in metabolite abundance.

\FloatBarrier

\subsection{Performance Comparison}
\label{subsec:supp-hcc-performance}

Table~\ref{tab:supp-hcc-performance} summarises coefficient sparsity, cumulative PVE, held-out reconstruction performance, and computation time across $250$ training--test fits. Both sparse methods produced highly sparse representations, whereas the estimated component functions from conventional FPCA remained dense. Mean coefficient sparsity was $83.8\%$ for VP--SFPCA and $88.6\%$ for SCAD--SFPCA, indicating a slightly sparser coefficient representation under SCAD--SFPCA. The mean cumulative adjusted PVE was $54.4\%$ for VP--SFPCA and $34.8\%$ for SCAD--SFPCA, while the mean conventional cumulative PVE for FPCA was $80.4\%$. Thus, VP--SFPCA represented a larger proportion of variation than SCAD--SFPCA under the adjusted PVE, while retaining a highly sparse representation.

Conventional FPCA yielded the lowest mean held-out ISE at $0.1325$, followed by VP--SFPCA at $0.1475$ and SCAD--SFPCA at $0.1658$. The mean held-out ISE for VP--SFPCA was approximately $11.0\%$ lower than that for SCAD--SFPCA, although it remained higher than the dense FPCA benchmark. The sparse methods therefore showed some loss in held-out reconstruction relative to conventional FPCA, while VP--SFPCA retained lower reconstruction error than SCAD--SFPCA.

\begin{table}[H]
	\centering
	\caption{Performance comparison of VP--SFPCA, SCAD--SFPCA, and conventional FPCA for the serum dataset across $250$ training--test fits.}
	\label{tab:supp-hcc-performance}
	\begin{tabular*}{\textwidth}{@{\extracolsep{\fill}}lcccc}
		\toprule
		\textbf{Method}
		& \textbf{Sparsity (\%)}
        & \textbf{Cumulative PVE (\%)}
        & \textbf{Held-out ISE}
		& \textbf{Runtime (s)} \\
		\midrule
		VP--SFPCA   & 83.8 & 54.4 & 0.1475 & 6.67 \\
		SCAD--SFPCA & 88.6 & 34.8 & 0.1658 & 801.23 \\
		FPCA        & 0.0  & 80.4 & 0.1325 & 0.22 \\
		\bottomrule
	\end{tabular*}
	\begin{tablenotes}[flushleft]
		\footnotesize
		\item[] \textit{Note:} A common rank of $K = 4$, selected as the smallest number of components accounting for at least $80\%$ of the total variation under conventional FPCA, was retained for all methods. All reported values are means across $250$ training--test fits. For VP--SFPCA and SCAD--SFPCA, cumulative adjusted PVE is reported to account for correlation among the sparse component scores, whereas the conventional cumulative PVE is reported for FPCA.
	\end{tablenotes}
\end{table}

\begin{figure*}[htbp]
	\centering
	\includegraphics[width=0.80\textwidth]{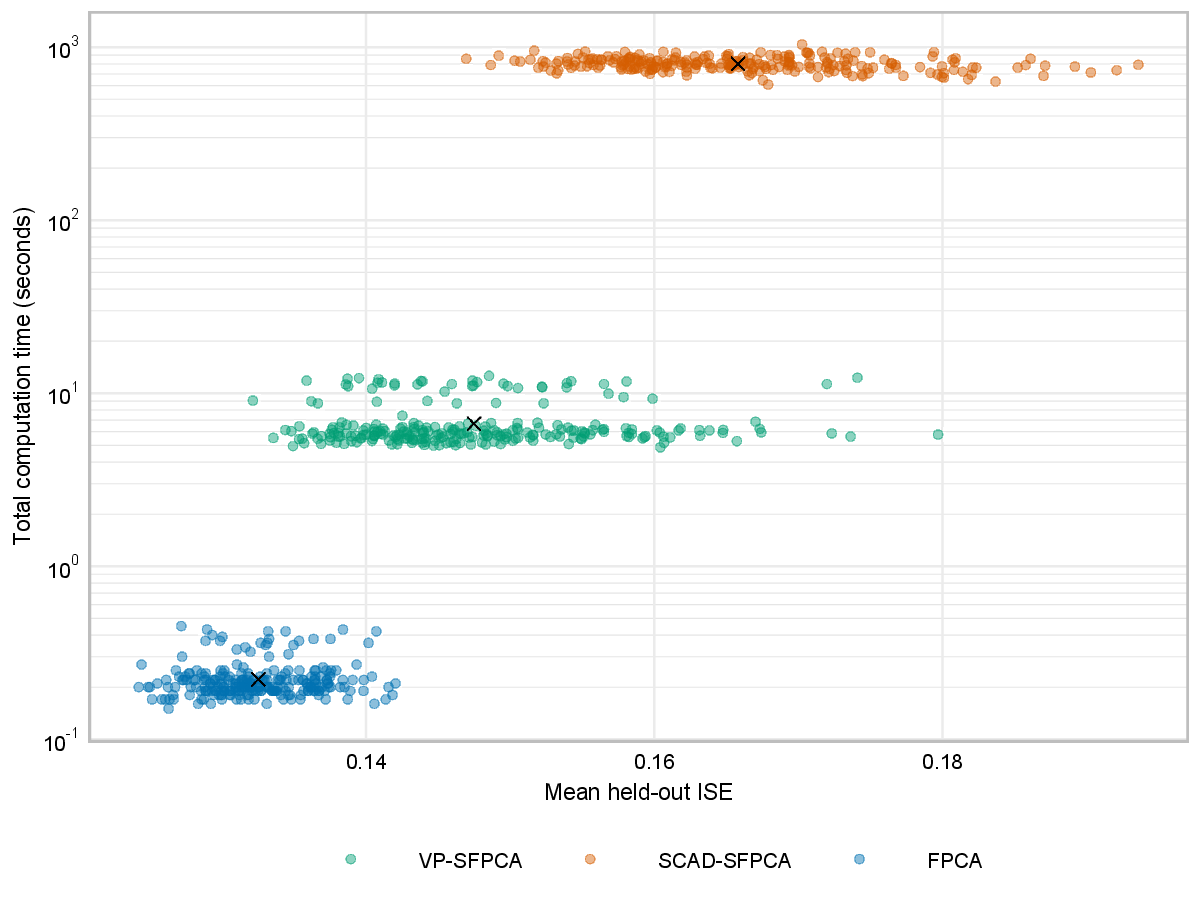}
    \caption{Mean held-out ISE versus computation time for the serum dataset. Each coloured point represents one training--test fit, and black crosses denote method-specific means across $250$ fits.}
	\label{fig:supp-hcc-accuracy-time}
\end{figure*}

Figure~\ref{fig:supp-hcc-accuracy-time} further illustrates the relationship between held-out reconstruction and total computation time. As in the adenine application, conventional FPCA was the least computationally demanding method and achieved the lowest reconstruction error. Among the sparse methods, VP--SFPCA combined lower held-out ISE with substantially shorter computation time than SCAD--SFPCA. The mean total runtime was $6.67$ seconds for VP--SFPCA and $801.23$ seconds for SCAD--SFPCA, indicating a substantial computational advantage for the proposed method under the present implementations and computational settings.

\FloatBarrier

\subsection{Stability of Localisation}
\label{subsec:supp-hcc-localisation}

\begin{figure*}[htbp]
	\centering
	\includegraphics[width=0.80\textwidth]{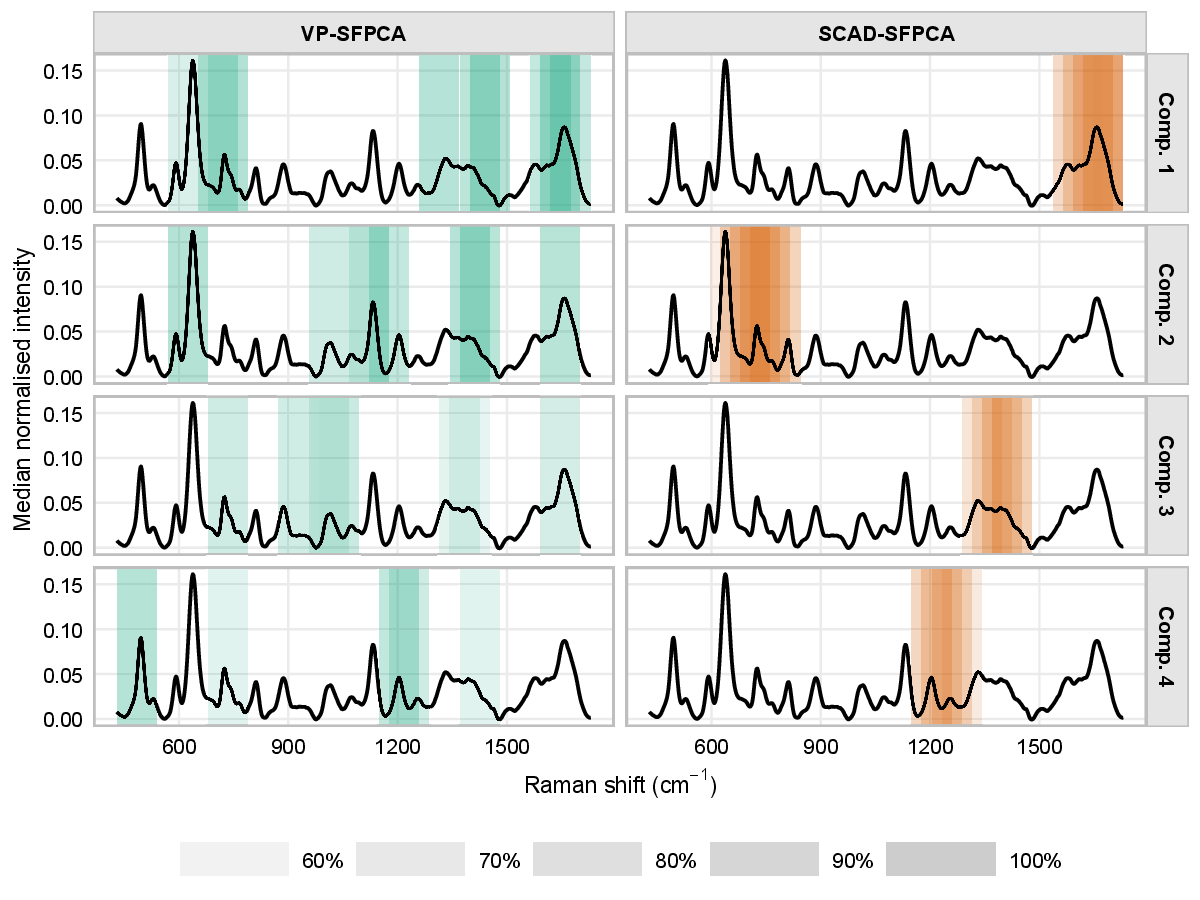}
    \caption{Selection frequencies of localised regions identified by VP--SFPCA and SCAD--SFPCA across $250$ training--test fits for the serum dataset. Shaded regions denote Raman-shift intervals selected in at least $60\%$ of the fits, with darker shading representing higher selection frequencies. Black curves show the median preprocessed serum spectrum.}
	\label{fig:supp-hcc-localisation}
\end{figure*}

The stability of localisation was assessed using component-specific selection frequencies across $250$ training--test fits. Before aggregation, the same within-method component mapping used for the estimated-function summaries in Figure~\ref{fig:supp-hcc-estimated-functions} was applied to the component labels. Figure~\ref{fig:supp-hcc-localisation} shows the frequency with which Raman-shift regions were selected by the two sparse methods, together with the median preprocessed serum spectrum for each component. Darker shading indicates more frequent selection. Both sparse methods identified regions that were repeatedly selected across the cross-validation fits, although the extent and location of these regions differed across methods and components.

\FloatBarrier

\subsection{Computation Time by Tuning Stage}
\label{subsec:supp-hcc-time-stage}

\begin{figure*}[htbp]
	\centering
	\includegraphics[width=0.80\textwidth]{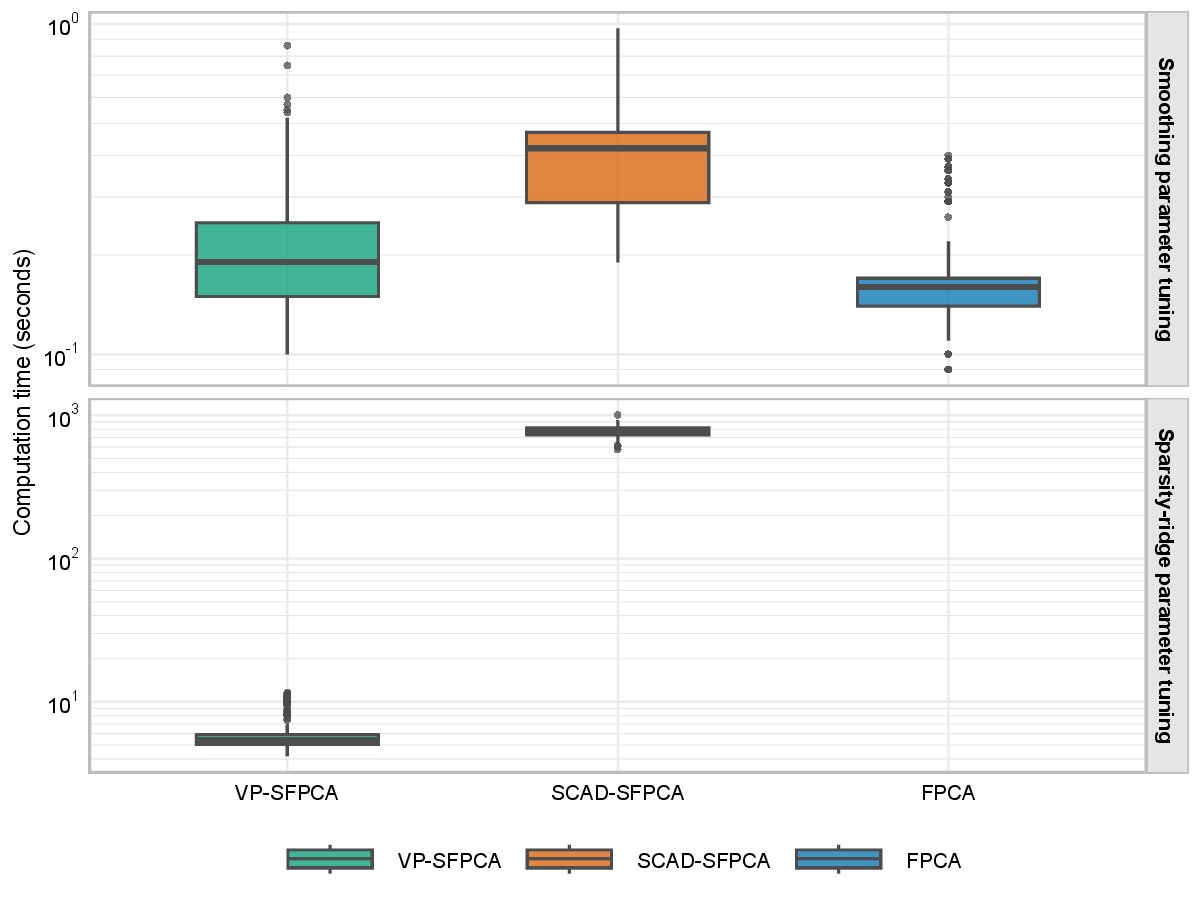}
    \caption{Distributions of computation time by tuning stage across $250$ fits of the repeated stratified five-fold cross-validation analysis for the serum dataset.}
	\label{fig:supp-hcc-time-stage}
\end{figure*}

Figure~\ref{fig:supp-hcc-time-stage} summarises computation time by tuning stage across the $250$ fits. Smoothing-parameter selection required comparatively little computation for all three methods. Conventional FPCA does not involve sparsity--ridge tuning and is therefore shown only for smoothing-parameter selection. A more pronounced computational difference between the two sparse methods occurred during the joint tuning of $\lambda$ and $\tau$. VP--SFPCA required a mean of $6.08$ seconds for this stage, compared with $771.75$ seconds for SCAD--SFPCA. The stage-specific timings indicate that the overall runtime difference between the sparse methods arose predominantly during the joint tuning of $\lambda$ and $\tau$. These timings pertain to the present implementations and computational settings.

\FloatBarrier


\bibliographystyle{plainnat}
\bibliography{referencesSupp}